\documentstyle[preprint,prd,aps,eqsecnum,psfig]{revtex}
\tightenlines

\draft
\newcommand{\be}{\begin{eqnarray}}
\newcommand{\ee}{\end{eqnarray}}
\newcommand{\ba}{\begin{array}}
\newcommand{\ea}{\end{array}}
\newcommand{\no}{\nonumber}

\begin{document}
\draft
\date{\today}
\preprint{
$\ba{r}
\mbox{HUPD-0001}\\
\mbox{KEK-CP-101}
\ea$}
\title{
  Renormalization of the $\Delta B$=2 four-quark operators
  in lattice NRQCD
  }
\author{
  S.~Hashimoto$^a$,
  K-I.~Ishikawa$^a$
  T.~Onogi$^b$,
  M.~Sakamoto$^b$,
  N.~Tsutsui$^a$,
  N.~Yamada$^a$
  }
\address{
  High Energy Accelerator Research Organization (KEK),
  Tsukuba 305-0801, Japan$^a$\\
  Department of Physics, Hiroshima University,
  Higashi-Hiroshima 739-8526, Japan$^b$
  }
\maketitle

\begin{abstract}
  We calculate perturbative renormalization constants for the
  $\Delta B$=2 four-quark operators in lattice NRQCD.
  Continuum operators 
  $\bar{b}\gamma_{\mu}(1-\gamma_5)q~
   \bar{b}\gamma_{\mu}(1-\gamma_5)q$ and
  $\bar{b}(1-\gamma_5)q~\bar{b}(1-\gamma_5)q$,
  which are necessary in evaluating the
  mass and width differences in $B^0_{d(s)}-\bar{B}^0_{d(s)}$ 
  systems, are matched at
  one-loop with corresponding lattice operators constructed
  from the NRQCD heavy quarks and the ${\cal O}(a)$-improved
  light quarks. 
  Using these perturbative coefficients, we also reanalyse
  our previous simulation
  results for the matrix elements of the above operators.
  Our new results are free from the systematic
  error of ${\cal O}(\alpha_s/(aM_b))$ in contrast to the previous 
  ones with matching coefficients evaluated in the static
  limit. 
\end{abstract}

\section{Introduction}
\label{sec:Introduction}

The $B$ meson decay constant and the $B$ parameter in the
$B-\overline{B}$  mixing are crucial quantities for
determining the Cabbibo-Kobayashi-Maskawa (CKM) mixing
matrix elements $|V_{td}|$ and $|V_{ts}|$ from the
experimental values of the oscillation frequency 
$\Delta M_{d(s)}$. 
While the lattice calculation of the decay constant has
reached a satisfactory level where the systematic error
except for the quenching effect is about 10\%, the $B$
parameter $B_B$ still has a large uncertainty of about 30\%
even in the quenched approximation \cite{Hashimoto_lat99}. 
Further effort in the lattice calculation is required to
constrain the CKM matrix elements more tightly.

In the limit of infinitely heavy quark mass, lattice
calculation of the $B$ parameter has been performed by
several authors using the static action and the ${\cal
  O}(a)$-improved (or unimproved) light quark actions 
\cite{UKQCD_96,Gimenez_Martinelli_97,%
  Christensen_Draper_McNeile_97,Gimenez_Reyes_99},
for which the perturbative matching factor of relevant
four-quark operators in continuum and lattice definitions is
available 
\cite{Flynn_Hernandez_Hill_91,Borrelli_Pittori_92,%
  DiPierro_Sachrajda_98,Ishikawa_Onogi_Yamada_99}.
The problem of large one-loop coefficient raised in
Refs.~\cite{UKQCD_96,Gimenez_Martinelli_97} is not essential
when used with the tadpole improved perturbation theory
\cite{Lepage_Mackenzie_93} as discussed in
Refs.~\cite{Christensen_Draper_McNeile_97,Gimenez_Reyes_99},
where they find that the results of several groups are in
reasonable agreement.

The next step towards the final prediction is to incorporate
the correction from finite $b$ quark mass $M_b$, which can
be systematically included using $p/M_b$ expansion, where 
$p$ is the typical momentum of the gluons and quarks. 
A naive order counting suggests that the correction is about
$\Lambda_{QCD}/M_b \sim$ 10\%, when the size of the QCD
scale is assumed to be around 350 MeV.
On the lattice, the nonrelativistic QCD (NRQCD)
\cite{NRQCD_reference} provides a necessary formulation to
calculate the $p/M_b$ corrections, and an exploratory lattice
calculation of the $B$ meson $B$ parameter has already been
made \cite{Hiroshima_B_B_99,Yamada_lat99}. 
One of the main drawbacks in that calculation is, however, 
that the one-loop coefficients for the infinitely heavy
quark mass is used instead of those for lattice NRQCD.
This approximation introduces a large systematic uncertainty
of order $\alpha_s/(aM_b)$, which is as large as 10--20\%
for a typical value of inverse lattice spacing $1/a\sim$
2~GeV and almost equivalent to or even larger than the
physical size of the $p/M_b$ correction itself. 

Calculation of $B_B$ using the relativistic lattice actions
for heavy quark 
\cite{Bernard_et_al_88,Abada_et_al_92,Soni_lat95,%
  Gupta_Bhattacharya_Sharpe_97,Bernard_Blum_Soni_98,%
  UKQCD_lat98,UKQCD_HF8,Becirevic_et_al_00} 
is another possibility to study the finite heavy quark mass
correction. 
These calculations, however, may suffer from large 
${\cal O}(aM)$ (${\cal O}(\alpha_s aM)$ or ${\cal
  O}((aM)^2)$ for the ${\cal O}(a)$-improved actions)
systematic error and the uncertainty in the extrapolation to
the $b$ quark mass from lighter heavy quark masses, for
which simulations are performed.

In this paper, we compute the one-loop renormalization 
constants for $\Delta B$=2 four-quark operators constructed
with the NRQCD heavy quarks and with the ${\cal
  O}(a)$-improved light quarks on the lattice in order to
remove the error of the order $\alpha_s/(aM_b)$ in the
lattice calculation of $B_B$. 
We consider the leading dimension six operators and neglect 
dimension seven operators which would remove errors of
${\cal O}(\alpha_s \Lambda_{QCD}/M_b)$ or 
${\cal O}(\alpha_s a\Lambda_{QCD})$.
The one-loop coefficient for the dimension seven operators
which corresponds to ${\cal O}(\alpha_s a\Lambda_{QCD})$
corrections has been obtained in
Ref.~\cite{Ishikawa_Onogi_Yamada_99} in the infinitely heavy
quark mass limit.

Using the renormalization constants obtained in this work,
we reanalyze the simulation data of Ref. \cite{Hiroshima_B_B_99}
to obtain an improved result for $B_B$, which is free from
the large systematic uncertainty of 
${\cal O}(\alpha_s/(aM_b))$.
The central value is increased by about 12\% with this new
analysis, which is within the size of errors expected by a
naive order counting argument. 

Another important application of our perturbative
work is the lattice calculation of $B_S$, which is a
$B$-parameter necessary to evaluate the width difference in
the $B_{(s)}-\bar{B}_{(s)}$ mixing.
An exploratory lattice NRQCD study with the one-loop
matching in the infinitely heavy quark mass limit is found
in Ref.~\cite{Hiroshima_B_S_99}. 
We reanalyze the data in that work with the renormalization
constant containing the finite heavy quark mass effect.
We find that our new analysis resulted in a change of the value of 
the bag parameter, but it remains within the expected size of 
the error in the previous analysis as is also the case for $B_B$.

This paper is organized as follows.
We summarize the definition of the lattice NRQCD action 
in Sec.~\ref{sec:Lattice_formulation_of_NRQCD}, and the
heavy-light four-quark operators in
Sec.~\ref{sec:Operators}.
Sec.~\ref{sec:One-loop_calculation} is the main part of this
paper, where we present the results of the one-loop matching
calculations for bilinear operators
(\ref{sec:Bilinear_operators}) and for the four-quark
operators (\ref{sec:Four-quark_operators}).
The reanalysis of our previous NRQCD simulations are given
in Sec.~\ref{sec:Physics_results}.
Sec.~\ref{sec:Conclusions} is devoted to our conclusion.
Details of the one-loop calculations are collected in
Appendices.
The Feynman rules for the lattice NRQCD action is given in
Appendix~\ref{sec:Lattice_NRQCD_Feynman_rules}, while
several expressions of one-loop integrals and amplitudes are
summarized in Appendix~\ref{sec:One-loop_integrals} and
\ref{sec:One-loop_four-quark_amplitudes} respectively.

\section{Lattice formulation of NRQCD}
\label{sec:Lattice_formulation_of_NRQCD}

In this section we briefly summarize the definition of the
NRQCD action used in the following perturbative
calculations.
A complete formulation of the lattice NRQCD is found in Ref.
\cite{NRQCD_reference}. 

Our NRQCD action is defined by
\begin{equation}
  \label{eq:NRQCD_action}
  S_{\mbox{\scriptsize NRQCD}}= 
  \sum_{x,y}Q^{\dag}(x)( 1 - K_{Q})(x,y) Q(y)
  + \sum_{x,y}\chi^{\dag}(x)( 1 - K_{\chi})(x,y) \chi(y).
\end{equation}
The nonrelativistic two-component spinor fields $Q$ and
$\chi$ represent a heavy quark and an anti-quark
respectively. 
Their evolution is described by
\footnote{The evolution equations
  (\ref{eq:evolution_operator_Q}) and 
  (\ref{eq:evolution_operator_chi}) are slightly different
  from the definition used, for example, in Ref.
  \cite{Morningstar_93}, where the $(1-aH_0/2n)^n$ terms
  appear inside of the $(1-a\delta H/2)$ terms.}
\begin{eqnarray}
  \label{eq:evolution_operator_Q}
  K_{Q}(x,y) &=& \left[ 
    \left( 1-\frac{a H_{0}}{2 n} \right)^{n}
    \left( 1-\frac{a \delta H}{2} \right)
    \delta^{(-)}_{4}{U^{\dagger}_{4}}
    \left( 1-\frac{a \delta H}{2} \right)
    \left( 1-\frac{a H_{0}}{2 n} \right)^{n}
  \right](x,y), \\
  \label{eq:evolution_operator_chi}
  K_{\chi}(x,y) &=& \left[ 
    \left( 1-\frac{a H_{0}}{2 n} \right)^{n}
    \left( 1-\frac{a \delta H}{2} \right)
    \delta^{(+)}_{4}{U_{4}}
    \left( 1-\frac{a \delta H}{2} \right)
    \left( 1-\frac{a H_{0}}{2 n} \right)^{n}
  \right](x,y),
\end{eqnarray}
where $n$ denotes a stabilization parameter introduced in
order to remove an instability arising from unphysical
momentum modes in the evolution equation.  
Note that following Ref.~\cite{NRQCD_reference} all the link
variable $U_{\mu}$ in the NRQCD action is always divided by 
the mean field value $u_0$ determined from the plaquette
expectation value.
This tadpole improvement will give rise to ${\cal O}(g^2)$ counter
term in the Feynman rule.

The operator $\delta^{(\pm)}_4$ is defined as 
$\delta^{(\pm)}_4(x,y) \equiv \delta_{x_4\pm 1,y_4}
\delta_{\mathbf{x},\mathbf{y}}$,
and the Hamiltonians 
$H_{0}$ and $\delta H$ are 
\begin{eqnarray}
  H_{0}      &=& -\frac{\mathbf{\Delta}^{(2)}}{2 aM_{0}}, \\
  \delta{H}  &=& -c_{B} \frac{g}{2aM_{0}}
                 \mathbf{\sigma}\cdot\mathbf{B},
\end{eqnarray}
where $aM_0$ denotes a bare heavy quark mass in lattice unit.
The operator $\mathbf{\Delta^{(2)}}$ $\equiv$ 
$\sum_{i=1}^{3}\Delta^{(2)}_i$ 
is a Laplacian defined on the lattice through
$\Delta^{(2)}_{i}$, the second symmetric covariant
differentiation operator in the spatial direction $i$.
The space-time indices $x$ and $y$ are implicit in these
expressions. 
The Hamiltonian $\delta H$ represents the effect of the
spin-(chromo)magnetic interaction, in which $\mathbf{B}$ is
the chromomagnetic field defined as a standard clover-leaf
operator. 
$g$ is a gauge coupling, and $c_B$ is a constant to
parametrize the strength of the
$\mathbf{\sigma}\cdot\mathbf{B}$ interaction.
It should be tuned until the NRQCD action reproduces the
same dynamics as that of continuum relativistic action.
We take the tree level value $c_B=1$.
The relativistic four-component Dirac spinor field $b$
is related to the two-component nonrelativistic field $Q$
and $\chi$ appearing in the NRQCD action in
Eq.~(\ref{eq:NRQCD_action}) via the Foldy-Wouthuysen-Tani
(FWT) transformation 
\begin{equation}
  \label{eq:FWT_transformation}
  b(x) = \sum_{z} R(x,z)
  \left( \begin{array}{c}
      Q(z)\\ \chi^{\dag}(z)
    \end{array} \right),
\end{equation}
where $R$ is defined as
\begin{equation}
  \label{eq:R}
  R = 1 -
  \frac{\mathbf{\gamma}\cdot\mathbf{\Delta^{(\pm)}}}{2aM_0},
\end{equation}
where $\Delta^{(\pm)}_i$ is the first symmetric covariant
differentiaon operators in spatial direction.

The Feynman rules derived from the NRQCD action in
Eq.~(\ref{eq:NRQCD_action}) and the FWT transformation
(\ref{eq:FWT_transformation}) are given in the
Appendix~\ref{sec:Lattice_NRQCD_Feynman_rules}.
The light quark action is the 
${\cal O}(a)$-improved Wilson action
\cite{Sheikholeslami_Wohlert_85}, and the gluon action is 
the standard plaquette action. 
The Feynman rules for light quarks and gluons are also
summarized in the Appendix~\ref{sec:Lattice_NRQCD_Feynman_rules}.

\section{Operators}
\label{sec:Operators}

The $B$ parameters 
$B_L$\footnote{We use a notation $B_L$ instead of the usual
  $B_B$ in order to emphasize that it represents a matrix
  element of the ``LL'' operator.}
and $B_S$ are defined using the $\Delta B$=2 four-quark
operators
$\bar{b}\gamma_{\mu}(1-\gamma_5)q~
 \bar{b}\gamma_{\mu}(1-\gamma_5)q$ and
$\bar{b}(1-\gamma_5)q~\bar{b}(1-\gamma_5)q$ respectively.
In the perturbative matching, however, we have to consider
other operators which mix under the radiative correction. 
Since the lattice regularization violates the chiral
symmetry, some operators that do not appear in the matching
between continuum regularizations are also necessary.
We define the following set of operators.
\begin{eqnarray}
  \label{eq:O_VLL}
  O_{VLL} & = &
  \bar{b} \gamma_{\mu} P_L q~\bar{b} \gamma_{\mu} P_L q, \\
  \label{eq:O_VRR}
  O_{VRR} & = & 
  \bar{b} \gamma_{\mu} P_R q~\bar{b} \gamma_{\mu} P_R q, \\
  \label{eq:O_VLR}
  O_{VLR} & = & 
  \bar{b} \gamma_{\mu} P_L q~\bar{b} \gamma_{\mu} P_R q, \\
  \label{eq:O_SLL}
  O_{SLL} & = & 
  \bar{b} P_L q~\bar{b} P_L q, \\
  \label{eq:O_SLR}
  O_{SLR} & = & 
  \bar{b} P_L q~\bar{b} P_R q, \\
  \label{eq:O_tilde_VLL}
  \tilde{O}_{VLL} & = &
  \bar{b} \gamma_{\mu} P_L  T^{a} q~
  \bar{b} \gamma_{\mu} P_L  T^{a} q, \\
  \label{eq:O_tilde_VRR}
  \tilde{O}_{VRR} & = &
  \bar{b} \gamma_{\mu} P_R  T^{a} q~
  \bar{b} \gamma_{\mu} P_R  T^{a} q, \\
  \label{eq:O_tilde_VLR}
  \tilde{O}_{VLR} & = &
  \bar{b} \gamma_{\mu} P_L  T^{a} q~
  \bar{b} \gamma_{\mu} P_R  T^{a} q, \\
  \label{eq:O_tilde_SLL}
  \tilde{O}_{SLL} & = &
  \bar{b} P_L  T^{a} q~\bar{b} P_L  T^{a} q, \\
  \label{eq:O_tilde_SLR}
  \tilde{O}_{SLR} & = &
  \bar{b} P_L  T^{a} q~\bar{b} P_R  T^{a} q,
\end{eqnarray}
where $P_L$ and $P_R$ are chirality projection operators
$P_{L/R}=(1\mp\gamma_5)/2$, and $T^a$ is a generator of the
$SU(N)$ group.
The operators with a tilde contains a summation over the
$SU(N)$ generators $T^a$.
Fierz identities relate the `tilde' operators in
Eqs.~(\ref{eq:O_tilde_VLL})-(\ref{eq:O_tilde_SLR}) to those
without tilde in Eqs.~(\ref{eq:O_VLL})-(\ref{eq:O_SLR}) as 
\begin{eqnarray}
  \label{eq:Fierz_O_VLL}
  \tilde{O}_{VLL} & = & \frac{N-1}{2N} O_{VLL}, \\
  \label{eq:Fierz_O_VRR}
  \tilde{O}_{VRR} & = & \frac{N-1}{2N} O_{VRR}, \\
  \label{eq:Fierz_O_VLR}
  \tilde{O}_{VLR} & = & -\frac{1}{2N} O_{VLR} - O_{SLR}, \\
  \label{eq:Fierz_O_SLL}
  \tilde{O}_{SLL} & = & - \frac{N+1}{2N} O_{SLL}
                        - \frac{1}{4} O_{VLL}, \\
  \label{eq:Fierz_O_SLR}
  \tilde{O}_{SLR} & = & - \frac{1}{2N} O_{SLR}
                        - \frac{1}{4} O_{VLR}.
\end{eqnarray}
We use these relations to eliminate the `tilde' operators
from matching relations. 
We note that all equations except Eq.~(\ref{eq:Fierz_O_SLL})
are exact, whereas Eq.~(\ref{eq:Fierz_O_SLL}) is valid up to
${\cal O}(p/M_0)$ correction terms described by dimension seven
operators. 
When computing the matching of $O_{VLL}$ and $O_{SLL}$ 
operators, the neglected terms give errors of 
${\cal O}(\alpha_s p/M_0)$ through one-loop mixing.

In Sec.\ref{sec:Physics_results} we present our final result
using the following set of operators in more conventional
definitions
\begin{eqnarray}
  \label{eq:O_L}
  O_L & = &
  \bar{b}\gamma_{\mu}(1-\gamma_5)q~
  \bar{b}\gamma_{\mu}(1-\gamma_5)q, \\
  \label{eq:O_R}
  O_R & = &
  \bar{b}\gamma_{\mu}(1+\gamma_5)q~
  \bar{b}\gamma_{\mu}(1+\gamma_5)q, \\
  \label{eq:O_S}
  O_S & = &
  \bar{b}(1-\gamma_5)q~\bar{b}(1-\gamma_5)q, \\
  \label{eq:O_N}
  O_N & = &
  2 \bar{b}\gamma_{\mu}(1-\gamma_5)q~
    \bar{b}\gamma_{\mu}(1+\gamma_5)q 
  + 4 \bar{b}(1-\gamma_5)q~
      \bar{b}(1+\gamma_5)q, \\
  \label{eq:O_N'}
  O_M & = &
  2 \bar{b}\gamma_{\mu}(1-\gamma_5)q~
    \bar{b}\gamma_{\mu}(1+\gamma_5)q 
  - 4 \bar{b}(1-\gamma_5)q 
      \bar{b}(1+\gamma_5)q, \\
   \label{eq:O_S_tilde}
   \tilde{O}_S & = &
   \bar{b}^i (1-\gamma_5) q^j~
   \bar{b}^j (1-\gamma_5) q^i, \\
   \label{eq:O_P}
  O_P & = &
  2 \bar{b}\gamma_{\mu}(1-\gamma_5) q~
    \bar{b}\gamma_{\mu}(1+\gamma_5) q 
  + 4 N \bar{b}(1-\gamma_5) q~
        \bar{b}(1+\gamma_5) q, \\
  \label{eq:O_Q}
  O_Q & = &
  2N \bar{b}\gamma_{\mu}(1-\gamma_5) q~
     \bar{b}\gamma_{\mu}(1+\gamma_5) q 
  + 4 \bar{b}(1-\gamma_5) q~
      \bar{b}(1+\gamma_5) q, \\
  \label{eq:O_T}
  O_T & = &
  (2+N)  \bar{b}\gamma_{\mu}(1-\gamma_5) q~
     \bar{b}\gamma_{\mu}(1+\gamma_5) q 
  -2(3N^2-2N-4) \bar{b}(1-\gamma_5) q~
      \bar{b}(1+\gamma_5) q.
\end{eqnarray}
The indices $i$ and $j$, which appear in the definition of
$\tilde{O}_S$, run over color of quarks, while other
operators are products of color-singlet bilinear operators.
As is obvious from Eqs. (\ref{eq:O_VLL})-(\ref{eq:O_T}),
the set of operators in conventional definition are related to the
first set of operators as
\begin{eqnarray}
  \label{eq:O_L-O_VLL}
  O_L  & = & 4\ O_{VLL},\\
  \label{eq:O_R-O_VRR}
  O_R  & = & 4\ O_{VRR},\\
  \label{eq:O_S-O_SLL}
  O_S  & = & 4\ O_{SLL},\\
  \label{eq:O_N-O_VLR}
  O_N  & = & 8\ ( O_{VLR} + 2 O_{SLR} ), \\
  \label{eq:O_N'-O_VLR}
  O_M  & = & 8\ ( O_{VLR} - 2 O_{SLR} ), \\
  \label{eq:O_S^tilde-O_SLL}
  \tilde{O}_S 
       & = & 8\ \left( \tilde{O}_{SLL} 
         + \frac{1}{2N} O_{SLL} \right) , \\
  \label{eq:O_P-O_VLR}
  O_P  & = & 8\ (  O_{VLR} + 2N O_{SLR} ), \\
  \label{eq:O_Q-O_VLR}
  O_Q  & = & 8\ (N O_{VLR} + 2\ O_{SLR} ), \\
  O_T  & = & 4 (2+N) O_{VLR} - 8 (3N^2-2N-4) O_{SLR}.
\end{eqnarray}

\section{One-loop calculation}
\label{sec:One-loop_calculation}

In order to match the operators defined in the continuum
theory, say the $\overline{MS}$ scheme with the dimensional
regularization, to the lattice counterparts, we compute the
on-shell amplitude both in the continuum and on the lattice
at one-loop level.

Let $O_X^{\overline{MS}}(\mu)$ and $O_X^{lat}(1/a)$ be
certain continuum and lattice operators defined at scale
$\mu$ and $1/a$ respectively. 
The on-shell amplitude for a certain external state can be
expressed by a linear combination of tree-level amplitudes 
$\langle O_Y\rangle_0$, where the subscript $Y$ runs over
all possible operators which can mix with
$O_X^{\overline{MS}}$ and $O_X^{lat}$ at one-loop, namely
\begin{eqnarray}
  \langle O_X^{\overline{MS}}(\mu)\rangle
  & = & 
  \langle O_X\rangle_0
  + \frac{\alpha_s}{4\pi} 
  \sum_Y \rho^{\overline{MS}}_{X,Y}(\mu)
  \langle O_Y\rangle_0
  + {\cal O}(\alpha_s^2), \\
  \langle O_X^{lat}(1/a)\rangle
  & = & 
  \langle O_X\rangle_0
  + \frac{\alpha_s}{4\pi} 
  \sum_Y \rho^{lat}_{X,Y}(1/a)
  \langle O_Y\rangle_0
  + {\cal O}(\alpha_s^2),
\end{eqnarray}
where $\alpha_s=g^2/4\pi$, and
$\rho^{\overline{MS}}_{X,Y}(\mu)$ and
$\rho^{lat}_{X,Y}(1/a)$ represent the one-loop coefficients
in the $\overline{MS}$ and the lattice schemes.
We take zero spatial momentum on-shell free quarks for the
external state. 
This choice is the easiest and sufficient to obtain the
matching coefficients uniquely, since we restrict ourselves
to the matching at lowest operator dimension, for which no
derivative operator appears.

Requiring that the both operators give identical one-loop
on-shell amplitudes, we obtain the following matching
relation 
\begin{eqnarray}
  O_X^{\overline{MS}}(\mu)
  & = & \sum_Y
  \left[ \delta_{X,Y} 
    + \frac{\alpha_s}{4\pi} 
    \left(
      \rho_{X,Y}^{\overline{MS}}(\mu) -
      \rho_{X,Y}^{lat}(1/a)
    \right)
    + {\cal O}(\alpha_s^2) \right]
  O_Y^{lat}(1/a).
\end{eqnarray}

In the following we compute the coefficients 
$\rho^{\overline{MS}}_{X,Y}(\mu)$ and
$\rho^{lat}_{X,Y}(1/a)$ for the heavy-light bilinear
operators and $\Delta B$ = 2 four-quark operators.

\subsection{Bilinear operators}
\label{sec:Bilinear_operators}

First of all, we give the expression for the matching of the
bilinear operators for completeness. 
Although the one-loop coefficients for the matching of 
the heavy-light vector and axial vector currents have already 
been obtained by Morningstar and Shigemitsu \cite{Mor_Shi_98_99} 
even through ${\cal O}(\alpha p/M_0)$ and ${\cal O}(\alpha a p)$,
we present the one-loop  matching coefficients for the 
general heavy-light bilinear operators for completeness.

The one-loop expression of the perturbative on-shell
amplitudes of the heavy-light bilinear operator
$\bar{b}\Gamma q$ with arbitrary Dirac structure $\Gamma$ 
is given as
\begin{eqnarray}
  \langle(\bar{b}\Gamma q)(\mu)\rangle
  & = &  
  \left[
    1 + \frac{\alpha_s}{4\pi}\rho_{\Gamma}(\mu) 
    + {\cal O}(\alpha_s^2) 
  \right]
  \langle \bar{b}\Gamma q\rangle_0,
\end{eqnarray}
for both continuum and lattice operators.
There is no operator mixing in the lowest dimension bilinear 
operators. 
In the continuum (the $\overline{MS}$ scheme with totally
anti-commuting $\gamma_5$), the coefficient 
$\rho^{\overline{MS}}_{\Gamma}(\mu)$ is obtained
\cite{Borrelli_Pittori_92,A4GH,A4D}  as
\begin{equation}
  \rho^{\overline{MS}}_{\Gamma}(\mu)
  = C_F
  \left[
    \frac{H^2-4}{4} \ln\frac{\mu^2}{M_0^2}
    - \frac{3}{2}\ln\frac{\lambda^2}{M_0^2}
    +\frac{3H^2}{4}- H H^{\prime} 
    -\frac{G H}{2} - \frac{11}{4} 
  \right],
\end{equation}
where $C_F=(N^2-1)/2N$, and $\lambda$ denotes a gluon mass
introduced to regularize the infrared divergence.
The constants $H$, $H^{\prime}$ and $G$
are defined through the following equations
\begin{equation}
  H\Gamma \equiv 
  \sum_{\mu=1}^{D}\gamma_{\mu}\Gamma\gamma_{\mu}, 
  ~~ 
  H^{\prime} \equiv \frac{dH}{dD}, 
  ~~ 
  G \Gamma \equiv  \gamma_4 \Gamma \gamma_4,
\end{equation}
with space-time dimension $D$=4.
The corresponding one-loop expression for the lattice
operator is
\begin{eqnarray}
  \rho^{lat}_{\Gamma}(1/a) & = &
  C_F \Biggl[ -\frac{3}{2} \ln(a^2\lambda^2)
  +\frac{1}{2}(C_l+C_h) \nonumber\\
  &&
  +(4\pi)^2 [ I_A+G I_B+(H-G)^2 I_C
  +(H-G)(I_D+I_F)+(H-G) G I_E ]
  \Biggr].
\end{eqnarray}
The infrared divergence of form $\frac{3}{2}C_F\ln\lambda^2$
is canceled between continuum and lattice expressions for
any bilinear operator in the combination of the matching
coefficient 
$\rho_{\Gamma}^{\overline{MS}}-\rho_{\Gamma}^{lat}$. 
Numerical value of the light quark wave function
renormalization factor $C_l$ is 9.076 for the 
${\cal O}(a)$-improved action.
If one uses the normalization $1/\sqrt{u_0}$ for the light
quark field motivated by the tadpole improvement
\cite{Lepage_Mackenzie_93}, the number becomes $-$0.164 for 
$u_0\equiv\langle \frac{1}{3} \mbox{Tr} U_P\rangle^{1/4}$
(average plaquette), or 1.7106 for 
$u_0\equiv 1/8\kappa_{crit}$ (critical hopping parameter). 
The heavy quark wave function renormalization $C_h$ depends
on the heavy quark mass $aM_0$, and its numerical values
are summarized in Table \ref{tab:C_h}.
The constants $I_A$, $I_B$, $I_C$, $I_D$, $I_E$ and $I_F$
are one-loop integrals from the vertex corrections shown in
Figure \ref{fig:vertex2}. 
Their explicit expressions are given in the Appendix
\ref{sec:One-loop_integrals}
(Eqs.(\ref{eq:I_A})-(\ref{eq:I_F})),
and their numerical values are given in Table~\ref{tab:AF}. 

In the following, we present the expressions of the
matching factors for the temporal component of the axial
current $A_4$ and for the pseudoscalar density $P$.

\subsubsection{Axial vector current}
For the axial-vector current $A_4$ with 
$\Gamma$=$\gamma_5\gamma_4$, we obtain $H$=2,
$H^{\prime}$=1, $G$=$-$1, for which the matching
coefficients are 
\begin{eqnarray}
  \rho^{\overline{MS}}_{A_4}
  & = & C_F \left[
    -\frac{3}{2}\ln\frac{\lambda^2}{M_0^2}
    -\frac{3}{4}
  \right], \\
  \rho^{lat}_{A_4}(1/a)
  & = & C_F \Biggl[  
    -\frac{3}{2} \ln(a^2\lambda^2) 
    +\frac{1}{2} (C_h+C_l) \nonumber\\
    & &  +(4\pi)^2(I_A-I_B+9I_C+3I_D-3I_E+3I_F)
    \Biggl].
\end{eqnarray}
$\rho^{\overline{MS}}_{A_4}$ does not have the logarithmic 
scale dependence because of the (partial-)conservation of
the axial vector current.
Combining the two expressions we obtain the matching
relation 
\begin{equation}
  A_4^{\overline{MS}} = 
  \left[ 
    1 + \frac{\alpha_s}{4\pi} 
    \left( 2\ln(a^2M_0^2) + \zeta_A
    \right)
    + {\cal O}(\alpha_s^2)
  \right] A_4^{lat}(1/a),
\end{equation}
where
\begin{equation}
  \label{eq:zeta_A}
  \zeta_A = C_F \left[
    -\frac{3}{4} 
    -\frac{1}{2}(C_h+C_l)
    -(4\pi)^2(I_A-I_B+9I_C+3I_D-3I_E+3I_F)
  \right].
\end{equation}
Numerical values of the coefficient $\zeta_A$ are listed in
Table \ref{tab:zeta_A,P}.

\subsubsection{Pseudoscalar density}
For the pseudo-scalar density $P$ with $\Gamma$=$\gamma_5$,
we obtain $H$=$-$4, $H^{\prime}$=$-$1 and $G$=$-$1.
The matching coefficients are
\begin{eqnarray}
  \rho^{\overline{MS}}_P(\mu)
  & = & C_F
  \left[ 3 \ln\frac{\mu^2}{M_0^2}
    - \frac{3}{2} \ln\frac{\lambda^2}{M_0^2}
    + \frac{13}{4} 
  \right], \\
  \rho^{lat}_P(1/a)
  & = & C_F
  \Biggl[ - \frac{3}{2}\ln(a^2\lambda^2) 
    + \frac{1}{2}(C_h+C_l) \nonumber \\
    & & + (4\pi)^2( I_A-I_B+9I_C-3I_D+3I_E-3I_F) 
  \Biggl].
\end{eqnarray}
Combining these expressions we obtain the matching relation
\begin{equation}
  P^{\overline{MS}}(\mu) =
  \left[ 1 + \frac{\alpha_s}{4\pi}
    \left( 
      4\ln\frac{\mu^2}{M_0^2}
      + 2\ln(a^2M_0^2) + \zeta_P
    \right)
    + {\cal O}(\alpha_s^2)
  \right] P^{lat}(1/a),
\end{equation}
where
\begin{equation}
  \label{eq:zeta_P}
  \zeta_P = C_F \left[
    \frac{13}{4} 
    -\frac{1}{2}(C_h+C_l)
    -(4\pi)^2( I_A-I_B+9I_C-3I_D+3I_E-3I_F)
  \right].
\end{equation}
Numerical values of the coefficient $\zeta_P$ are listed in
Table \ref{tab:zeta_A,P}.

\subsection{Four-quark operators}
\label{sec:Four-quark_operators}

We present the one-loop matching calculation of the
four-quark operators $O_{VLL}$ and $O_{SLL}$, which appear
in the evaluation of the mass and width differences in the
$B_{d(s)}-\bar{B}_{d(s)}$ systems. 

\subsubsection{$O_{VLL}$}
In the continuum theory preserving the chiral symmetry, the
four-quark operator $O_{VLL}$ mixes with $O_{SLL}$ under the
radiative correction.
At one-loop level, the on-shell amplitude of $O_{VLL}(\mu)$
defined at scale $\mu$ is written as 
\begin{equation}
  \langle O_{VLL}^{\overline{MS}}(\mu)\rangle =
  \left[ 1+\frac{\alpha_s}{4\pi}
    \rho^{\overline{MS}}_{VLL,VLL}(\mu)
  \right] \langle O_{VLL}\rangle_0 +
  \left[ \frac{\alpha_s}{4\pi}
    \rho^{\overline{MS}}_{VLL,SLL}
  \right] \langle O_{SLL}\rangle_0 ,
\end{equation}
where
\begin{eqnarray}
  \rho^{\overline{MS}}_{VLL,VLL}(\mu) & = & 
   2\ln\frac{M_0^2}{\mu^2}
  -4\ln\frac{\lambda^2}{M_0^2} - \frac{35}{3},\\ 
  \rho^{\overline{MS}}_{VLL,SLL} & = & -8.
\end{eqnarray}
The mixing coefficient $\rho^{\overline{MS}}_{VLL,SLL}$ does 
not have the scale dependence at one-loop level.

The same operator mixes with four operators $O_{VLR}$,
$O_{SLR}$, $O_{SLL}$ and $O_{VRR}$ on the lattice due to the
lack of the chiral symmetry.
We obtain the following expression for the on-shell
amplitude of the lattice operator $O_{VLL}(1/a)$:
\begin{eqnarray}
  \label{eq:VLL_lat}
  \langle O_{VLL}^{lat}(1/a)\rangle 
  & = &
  \left[ 1+\frac{\alpha_s}{4\pi}\rho^{lat}_{VLL,VLL}(1/a)
  \right] \langle O_{VLL}\rangle_0 \nonumber \\
  & & +
  \left[ \frac{\alpha_s}{4\pi}\rho^{lat}_{VLL,VLR}
  \right] \langle O_{VLR}\rangle_0 +
  \left[ \frac{\alpha_s}{4\pi}\rho^{lat}_{VLL,SLR}
  \right] \langle O_{SLR}\rangle_0 \nonumber \\
  & & +
  \left[ \frac{\alpha_s}{4\pi}\rho^{lat}_{VLL,SLL}
  \right] \langle O_{SLL}\rangle_0 +
  \left[ \frac{\alpha_s}{4\pi}\rho^{lat}_{VLL,VRR}
  \right] \langle O_{VRR}\rangle_0,
\end{eqnarray}
where
\begin{eqnarray}
  \rho^{lat}_{VLL,VLL}(1/a) & = &
  -4 \ln(a^2\lambda^2)+ C_F[C_l+C_h] \nonumber \\
  & & + (4\pi)^2 \Biggl[ \frac{10}{3}I_A + 2I_C+ 4I_E
  \nonumber\\
  & & \;\;\;\; 
  +\frac{1}{3}\left(I_G-3(I_H+I_I+I_J+2I_K)+16I_L+I_N\right)
  \Biggr],\\
  \rho^{lat}_{VLL,VLR} & = &
  (4\pi)^2 \left[ 2 
    \left( -2I_B - \frac{5}{3}(I_D+I_F)\right)
  \right],\\
  \rho^{lat}_{VLL,SLR} & = &
  (4\pi)^2 \left[ 2 
    \left( -4I_B + \frac{10}{3}(I_D+I_F) \right)
  \right],\\
  \rho^{lat}_{VLL,SLL} & = &  
  (4\pi)^2 \left[ - 16 (2I_C-I_E) \right], \\
  \rho^{lat}_{VLL,VRR} & = &
  (4\pi)^2 \left[ - \frac{4}{3} I_M \right].
\end{eqnarray}
The integrals $I_A$, $I_B$, $I_C$, $I_D$, $I_E$ and $I_F$
come from the diagrams in which a gluon mediates between
heavy and light quark lines as shown in Figures
\ref{fig:vertex4_12} and \ref{fig:vertex4_34}.
These are the same integrals as in the vertex correction of
the bilinear operators, whose numerical values are given in
Table~\ref{tab:AF}. 
Other integrals $I_G$, $I_H$, $I_I$, $I_J$, $I_K$, $I_L$,
$I_M$, $I_N$ are characteristic of the corrections of the
four-quark operators. 
The diagrams in which a gluon line mediates between two
heavy quark lines (Figure \ref{fig:vertex4_5}) produce the
five integrals  $I_G$, $I_H$, $I_I$, $I_J$ and $I_K$, whose
expressions are given in the
Appendix~\ref{sec:One-loop_integrals} 
(Eqs.(\ref{eq:I_G})-(\ref{eq:I_K})). 
Their heavy quark mass dependence is summarized in
Table~\ref{tab:GK}. 
Other three, $I_L$, $I_M$ and $I_N$ defined in
Eqs.~(\ref{eq:I_L})-(\ref{eq:I_N}) correspond to the
diagrams in which the two light quark lines are connected by
a gluon line. 
These do not depend on the heavy quark mass, and their
numerical values are
\begin{equation}
  \label{eq:LMN}
  I_L = -0.004635(3), ~~ 
  I_M = -0.002433(1), ~~
  I_N = -0.012204(6).
\end{equation}
In Appendix \ref{sec:One-loop_four-quark_amplitudes},
one-loop expressions of the lattice on-shell amplitudes with 
general four-quark operators 
$\bar{b}\Gamma q~\bar{b}\Gamma q$ are presented.
The above result in Eq. (\ref{eq:VLL_lat}) is obtained by
applying the Fierz transformation for the color and spinor
indices on the expressions 
(\ref{eq:X_heavy-light^singlet})-(\ref{eq:X_light-light^octet}).

Combining the continuum and the lattice results we obtain 
\begin{eqnarray}
  O^{\overline{MS}}_{VLL}(\mu) & = &
  \left[ 1+\frac{\alpha_s}{4\pi}
    \left(\rho^{\overline{MS}}_{VLL,VLL}(\mu) -
     \rho^{lat}_{VLL,VLL}(1/a) \right)
  \right] O^{lat}_{VLL}(1/a) \nonumber \\
  & & +
  \left[ - \frac{\alpha_s}{4\pi}
    \rho^{lat}_{VLL,VLR}
  \right] O^{lat}_{VLR}(1/a)
  + 
  \left[ - \frac{\alpha_s}{4\pi}
    \rho^{lat}_{VLL,SLR}
  \right] O^{lat}_{SLR}(1/a) \nonumber \\
  & & +
  \left[ \frac{\alpha_s}{4\pi}
    \left(\rho^{\overline{MS}}_{VLL,SLL} -
      \rho^{lat}_{VLL,SLL}\right)  
  \right] O^{lat}_{SLL}(1/a) 
  +
  \left[ - \frac{\alpha_s}{4\pi}
    \rho^{lat}_{VLL,VRR}
  \right] O^{lat}_{VRR}(1/a).
\end{eqnarray}
The numerical values of $\rho^{lat}_{VLL,Y}$ are listed in
Tables \ref{tab:rho_VLL}.

The matching relation for $O_L$ is obtained using the
conversion formula (\ref{eq:O_L-O_VLL})-(\ref{eq:O_Q-O_VLR}) 
as follows
\begin{eqnarray}
  \label{eq:O_L_matching}
  O_L^{\overline{MS}}(\mu) & = &
  \left[ 1 + \frac{\alpha_s}{4\pi} 
    \left(
      -2\ln\frac{\mu^2}{M_0^2}
      +4\ln(a^2 M_0^2) + \zeta_{L,L}
    \right)
  \right] O_L^{lat}(1/a)
  \nonumber \\
  & & 
  + \frac{\alpha_s}{4\pi} \zeta_{L,S} O_S^{lat}(1/a)
  + \frac{\alpha_s}{4\pi} \zeta_{L,R} O_R^{lat}(1/a)
  + \frac{\alpha_s}{4\pi} \zeta_{L,N} O_N^{lat}(1/a)
  \nonumber \\
  & &
  + \frac{\alpha_s}{4\pi} \zeta_{L,M} O_M^{lat}(1/a).
\end{eqnarray}
The coefficients $\zeta_{L,S}$, $\zeta_{L,R}$,
$\zeta_{L,N}$ and $\zeta_{L,M}$ are listed in Table
\ref{tab:zeta_L}.
The coefficient $\zeta_{L,M}$ of $O_M^{lat}$ vanishes in
the static limit, and other coefficients agree with the
previous work
\cite{Hiroshima_B_B_99,Ishikawa_Onogi_Yamada_99} in the same 
limit. 

\subsubsection{$O_{SLL}$}
The matching relation for the operator $O_{SLL}$ is obtained
in a similar manner.

The operator $O_{SLL}$ mixes with $O_{VLL}$ with the
radiative correction in the continuum.
The on-shell amplitude with $O_{SLL}$ for the zero momentum
external state is written as
\begin{equation}
  \langle O_{SLL}^{\overline{MS}}(\mu)\rangle =
  \left[ 1 +
    \frac{\alpha_s}{4\pi}
    \rho_{SLL,SLL}^{\overline{MS}}(\mu)
  \right] \langle O_{SLL}\rangle_0 +
  \frac{\alpha_s}{4\pi} 
  \rho_{SLL,VLL}^{\overline{MS}}(\mu)
  \langle O_{VLL}\rangle_0,
\end{equation}
at one-loop level.
The coefficients are
\begin{eqnarray}
  \rho_{SLL,SLL}^{\overline{MS}}(\mu) & = &
  \frac{16}{3} \ln\frac{\mu^2}{M_0^2}
  -\frac{4}{3} \ln\frac{\lambda^2}{M_0^2}
  +10, \\
  \rho_{SLL,VLL}^{\overline{MS}}(\mu) & = &
   \frac{1}{3} \ln\frac{\mu^2}{M_0^2}
  +\frac{2}{3} \ln\frac{\lambda^2}{M_0^2}
  +\frac{3}{2}.
\end{eqnarray}

The lattice operator $O_{SLL}^{lat}$ mixes with five
operators in the on-shell amplitude as
\begin{eqnarray}
  \langle O_{SLL}^{lat}\rangle & = &
  \left[ 1+\frac{\alpha_s}{4\pi}
    \rho_{SLL,SLL}^{lat}(1/a) 
  \right] \langle O_{SLL}\rangle_0 \nonumber \\
  & & +
  \left[ \frac{\alpha_s}{4\pi}
    \rho_{SLL,VLL}^{lat}(1/a)
  \right] \langle O_{VLL}\rangle_0 +
  \left[ \frac{\alpha_s}{4\pi} 
    \rho_{SLL,SLR}^{lat}
  \right] \langle O_{SLR}\rangle_0 \nonumber \\
  & & +
  \left[ \frac{\alpha_s}{4\pi}
    \rho_{SLL,VLR}^{lat} 
  \right] \langle O_{VLR}\rangle_0 +
  \left[ \frac{\alpha_s}{4\pi}
    \rho_{SLL,VRR}^{lat} 
  \right] \langle O_{VRR}\rangle_0,
\end{eqnarray}
where
\begin{eqnarray}
  \rho_{SLL,SLL}^{lat}(1/a) & = &  
  -\frac{4}{3} \ln(a^2\lambda^2) + C_F [C_l+C_h]
  \nonumber\\ & &
  + (4\pi)^2 
  \Biggl[ \frac{1}{3} (4I_A+52I_C+20I_E
    \nonumber\\ & &
    -2(I_G+I_H+I_I+I_J+2I_K+I_N) )
  \Biggl],\\
  \rho_{SLL,VLL}^{lat}(1/a) & = &  
  \frac{2}{3} \ln(a^2\lambda^2) + (4\pi)^2
  \Biggl[ \frac{1}{12} (
  2 ( -3I_A-7I_C+I_E) \nonumber\\ & &
    -3I_G+I_H+I_I+I_J+2I_K-16I_L-3I_N )
  \Biggr], \\
  \rho_{SLL,SLR}^{lat} & = & (4\pi)^2 
  \left[ 2 [ \frac{3}{2}I_B + \frac{17}{6}(I_D+I_F) ] \right], \\
  \rho_{SLL,VLR}^{lat} & = & (4\pi)^2 
  \left[ 2 [ \frac{1}{4}I_B - \frac{5}{12} (I_D+I_F) ] \right], \\
  \rho_{SLL,VRR}^{lat} & = & (4\pi)^2 
  \left[ \frac{1}{3} I_M \right],
\end{eqnarray}
Numerical values of $\rho_{SLL,Y}^{lat}$ are given in
Table~\ref{tab:rho_SLL}. 

Combining the continuum and the lattice results we obtain
\begin{eqnarray}
  O_{SLL}^{\overline{MS}}(\mu) & = &
  \left[ 1+\frac{\alpha_s}{4\pi}
    \left (\rho_{SLL,SLL}^{\overline{MS}}(\mu) - 
           \rho_{SLL,SLL}^{lat}(1/a) \right)
   \right] O_{SLL}^{lat}(1/a) \nonumber \\
   & & +
   \left[ \frac{\alpha_s}{4\pi}
     \left(\rho_{SLL,VLL}^{\overline{MS}}(\mu) - 
           \rho_{SLL,VLL}^{lat}(1/a) \right)
   \right] O_{VLL}^{lat}(1/a) \nonumber \\
   & & +
   \left[ - \frac{\alpha_s}{4\pi}
     \rho_{SLL,SLR}^{lat}
   \right] O_{SLR}^{lat}(1/a) \nonumber \\
   & & +
   \left[ - \frac{\alpha_s}{4\pi}
     \rho_{SLL,VLR}^{lat}
   \right] O_{VLR}^{lat}(1/a) +
   \left[ - \frac{\alpha_s}{4\pi}
     \rho_{SLL,VRR}^{lat} 
   \right] O_{SLL,VRR}^{lat}(1/a).
\end{eqnarray}

The matching relation for $O_S$ is obtained using the
conversion formula (\ref{eq:O_L-O_VLL})-(\ref{eq:O_Q-O_VLR})
as follows
\begin{eqnarray}
  \label{eq:O_S_matching}
  O_S^{\overline{MS}}(\mu) & = &
  \left[ 1 + \frac{\alpha_s}{4\pi} 
    \left(
      \frac{16}{3}\ln\frac{\mu^2}{M_0^2}
      +\frac{4}{3}\ln(a^2 M_0^2) + \zeta_{S,S}
    \right)
  \right] O_S^{lat}(1/a)
  \nonumber \\
  & & 
  + \frac{\alpha_s}{4\pi}
  \left[
    \frac{1}{3}\ln\frac{\mu^2}{M_0^2}
    -\frac{2}{3}\ln(a^2 M_0^2) + \zeta_{S,L}
  \right] O_L^{lat}(1/a)
  \nonumber \\
  & &
  + \frac{\alpha_s}{4\pi} \zeta_{S,R} O_R^{lat}(1/a)
  + \frac{\alpha_s}{4\pi} \zeta_{S,P} O_P^{lat}(1/a)
  + \frac{\alpha_s}{4\pi} \zeta_{S,T} O_T^{lat}(1/a).
\end{eqnarray}
The coefficients $\zeta_{S,S}$, $\zeta_{S,L}$,
$\zeta_{S,R}$, $\zeta_{S,P}$ and $\zeta_{S,T}$ are listed in
Table \ref{tab:zeta_S}.
The coefficient $\zeta_{S,T}$ of $O_T^{lat}$ vanishes in the
static limit, and other coefficients agree with the previous
work \cite{Hiroshima_B_S_99,B_S_lat99} in the same limit.

\section{Physics results}
\label{sec:Physics_results}
Using the one-loop coefficients obtained in this work, we
reanalyze the lattice NRQCD calculations of $B_L$
\cite{Hiroshima_B_B_99} and of $B_S$
\cite{Hiroshima_B_S_99}. 
These previous calculations were performed with the
lattice NRQCD action for heavy quark, but the perturbative
matching of the four-quark operators were done using the
coefficients in the infinitely heavy quark mass limit.
Due to this approximation for the matching coefficients, 
the previous results contain errors of order
$\alpha_s/(aM)$, which is one of the largest uncertainties 
among all the systematic errors.

\subsection{$B_L$}
The $B$ parameter $B_L$ is defined through
\begin{equation}
  \label{eq:B_L^cont}
  B_L(\mu) \equiv 
  \frac{\langle \bar{B}^0| 
    O_L^{\overline{MS}}(\mu) |B^0\rangle}{
    \frac{8}{3} 
    \langle \bar{B}^0| A_4^{\overline{MS}} |0\rangle
    \langle 0|A_4^{\overline{MS}}|B^0 \rangle},
\end{equation}
where the scale $\mu$ is usually set at the $b$ quark mass
$M_b$. 
In the following analysis we use $\mu$ = 4.8 GeV.
On the lattice we measure the `$B$ parameters'
\begin{equation}
  \label{eq:B_L^lat}
  B_X^{lat}(1/a) \equiv 
  \frac{\langle \bar{B}^0| 
    O_X^{lat}(1/a) |B^0\rangle}{
    \frac{8}{3} 
    \langle \bar{B}^0| A_4^{lat}(1/a) |0\rangle
    \langle 0|A_4^{lat}(1/a)|B^0 \rangle},
\end{equation}
for four-quark operators $O_X$=$O_L$, $O_S$, $O_R$, $O_N$
and $O_M$ using the NRQCD action in Eq. (\ref{eq:NRQCD_action}).
We performed the simulations on a quenched 16$^3\times$48
lattice at $\beta$=5.9.
Other details of the lattice calculations are found in Ref. 
\cite{Hiroshima_B_B_99}.

\begin{figure}
  \begin{center}
    \leavevmode\psfig{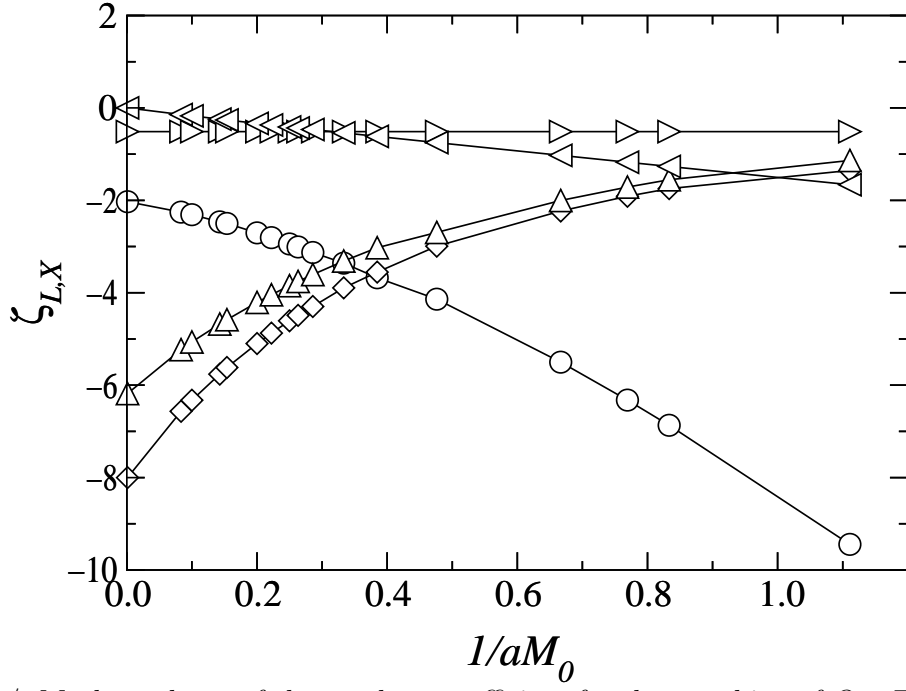}
    \caption{$1/aM_0$ dependence of the one-loop coefficient
      for the matching of $O_L$.
      For $\zeta_L$ we plot $\zeta_{L,L}-2\zeta_A$ (cicle),
      and others are $\zeta_{L,S}$ (diamond),
      $\zeta_{L,R}$ (triangle right),
      $\zeta_{L,N}$ (triangle up) and
      $\zeta_{L,M}$ (triangle left).
      }
    \label{fig:zeta_L}
  \end{center}
\end{figure}

The perturbative matching relation for the continuum
operator $O_L$ in Eq. (\ref{eq:O_L_matching}) may be used to obtain 
\begin{eqnarray}
  \label{eq:B_L_matching}
  B_L(\mu) & = & 
  Z_{L,L/A^2}(\mu,a) B_L^{lat}(1/a)
  + Z_{L,S/A^2} B_S^{lat}(1/a) 
  \nonumber \\
  & &
  + Z_{L,R/A^2} B_R^{lat}(1/a)
  + Z_{L,N/A^2} B_N^{lat}(1/a)
  + Z_{L,M/A^2} B_M^{lat}(1/a), 
\end{eqnarray}
where the matching factors are
\begin{eqnarray}
  Z_{L,L/A^2}(\mu,a) & = & 
  1 + \frac{\alpha_s}{4\pi} 
  \left( 2\ln\frac{M_0^2}{\mu^2}
    + \zeta_{L,L} - 2 \zeta_A \right), \\
  Z_{L,S/A^2} & = & \frac{\alpha_s}{4\pi}\zeta_{L,S}, \\
  Z_{L,R/A^2} & = & \frac{\alpha_s}{4\pi}\zeta_{L,R}, \\
  Z_{L,N/A^2} & = & \frac{\alpha_s}{4\pi}\zeta_{L,N}, \\
  Z_{L,M/A^2} & = & \frac{\alpha_s}{4\pi}\zeta_{L,M}.
  \label{eq:Z_L}
\end{eqnarray}
The one-loop coefficients $\zeta_{L,X}$ are defined in 
Eq. (\ref{eq:O_L_matching}) and plotted in
Figure~\ref{fig:zeta_L} as a function of $1/aM_0$.
The coefficient of the leading contribution
$\zeta_{L,L}-2\zeta_A$ becomes larger in magnitude toward 
lighter heavy quark. 
The mass dependence of $\zeta_{L,L}-2\zeta_A$ is relatively
smaller than that of $\zeta_{L,L}$ itself, due to the
cancellation of singlet diagrams
(Figure~\ref{fig:vertex4_12}) against the contribution of
the denominator $-2\zeta_A$. 
Two mixing coefficients $\zeta_{L,S}$ and $\zeta_{L,N}$
become smaller when $1/M$ correction is incorporated. 
It is also important that $\zeta_{L,M}$, which vanishes in
the static limit, becomes non-zero for finite heavy quark
mass. 

\begin{figure}
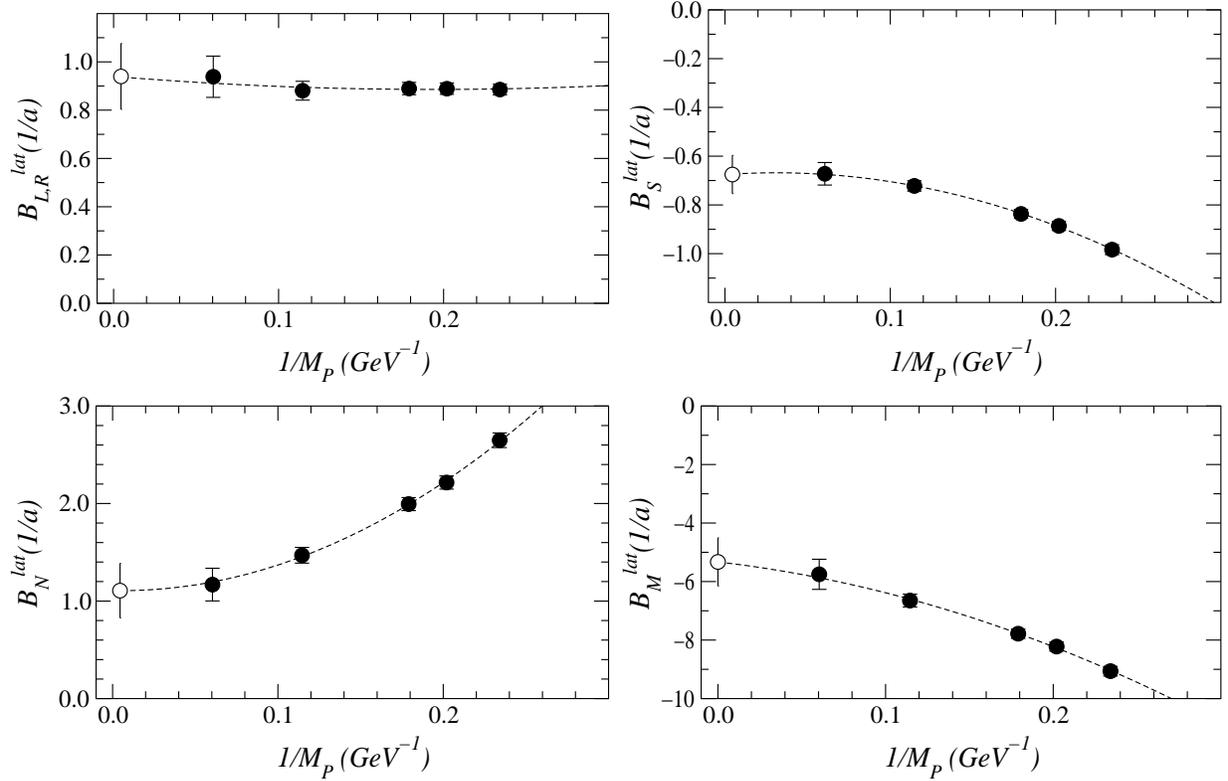

  \begin{center}
    \begin{tabular}{rr}
      \psfig{file=fig_B_lat_L/mdep.BL.eps,width=8cm,clip=} &
      \psfig{file=fig_B_lat_L/mdep.BS.eps,width=8cm,clip=} \\
      \psfig{file=fig_B_lat_L/mdep.BN.eps,width=8cm,clip=} &
      \psfig{file=fig_B_lat_L/mdep.BM.eps,width=8cm,clip=}
    \end{tabular}
    \caption{Matrix elements $B_X^{lat}(1/a)$ in Eq. 
      (\ref{eq:B_L^lat}) measured on the lattice.
      Dashed curves represent a quadratic fit to the data,
      and extrapolated static limit is shown by a open
      symbol. 
      }
    \label{fig:B_lat_L}
  \end{center}
\end{figure}

The matrix elements on the lattice $B^{lat}_X(1/a)$ measured
in Ref. \cite{Hiroshima_B_B_99} are shown in
Figure~\ref{fig:B_lat_L} as a function of inverse meson mass 
$1/M_P$.
Their mass dependence is qualitatively well described by the
vacuum saturation approximation.

For the coupling constant $\alpha_s$ we choose the
$V$-scheme coupling $\alpha_V(q^*)$
\cite{Lepage_Mackenzie_93} with $q^*$ = $2/a$.
To estimate the size of higher order perturbative errors we
also analyze with $q^*$ = $1/a$ and $\pi/a$.

Combining $Z_{L,X/A^2}$ and $B^{lat}_X$ we obtain the
contribution of each term $Z_{L,X/A^2} B_X^{lat}(1/a)$ in
Eq. (\ref{eq:B_L_matching}) as shown in 
Figure \ref{fig:Z_L,X*B_X}.
In spite of the large mass dependence of the coefficients 
$\zeta_{L,S}$ and $\zeta_{L,N}$, there is no significant
mass dependence in the corresponding combined quantities
$Z_{L,X/A^2} B_X^{lat}(1/a)$, since those are canceled by
the large and opposite mass dependence in $B_S^{lat}(1/a)$
and $B_N^{lat}(1/a)$. 
Small increase of $Z_{L,M/A^2} B_M^{lat}(1/a)$ from zero in
the static limit is observed, which reflects the increasing
trend of both $Z_{L,M/A^2}$ and $B_M^{lat}(1/a)$.

\begin{figure}
  \begin{center}
    \leavevmode\psfig{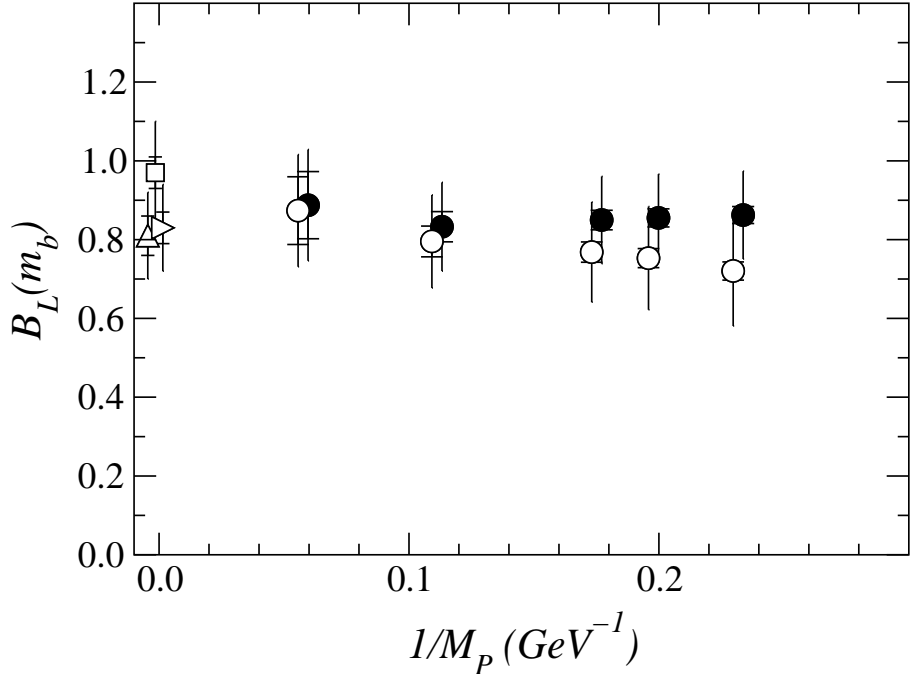}
    \caption{$1/M_P$ dependence of $B_L(m_b)$ obtained with
      the NRQCD action (filled circles).
      The same calculation but with the matching coefficient
      in the infinite mass limit is shown by open circles.
      Other symbols in the static limit are calculations
      with the static action by
      UKQCD \protect\cite{UKQCD_96} (triangle up),
      Gim\'enez-Martinelli 
      \protect\cite{Gimenez_Martinelli_97}
      (triangle right) and
      Christensen \textit{et al.}
      \protect\cite{Christensen_Draper_McNeile_97}
      (square). 
      We reanalysed their raw data with the same method as
      ours, namely we used Eqs. 
      (\ref{eq:B_L_matching})-(\ref{eq:Z_L}) with
      $\alpha_V(2/a)$ and $\zeta$'s in the static limit.
      }
    \label{fig:B_L}
  \end{center}
\end{figure}

Total result for $B_L(m_b)$ is presented in
Figure~\ref{fig:B_L} by filled circles.
Because of the cancellation of the large mass dependences in
$\zeta_{L,X}$ and in $B_X^{lat}(1/a)$, there is little
$1/M_P$ dependence in our final result (filled circles).
A small increase toward larger $1/M_P$ comes from the
contribution of $O_M$.  
In this plot our estimate of systematic uncertainty is shown
by error bars.
The horizontal ticks attached to the error bars represent
the size of statistical error, which is much smaller than
the systematic errors especially for large $1/M_P$ points.

We also plot our previous analysis with the same NRQCD
action but using the perturbative matching in the static
limit (open circles in Figure~\ref{fig:B_L}).
There is a small negative slope in $1/M_P$ so that the
previous result is about 12\% smaller than our new result at
the $B$ meson mass.
An estimation of ${\cal O}(\alpha_s/(aM_b))$ error in our previous
analysis is around 10\%, when we assume an order counting
argument with typical value of the strong coupling constant
$\alpha_s\sim$ 0.3.
In addition to this error, there are also other errors of
${\cal O}(\alpha_s^2)$, ${\cal O}(a^2 \Lambda_{QCD}^2)$ and ${\cal O}(\alpha_s a
\Lambda_{QCD})$. 
Thus the shift of our result does not exceed the systematic
uncertainty discussed in Ref.~\cite{Hiroshima_B_B_99}. 

Our final numerical result is
\begin{equation}
  \label{eq:B_L_result}
  B_L(m_b) = 0.85 \pm 0.03 \pm 0.11,
\end{equation}
where the first error is statistical and the second is
systematic. 
The systematic error is estimated using the order counting
of neglected contributions.
In our new analysis in which the ${\cal O}(\alpha_s/(aM_b))$ error
is removed, the remaining sources of uncertainty are the
dicretization errors ${\cal O}(a^2\Lambda_{QCD}^2)$ ($\sim$ 5\%)
and ${\cal O}(\alpha_s a\Lambda_{QCD})$ ($\sim$ 5\%), as well as
higher order perturbative error ${\cal O}(\alpha_s^2)$ 
($\sim$ 10\%). 
Higher order contributions in the nonrelativistic expansion
are ${\cal O}(\alpha_s\Lambda_{QCD}/M_b)$ ($\sim$ 2\%) and
${\cal O}(\Lambda_{QCD}^2/M_b^2)$ ($<$ 1\%), which are not dominant
uncertainties. 
We assume $\alpha_s\sim$ 0.3 and $\Lambda_{QCD}\sim$ 350 MeV 
when we estimate the errors listed above, which are added in
quadrature to give the systematic error of about 13\% in the
final result.

The result (\ref{eq:B_L_result}) may be compared with the
recent lattice calculations with relativistic heavy quark
actions: 
0.92(4)$^{+3}_{-0}$ \cite{UKQCD_HF8} and 
0.93(8)$^{+0.0}_{-0.6}$ \cite{Becirevic_et_al_00},
where the first error is statistical and the second is their 
estimate of systematic errors.
It is encouraging that our result agrees with these
relativistic calculations within the large systematic
uncertainty in (\ref{eq:B_L_result}). 
Although the systematic error in the relativistic results
seem much smaller, it should be noted that the quoted
systematic uncertainty could be underestimated.
They extrapolate their simulation results performed around
charm mass regime assuming the $1/M$ scaling without
considering $O((aM_0)^2)$ errors.  
However, the $1/M$ dependence of the simulation results
could be distorted by the ${\cal O}((aM_0)^2)$ error, which
can be as large as 30\% toward the heavier side in the naive
order counting. 
In order to have a reliable prediction of $B_L(m_b)$,
one has to at least include $O((aM_0)^2$ error when 
extrapolating in $1/M$, or take careful contiuum limit before 
doing $1/M$ extrapolation. 
Furthermore, the heavy quark expansion becomes questionable
to describe the heavy quark in the charm quark mass regime 
when truncated at $1/M$ or at $1/M^2$.
Therefore, an alternative method to fit the results would be
to include the result in the static limit in order to
constrain at least the leading term in the $/1M$ expansion.

Finally, we also obtain chiral breaking effect on the ratio
of $B_L$ as 
\begin{eqnarray}
\frac{B_{B_s}(m_{B_s})}{B_{B_d}(m_{B_d})} 
& = & 1.01 \pm 0.01 \pm 0.03,
\end{eqnarray}
for which the relativistic results are
0.98(3) \cite{UKQCD_HF8} and 
0.98(5) \cite{Becirevic_et_al_00}.

\subsection{$B_S$}
The $B$ parameter $B_S$ is defined as
\begin{equation}
  \label{eq:B_S^cont}
  B_S(\mu) \equiv 
  \frac{\langle\bar{B}^0|O_S^{\overline{MS}}(\mu)|B^0\rangle}{
    \frac{5}{3}
    \langle\bar{B}^0|P^{\overline{MS}}(\mu)|0\rangle
    \langle 0|P^{\overline{MS}}(\mu)|B^0\rangle }.
\end{equation}
We also define a ratio of the matrix elements of bilinear
operators 
\begin{equation}
  \label{eq:R^cont}
  {\cal R}(\mu) \equiv
  \left|
    \frac{\langle 0| A_4^{\overline{MS}} |B^0\rangle}{
      \langle 0| P^{\overline{MS}}(\mu)  |B^0\rangle}
  \right|,
\end{equation}
and calculate
\begin{equation}
  \label{eq:B_S/R^2}
  \frac{B_S(\mu)}{{\cal R}(\mu)^2} =
  \frac{\langle\bar{B}^0|O_S^{\overline{MS}}(\mu)|B^0\rangle}{
    - \frac{5}{3}
    \langle\bar{B}^0|A_4^{\overline{MS}}|0\rangle
    \langle 0|A_4^{\overline{MS}}|B^0\rangle },
\end{equation}
which is necessary in evaluating the $B_s$ meson width
difference \cite{Hiroshima_B_S_99}. 

\begin{figure}
  \begin{center}
    \leavevmode\psfig{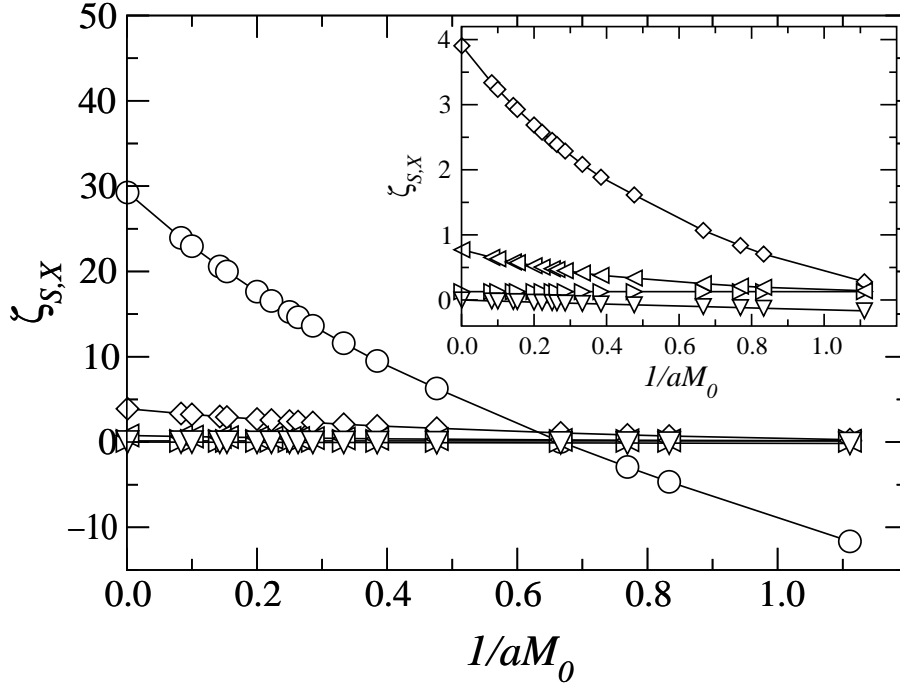}
    \caption{$1/aM_0$ dependence of the one-loop coefficient
      for the matching of $O_S$.
      For $\zeta_S$ we plot $\zeta_{S,S}-2\zeta_A$ (cicle),
      and others are $\zeta_{S,L}$ (diamond),
      $\zeta_{S,R}$ (triangle right),
      $\zeta_{S,P}$ (triangle left) and
      $\zeta_{S,T}$ (triangle down).
      }
    \label{fig:zeta_S}
  \end{center}
\end{figure}

In the lattice simulation we measure 
\begin{equation}
  \label{eq:B_S^lat}
  B_X^{\prime~ lat}(1/a) \equiv 
  \frac{\langle \bar{B}^0| 
    O_X^{lat}(1/a) |B^0\rangle}{
    -\frac{5}{3} 
    \langle \bar{B}^0| A_4^{lat}(1/a) |0\rangle
    \langle 0|A_4^{lat}(1/a)|B^0 \rangle},
\end{equation}
for four quark operators $O_X$ = $O_S$, $O_L$, $O_R$, $O_P$
and $O_T$.
Note that the denominator of $B_X^{\prime~ lat}$ is different from 
Eq.~(\ref{eq:B_S^cont}).
The perturbative matching relation for the continuum
operator $O_S^{\overline{MS}}$ in Eq. (\ref{eq:O_S_matching}) may
be used to obtain  
\begin{eqnarray}
  \label{eq:B_S_matching}
  \frac{B_S(\mu)}{{\cal R}(\mu)^2} & = &
  Z_{S,S/A^2}(\mu,a) B_S^{\prime~ lat}(1/a) +
  Z_{S,L/A^2}(\mu,a) B_L^{\prime~ lat}(1/a) 
  \nonumber \\
  & &
  + Z_{S,R/A^2} B_R^{\prime~ lat}(1/a) +
  Z_{S,P/A^2} B_P^{\prime~ lat}(1/a) +
  Z_{S,T/A^2} B_T^{\prime~ lat}(1/a),
\end{eqnarray}
where
\begin{eqnarray}
  \label{eq:Z_S}
  Z_{S,S/A^2}(\mu,a) & = &
  1 + \frac{\alpha_s}{4\pi}
  \left(
    \frac{16}{3}\ln\frac{\mu^2}{M_0^2}
    -\frac{8}{3}\ln(a^2 M_0^2)
    + \zeta_{S,S} - 2\zeta_A
  \right), \\
  Z_{S,L/A^2}(\mu,a) & = &
  \frac{\alpha_s}{4\pi}
  \left(
    \frac{1}{3}\ln\frac{\mu^2}{M_0^2}
    -\frac{2}{3}\ln(a^2 M_0^2)
    + \zeta_{S,L}
  \right), \\
  Z_{S,R/A^2} & = & \frac{\alpha_s}{4\pi}\zeta_{S,R}, \\
  Z_{S,P/A^2} & = & \frac{\alpha_s}{4\pi}\zeta_{S,P}, \\
  Z_{S,T/A^2} & = & \frac{\alpha_s}{4\pi}\zeta_{S,T}.
\end{eqnarray}
The one-loop coefficients $\zeta_{S,X}$ defined in
Eq.~(\ref{eq:O_S_matching}) are plotted in
Figure~\ref{fig:zeta_S} as a function of $1/aM_0$. 
The $1/M_0$ dependence is quite large for the leading
contribution $\zeta_{S,S}-2\zeta_A$, because the denominator
in Eq.~(\ref{eq:B_S/R^2}) is not a vacuum saturation of the
numerator and the cancellation of color singlet diagrams 
(Figure \ref{fig:vertex4_12}) does not take place.
Among other coefficients, $\zeta_{S,L}$ has relatively large
$1/M_0$ dependence and the mixing of operator $O_L$ becomes
smaller as $1/aM_0$ increases.

\begin{figure}
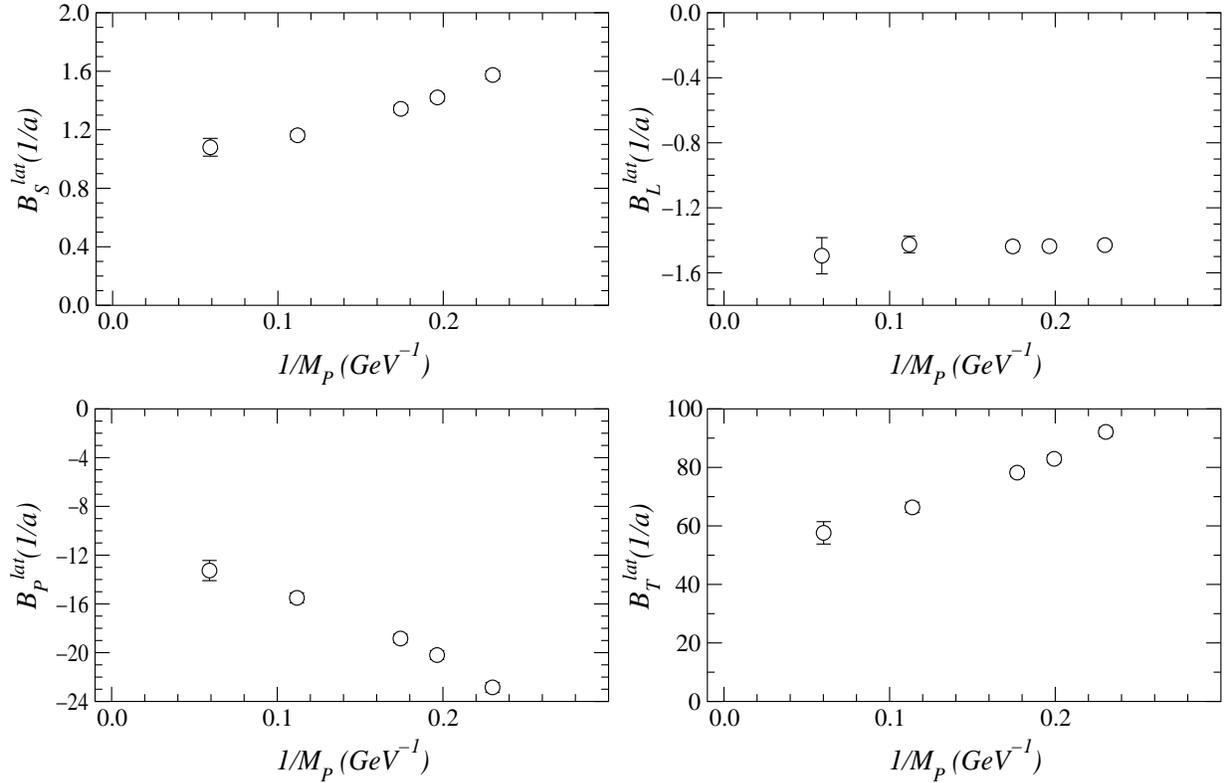

  \begin{center}
    \begin{tabular}{rr}
      \psfig{file=fig_B_lat_S/mdep.BS_S.eps,width=8cm,clip=} &
      \psfig{file=fig_B_lat_S/mdep.BL_S.eps,width=8cm,clip=} \\
      \psfig{file=fig_B_lat_S/mdep.BP_S.eps,width=8cm,clip=} &
      \psfig{file=fig_B_lat_S/mdep.BT_S.eps,width=8cm,clip=}
    \end{tabular}
    \caption{Matrix elements $B_X^{lat}(1/a)$ in Eq.
      (\ref{eq:B_S^lat}) measured on the lattice.
      }
    \label{fig:B_lat_S}
  \end{center}
\end{figure}

The matrix elements $B_X^{lat}(1/a)$ are shown in
Figure~\ref{fig:B_lat_S}.
The $1/M_P$ dependence in $B_S$, $B_P$ and $B_T$ is
significant, which is well described by the vacuum
saturation approximation as discussed in Ref.
\cite{Hiroshima_B_S_99}.
The contribution of each term
$Z_{S,X/A^2}B_X^{lat}(1/a)$ in Eq.~(\ref{eq:B_S_matching}) to
$B_S(m_b)/{\cal R}(m_b)^2$ is plotted in
Figure~\ref{fig:Z_S,X*B_X}, in which no clear $1/M_P$
dependence is observed. 

\begin{figure}
  \begin{center}
    \leavevmode\psfig{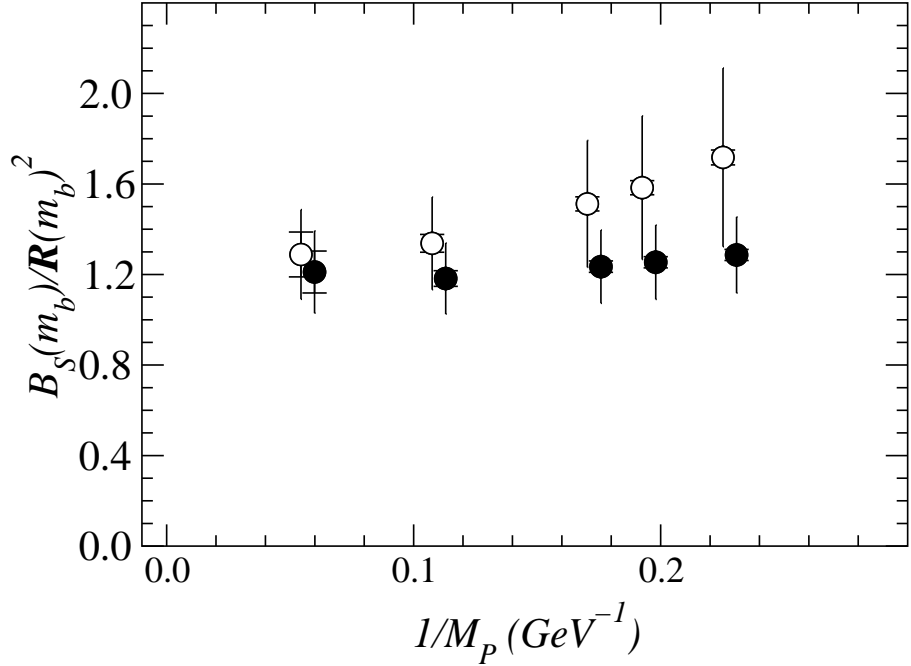}
    \caption{$1/M_P$ dependence of 
      $B_S(m_b)/{\cal R}(m_b)^2$ obtained with the NRQCD
      action (filled circles).
      The same calculation but with the matching coefficient
      in the infinite mass limit is shown by open circles.
      }
    \label{fig:B_S}
  \end{center}
\end{figure}

Total result for $B_S(m_b)/{\cal R}(m_b)^2$ is presented in 
Figure~\ref{fig:B_S} by filled circles.
As is evident from the plot of each term 
$Z_{S,X/A^2}B_X^{lat}(1/a)$ (Figure~\ref{fig:Z_S,X*B_X}),
there is no clear trend in the $1/M_P$ dependence of 
$B_S(m_b)/{\cal R}(m_b)^2$. 
Our previous analysis \cite{Hiroshima_B_S_99} with matching
coefficients in the static limit is plotted with open
symbols.
Reduction of the result with the correction is quite large
($\sim$ 20\%), but consistent with our estimate 
for the collection of systematic errors of ${\cal O}(\alpha_s/(aM))$,
${\cal O}(\alpha_s^2)$, ${\cal O}( a^2  \Lambda_{QCD}^2)$, ${\cal O}( \alpha a
\Lambda_{QCD})$ ($\sim$ 20\%).
Main effect comes from the large $1/aM_0$ dependence of
$\zeta_{S,S}-2\zeta_A$ shown in Figure~\ref{fig:zeta_S}.

Our final numerical result is 
\begin{equation}
  \label{eq:B_S/R^2_result}
  \frac{B_S(m_b)}{{\cal R}(m_b)^2} =
  1.24 \pm 0.03 \pm 0.16,
\end{equation}
where the first error is statistical one, and the second is
our estimate of systematic uncertainty obtained as in the
analysis of $B_L$.
For the width difference we obtain
\begin{equation}
  \label{eq:width_difference_result}
  \left(\frac{\Delta\Gamma}{\Gamma}\right)_{B_s} =
  0.107 \pm 0.026 \pm 0.014 \pm 0.017
\end{equation}
using Eq.(9) of Ref. \cite{Hiroshima_B_S_99}.
Errors are from the $B_s$ meson decay constant $f_{B_s}$ = 
245(30)~MeV, which is taken from the current world average
of unquenched lattice calculations \cite{Hashimoto_lat99},
from $B_S(m_b)/{\cal R}(m_b)^2$, and from an estimate of
higher order contribution in the $1/m_b$ expansion
\cite{Beneke_Buchalla_Dunietz_96,Beneke_et_al_99}. 

\section{Conclusions}
\label{sec:Conclusions}
In this paper we have performed one-loop calculations of
matching coefficients for $\Delta B$=2 four-quark operators
defined using lattice NRQCD.
This calculation allows to remove one of the dominant
systematic errors characterized by ${\cal O}(\alpha_s/(aM_b))$ from
the lattice simulation of the $B$ parameters $B_L$ and
$B_S$. 
We find sizable $1/aM_0$ dependence in several one-loop
coefficients, which affects the mass dependence of the
$B$ parameters as well as their absolute values at the $b$
quark mass. 

We have also presented a reanalysis of our previous
simulations and obtained results for $B_L$ and for
$B_S/{\cal R}^2$ with reduced systematic error.
The difference from our previous results is consistent with
the estimate obtained with order counting argument.
Remaining systematic uncertainty is dominated by unknown
two-loop matching coefficients.

\section*{Acknowledgment}
S.H. and T.O. are supported by the Grants-in-Aid
of the Ministry of Education (Nos. 10740125, 11740162).
K-I.I. and N.Y. would like to thank the JSPS for Young Scientists for
a research fellowship. 

\appendix
\section{Lattice NRQCD Feynman rules}
\label{sec:Lattice_NRQCD_Feynman_rules}
In order to simply the expression, we set the lattice
spacing $a$ = 1 throughout Appendix A and B.
When deriving the Feynman rules from the NRQCD action,
we followed the method which is explained in Ref.~\cite{Morningstar_93}. 
We also note that the Feynman rules for $O(1/M)$ NRQCD action with
slightly different definition from ours are given in 
Ref.~\cite{Mor_Shi_98_99}.  
\subsection{Functions}
We define the following functions which appear in the
Feynman rules below. 
\begin{eqnarray}
  \tilde{l}^2 
  & \equiv &
  \sum_{\mu=1}^4 4 \sin^2\frac{l_{\mu}}{2}
  \\
  A^{(0)}(l)
  & \equiv &
  1 - \frac{1}{nM_0}\sum_{i=1}^3 \sin^2\frac{l_i}{2},
  \\
  C(l',l) 
  & \equiv &  
  e^{-i l_4'} + e^{-i l_4},
  \\
  f^{A}_{\mu\nu}(q)
  & \equiv &
  \sin q_{\mu} \cos\frac{q_{\nu}}{2},
  \\
  f^{B}_{\mu\nu}(q_{1},q_{2})
  & \equiv &
  \cos\left(\frac{q_{1}+q_{2}}{2}\right)_{\mu}
  \sin\left(\frac{q_{1}+q_{2}}{2}\right)_{\nu}
  \sin\left(\frac{q_{1}-q_{2}}{2}\right)_{\mu},
  \\
  f^{C}_{\mu\nu}(q_{1},q_{2})
  & \equiv &
  \frac{1}{2}\left[ 
     \cos\left(\frac{q_{1}}{2}\right)_{\mu}
     \cos\left(q_{1}+\frac{q_{2}}{2}\right)_{\nu}
    +\cos\left(q_{2}+\frac{q_{1}}{2}\right)_{\mu}
     \cos\left(\frac{q_{2}}{2}\right)_{\nu}
  \right.
  \nonumber \\
  & & \left.
    +\cos\left(q_{2}+\frac{q_{1}}{2}\right)_{\mu}
     \cos\left(q_{1}+\frac{q_{2}}{2}\right)_{\nu}
    -\cos\left(\frac{q_{1}}{2}\right)_{\mu}
     \cos\left(\frac{q_{2}}{2}\right)_{\nu}
  \right].
\end{eqnarray}
We also define
\begin{eqnarray}
  A^{(1)}(l',l;p_1,\mu_1)
  & \equiv & 
  \frac{-1}{2nM_0} 
  \left[ 
    \sum_{i=0}^{n-1} A^{(0)}(l')^{i}A^{(0)}(l)^{n-1-i}   
  \right]
  \sin\left(\frac{l'+l}{2}\right)_{\mu_1},
  \\
  A^{(2)}_{1}(l',l;p_1,\mu_1,p_2,\mu_2)
  & \equiv &
  \frac{-1}{4nM_0}
  \left[
    \sum_{i=0}^{n-1} A^{(0)}(l')^{i}A^{(0)}(l)^{n-1-i}
  \right]
  \cos\left(\frac{l'+l}{2}\right)_{\mu_1},
  \\
  A^{(2)}_{2}(l',l,p_1 ,\mu_1,p_2,\mu_2)
  & \equiv &
  \frac{1}{(2 n M_0)^{2}}
  \left[ 
    \sum_{i=0}^{n-2} \sum_{j=0}^{n-2-i}
    A^{(0)}(l')^{j}A^{(0)}(l-p_2)^{i}A^{(0)}(l)^{n-2-i-j}
  \right]
  \nonumber \\
  &&  \mbox{}\times
  \sin\left(l'+\frac{p_1}{2}\right)_{\mu_1}
  \sin\left(l -\frac{p_2}{2}\right)_{\mu_2},
  \\
  dH^{(1)}_{1}(l',l;p_1,\mu_1)
  & \equiv &
  +i \frac{c_B}{4M_0}
  \sum_{i,j=1}^3 \epsilon_{ij\mu_1} \Sigma^{i}
  f^{A}_{j \mu_1}(p_1),
  \\
  dH^{(2)}_{2}(l',l;p_1,\mu_1,p_2,\mu_2)
  & \equiv &
  -i\frac{c_B}{4M_0}
  \sum_{i,j=1}^3 \epsilon_{ij\mu_1} \Sigma^{i}
  f^{B}_{j \mu_1}(p_1,p_2),
  \\
  dH^{(2)}_{3}(l',l;p_1,\mu_1,p_2,\mu_2)
  & \equiv &
  -i\frac{c_B}{4M_0}
  \sum_{i=1}^3 \epsilon_{i\mu_1\mu_2} \Sigma^{i}
  f^{C}_{\mu_1\mu_2}(p_1,p_2),
\end{eqnarray}
where $\Sigma^i$ denotes a four-by-four matrix
\begin{equation}
  \Sigma^i = 
  \left(
    \begin{array}{cc}
      \sigma^i & 0         \\
      0        & \sigma^i \\
    \end{array}
  \right). \\
\end{equation}
Using these functions the Fourier components of the
evolution operator in Eq. (\ref{eq:evolution_operator_Q}) is
written as
\begin{eqnarray}
  \lefteqn{\hspace*{-1em}
    v_K^{(1)}(l',l;p_{1},\mu_{1}) = 
    } \nonumber \\
  & &
  \left[
    \left(
        e^{-i l_4'}{A^{(0)}(l')}^n 
      + e^{-i l_4} {A^{(0)}( l)}^n
    \right) 
    A^{(1)}(l',l;p_{1},\mu_{1})
  \right.
  \nonumber \\
  & & \mbox{\hspace*{1em}}\left.
    + {A^{(0)}(l')}^n{A^{(0)}( l)}^n
    C(l',l) dH^{(1)}_{1}(l',l;p_{1},\mu_{1})
  \right]
  \hat{\delta}_{4 \mu_{1}} 
  \nonumber\\
  & &
  -i e^{-\frac{1}{2}(l_4'+l_4)} 
  {A^{(0)}(l')}^n{A^{(0)}( l)}^n \delta_{4 \mu_1},
  \\
  \lefteqn{\hspace*{-1em}
    v_K^{(2)}(l',l;p_{1},\mu_{1},p_{2},\mu_{2}) 
    = 
    -\frac{1}{2} e^{-\frac{i}{2}(l'_4+l_4)}
    \delta_{4\mu_1} \delta_{\mu_1\mu_2}
    } \nonumber \\
  & & + 
  \left[
    \left(
        e^{-i l_4'} {A^{(0)}(l')}^n
      + e^{-i l_4}  {A^{(0)}( l)}^n 
    \right)
    A_{2}^{(2)}(l',l;p_{1},\mu_{1},p_{2},\mu_{2})
  \right. \nonumber \\
  & & \mbox{\hspace*{1em}} + 
  e^{-i(l-p_2)_4}
  A^{(1)}(l',l - p_{2};p_{1},\mu_{1})
  A^{(1)}(l - p_{2}, l;p_{2},\mu_{2})
  \nonumber \\
  & & \mbox{\hspace*{1em}} + 
  A^{(0)}(l')^n C(l',l - p_{2})
  A^{(1)}(l - p_{2}, l;p_{2},\mu_{2})
  dH^{(1)}_{1}(l',l - p_{2};p_{1},\mu_{1})
  \nonumber \\
  & & \mbox{\hspace*{1em}} + 
  A^{(0)}( l)^n C(l - p_{2}, l)
  A^{(1)}(l',l - p_{2};p_{1},\mu_{1})
  dH^{(1)}_{1}(l - p_{2}, l;p_{2},\mu_{2})
  \nonumber \\
  & & \mbox{\hspace*{1em}} + 
  A^{(0)}(l')^n A^{(0)}(l)^n
  C(l',l) dH^{(2)}_{3}(l',l;p_{1},\mu_{1},p_{2},\mu_{2})
  \nonumber \\
  & & \mbox{\hspace*{1em}} + \left.
    e^{-i(l-p_2)_4}
    A^{(0)}(l')^n A^{(0)}(l)^n
    dH^{(1)}_{1}(l',l - p_{2};p_{1},\mu_{1})
    dH^{(1)}_{1}(l - p_{2}, l;p_{2},\mu_{2}) 
  \right] 
  \hat{\delta}_{4\mu_1}\hat{\delta}_{4\mu_2}
  \nonumber \\
  & & + 
  \left[
    \left(
        e^{-i l_4'} A^{(0)}(l')^n 
      + e^{-i l_4}  A^{(0)}(l)^n 
    \right)
    A_{1}^{(2)}(l',l;p_{1},\mu_{1},p_{2},\mu_{2})
  \right. \nonumber \\
  & & \mbox{\hspace*{1em}} \left. + 
    A^{(0)}(l')^n A^{(0)}(l)^n
    C(l',l) dH^{(2)}_{2}(l',l;p_{1},\mu_{1},p_{2},\mu_{2})
  \right] 
  \hat{\delta}_{4\mu_1} \delta_{\mu_1\mu_2},
\end{eqnarray}
where $\hat{\delta}_{4 \mu} \equiv 1 - \delta_{4 \mu}$.

\subsection{Feynman Rules}

We summarize the Feynman rules used in our calculation.

In our convention the arrows in the heavy (anti-)quark
propagator represent the flow of momentum irrespective of
whether it is particle $Q$ or anti-particle $\chi$.
Other notations are
\begin{itemize}
\item $a$, $b$, ...: color index of quarks,
\item $\alpha$, $\beta$, ...: spin index of quarks,
\item $l$, $l'$: momentum of quarks,
\item $A$, $B$, ...: color index of gluons,
\item $\mu$, $\nu$, ...: spin index of gluons,
\item $k$, $k_1$, ...: momentum of gluons.
\end{itemize}
Double lines denote the heavy (anti-)quark propagators.

We also need Feynman rules for heavy-light bilinear and
four-quark operators. 
In this appendix, we give the Feynman rules for $\Delta
B=-1$ heavy-light current as an illustration. 
Feynman rules for other operators can easily be deduced. 

In the diagrams involving a heavy-light current, heavy quark
is incoming into the current, and heavy anti-quark is
outgoing from the current.
Light quark represented by a single line is, on the other
hand, always outgoing.

\begin{itemize}
\item Gluon propagator:
  \begin{equation}
    \mbox{\hspace{-3.0cm}}
    \raisebox{-1em}{\leavevmode\psfig{file=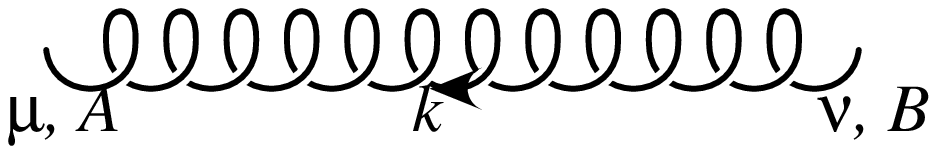,width=4cm}}
    \mbox{\hspace{2cm}} {\displaystyle
      \delta^{AB} \delta_{\mu\nu} G(k)
      \equiv
      \delta^{AB} \delta_{\mu\nu} \frac{1}{\tilde{k}^2+\lambda^2}
      }
  \end{equation}

\item Light quark propagator:
  \begin{equation}
    \mbox{\hspace{-2.0cm}}
    \raisebox{0em}{
      \leavevmode\psfig{file=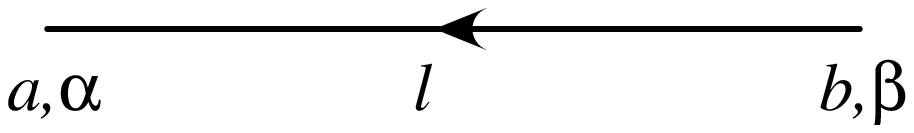,width=4cm}}
    \mbox{\hspace{2cm}}
    \begin{array}{l}
      {\displaystyle
        \delta_{ab}
        \left[ -i\sum_{\mu=1}^{4} \gamma_{\mu} \sin l_{\mu}
          +\frac{r}{2}\tilde{l}^2 \right]_{\alpha \beta} S(l),
        } \\
      {\displaystyle \mbox{where\ \ \ }
        S(l)^{-1} 
        = \sum_{\mu=1}^4 \sin^2 l_{\mu}
        + \left(\frac{r}{2}\tilde{l}^2\right)^2
        }
    \end{array}
  \end{equation}

\item ${\cal O}(g)$ vertex for light quark:
  \begin{equation}
    \mbox{\hspace{-0.3cm}}
    \raisebox{-5em}{\psfig{file=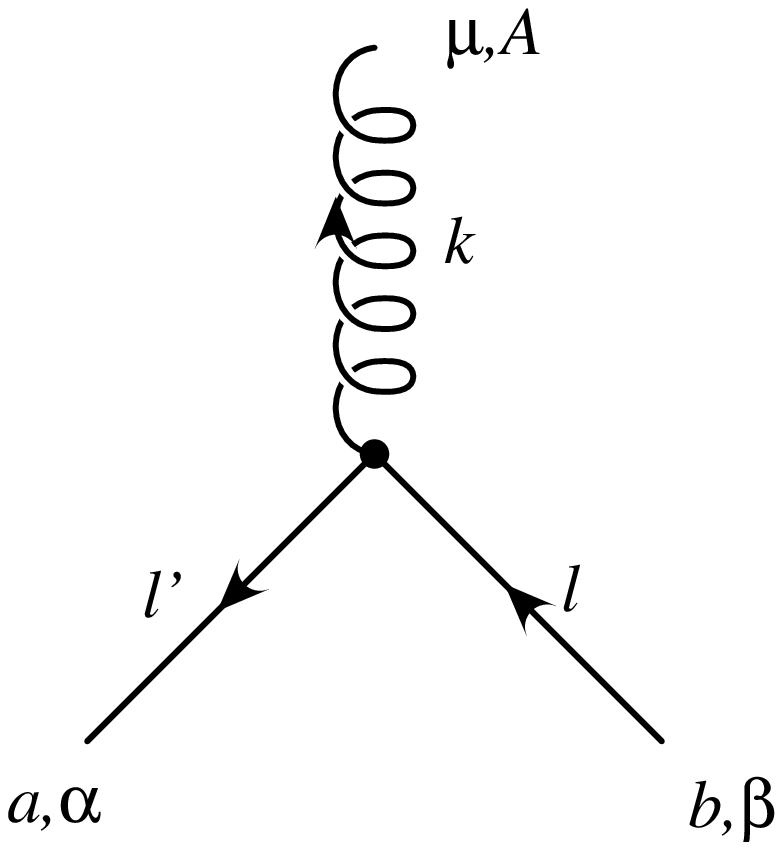,width=4cm}}
    \mbox{\hspace{2cm}}
    \begin{array}{l}
      {\displaystyle
        g (T^A)_{ab} 
        \left[
          -i \gamma_{\mu} \cos\left(\frac{l'+l}{2}\right)_{\mu}
          - r \sin\left(\frac{l'+l}{2}\right)_{\mu}
        \right.}\\
      \displaystyle{ \left.
          \;\;\;\;\;
          + \frac{i}{2} r c_{sw} \sum_{\lambda=1}^4 
          f^A_{\lambda \mu}(k) \sigma_{\lambda \mu} \right]_{\alpha\beta}}
    \end{array}
  \end{equation}

\item ${\cal O}(g^2)$ vertex for light quark:\\
  The vertex from the clover term does not give 
  any contribution to the diagrams we compute, thus 
  we do not give the explcit expression here.
  \begin{equation}
    \mbox{\hspace{-2.5cm}}
    \raisebox{-4em}{\psfig{file=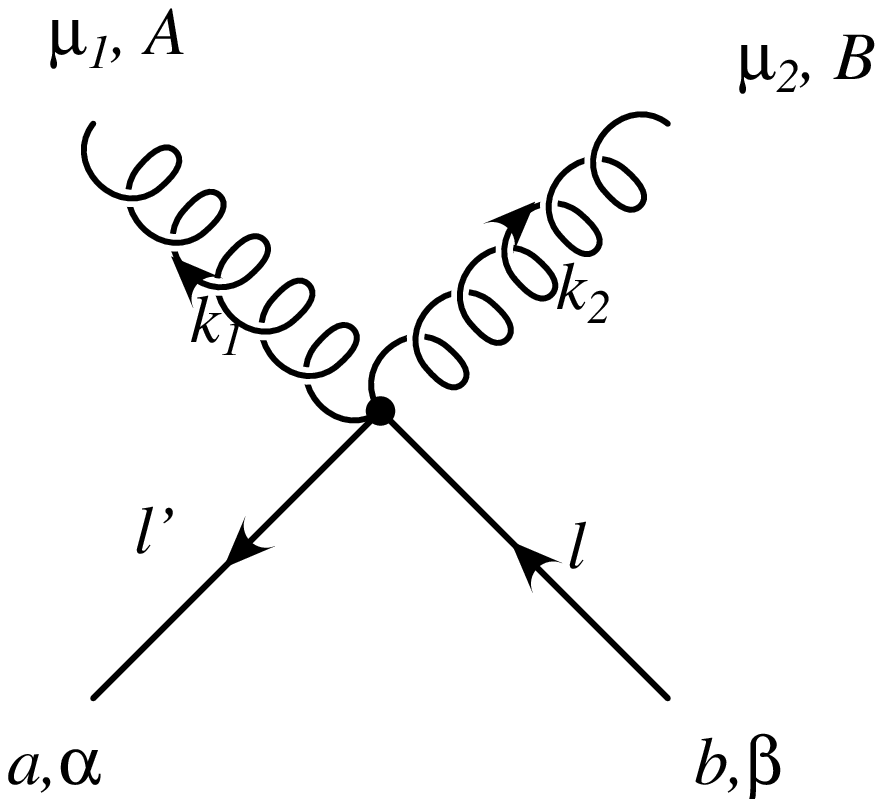,width=4.5cm}}
    \mbox{\hspace{1.5cm}}
    \begin{array}{l} 
      {\displaystyle
        -\frac{g^2}{2}\{T^A,T^B\}_{ab}
        \left[
          r \cos\left(\frac{l'+l}{2}\right)_{\mu_1}
        \right.}\\
      {\displaystyle
        \left.
          \;\;\;\;\;
          - i \gamma_{\mu_1} \sin\left(\frac{l'+l}{2}\right)_{\mu_1}
        \right]_{\alpha\beta}
        \delta_{\mu_1\mu_2}} \\
        + \mbox{contribution from the clover term}
    \end{array}
  \end{equation}

\item Heavy quark propagator:
  \begin{equation}
    \mbox{\hspace{-2.5cm}}
    \raisebox{-1em}{\psfig{file=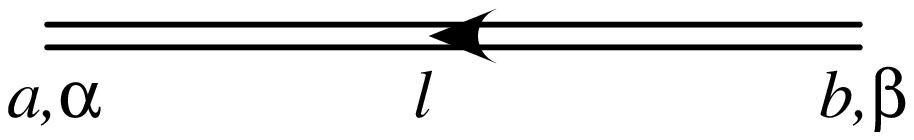,width=4cm}}
    \mbox{\hspace{2cm}}
    \begin{array}{l}
      {\displaystyle
        \delta_{ab} 
        \left(\frac{1+\gamma_4}{2}\right)_{\alpha\beta} 
        Q(l)
        \equiv
        } \\
      {\displaystyle
        \;\;\;\;\;
        \left(\frac{1+\gamma_4}{2}\right)_{\alpha\beta} 
        \frac{ \delta_{ab}}
        {1-e^{-i l_{4}} {A^{(0)}(l)}^{2n}}
        }
    \end{array}
  \end{equation}
  
\item ${\cal O}(g^2)$ counter term introduced for the tadpole
  improvement:\\
  This term appears because we devide all the link variables
  $U_{\mu}$ in the NRQCD action by the mean field value of 
  $u_0=1-g^2 u^{(2)}$. 
  \begin{equation}
    \mbox{\hspace{-1.5cm}}
    \raisebox{-1em}{\psfig{file=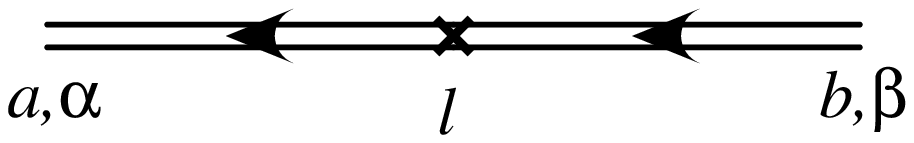,width=4cm}}
    \mbox{\hspace{1cm}}
    \begin{array}{l}
      {\displaystyle
        \frac{g^{2}u^{(2)}_{0}}{M_0}
        \delta_{ab}
        \left(\frac{1+\gamma_4}{2}\right)_{\alpha\beta}
        }\\[5mm]
      {\displaystyle
        \;\;\;\; \times
        e^{-i l_{4}} A^{(0)}(l)^{2n-1}
        \left( 2\kappa_{2}(l)-3-M_0 A^{(0)}(l) \right)
        }
    \end{array}
  \end{equation}

\item ${\cal O}(g)$ vertex for heavy quark:
  \begin{equation}
    \mbox{\hspace{-3.2cm}}
    \raisebox{-4em}{\psfig{file=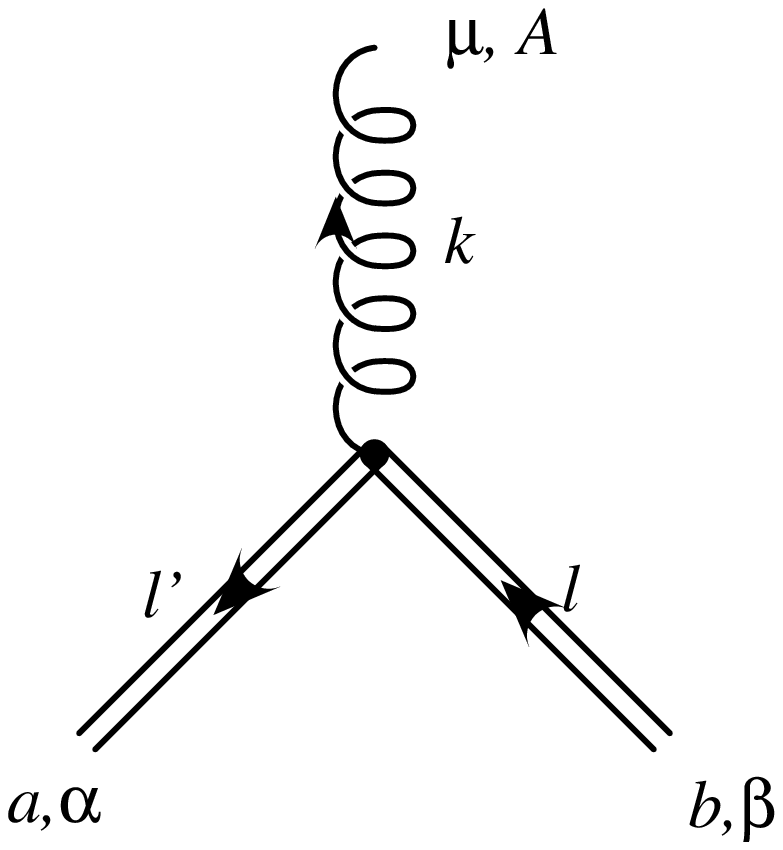,width=4cm}}
    \mbox{\hspace{1.2cm}}
    g(T^{A})_{ab}
    \left[
      v_K^{(1)}(l',l;k,\mu)
      \frac{1+\gamma_4}{2}
    \right]_{\alpha\beta}
  \end{equation}

\item ${\cal O}(g^2)$ vertex for vertex:
  \begin{equation}
    \mbox{\hspace{-0.5cm}}
    \raisebox{-3em}{\psfig{file=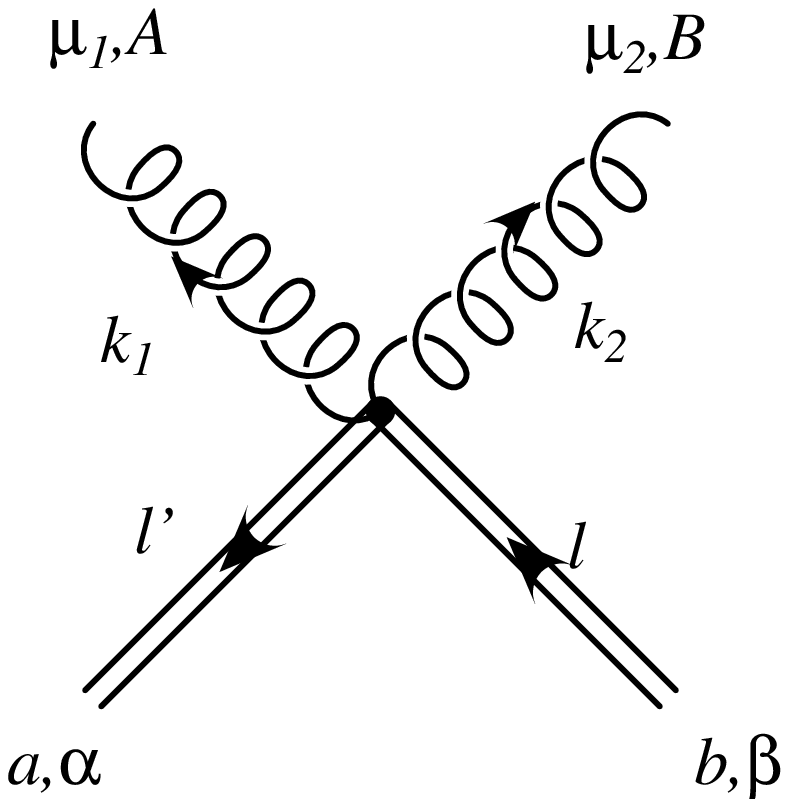,width=4cm}}
    \mbox{\hspace{1.2cm}}
    \begin{array}{l}
      {\displaystyle
        g^{2} 
        \left[\
          \left(
            (T^{A}T^{B})_{ab}
            v_K^{(2)}(l',l;k_1,\mu_1,k_2,\mu_2)
          \right. \right.
        } \\
      {\displaystyle 
        \left. \left. \;\;\;\; + 
            (T^{B}T^{A})_{ab}
            v_K^{(2)}(l',l;k_2,\mu_2,k_1,\mu_1) 
          \right)
          \frac{1+\gamma_4}{2}
        \right]_{\alpha\beta}
        }
    \end{array}
  \end{equation}

\item Heavy anti-quark propagator:
  \begin{equation}
    \mbox{\hspace{-3.5cm}}
    \raisebox{-1em}{\psfig{file=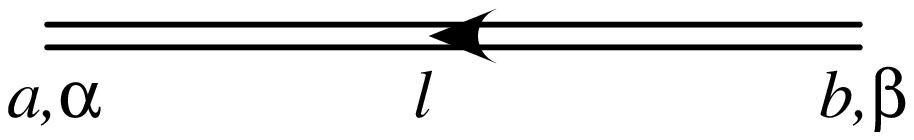,width=4cm}}
    \mbox{\hspace{1cm}}
    \begin{array}{l}
      {\displaystyle
        \delta_{ab} 
        \left(\frac{1-\gamma_4}{2}\right)_{\alpha \beta}
        Q(l)
        \equiv
        } \\
      {\displaystyle
        \left(\frac{1-\gamma_4}{2}\right)_{\alpha \beta}
        \frac{ \delta_{ab}}
        {1-e^{-i l_{4}} {A^{(0)}(l)}^{2n}}
        }
    \end{array}
  \end{equation}

\item ${\cal O}(g)$ vertex for heavy anti-quark:
  \begin{equation}
    \mbox{\hspace{-2cm}}
    \raisebox{-3em}{\psfig{file=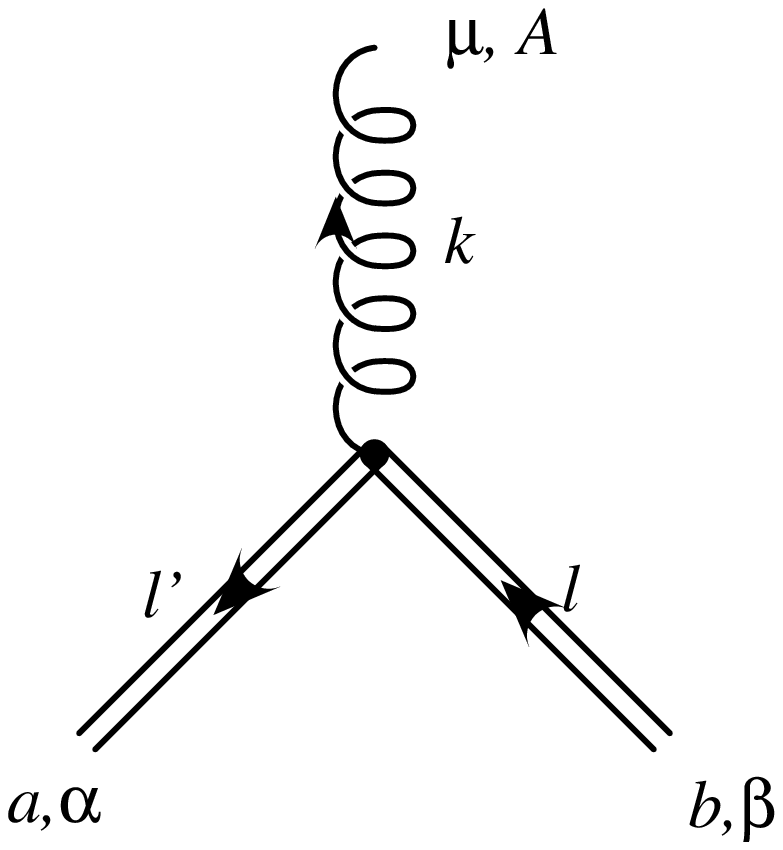,width=4cm}}
    \mbox{\hspace{1cm}}
    - g \left[
      \Sigma^2 v_K^{(1)}(l',l;p,\mu)
      \Sigma^2 \frac{1-\gamma_4}{2} 
    \right]_{\beta \alpha} (T^{A})_{ba}
  \end{equation}

\item heavy-light current:
  \begin{equation}
    \mbox{\hspace{-5cm}}
    \raisebox{-2em}{\psfig{file=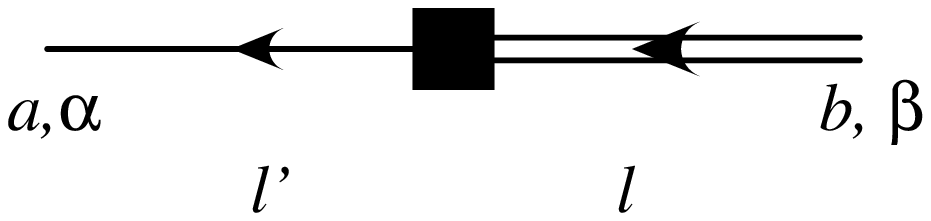,width=4cm}}
    \mbox{\hspace{2cm}}
    \left[
      \Gamma \frac{1+\gamma_4}{2}
    \right]_{\alpha \beta} \delta_{a b}
  \end{equation}

\item heavy-light current with the FWT rotation:
  \begin{equation}
    \mbox{\hspace{-2.2cm}}
    \raisebox{-2em}{\psfig{file=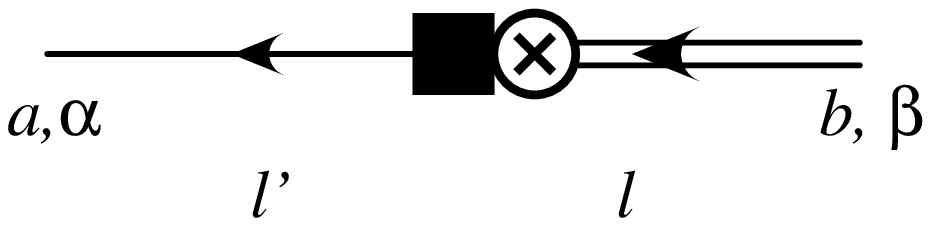,width=4cm}}
    \mbox{\hspace{2cm}}
    \left[
      \Gamma \frac{-i}{2M_0} \sum_{i=1}^3 \gamma_i \sin l_i  
      \frac{1+\gamma_4}{2} 
    \right]_{\alpha \beta} \delta_{a b}
  \end{equation}

\item vertex from rotated heavy-light current:
  \begin{equation}
    \mbox{\hspace{-0.5cm}}
    \raisebox{-2em}{\psfig{file=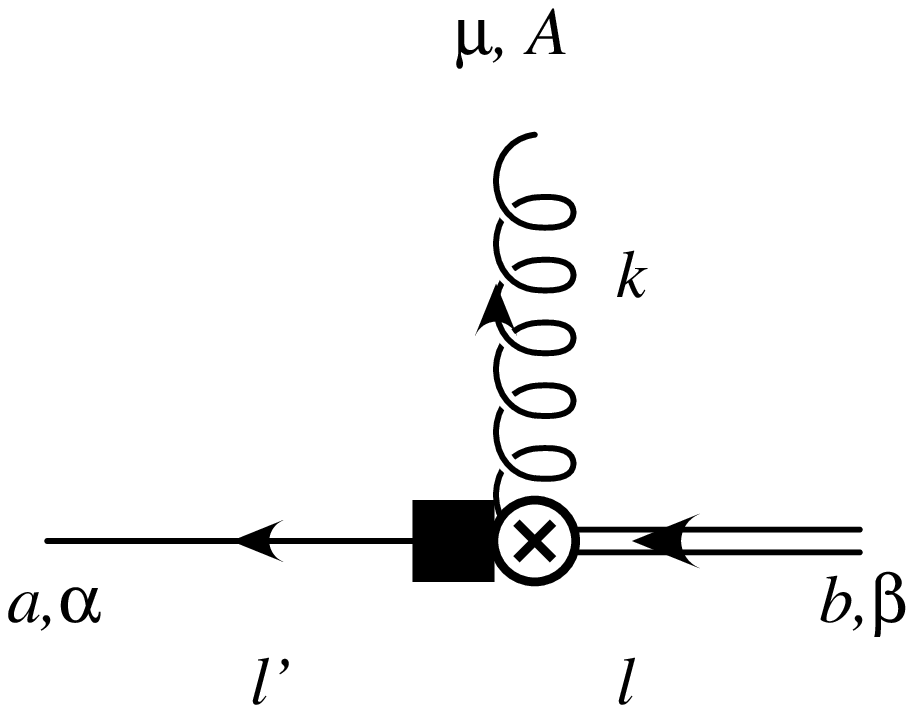,width=4cm}}
    \mbox{\hspace{2cm}}
    \left[
      \Gamma \frac{-i g}{2M_0} 
      \gamma_{\mu}
      \cos\left( l-\frac{k}{2}\right)_{\mu}
      \frac{1+\gamma_4}{2} 
    \right]_{\alpha \beta} (T^A)_{a b}
    \hat{\delta}_{\mu_4}
  \end{equation}

\item (anti-)heavy-light current:
  \begin{equation}
    \mbox{\hspace{-5.5cm}}
    \raisebox{-2em}{\psfig{file=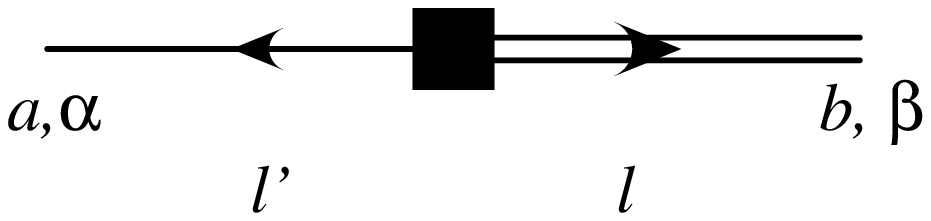,width=4cm}}
    \mbox{\hspace{2cm}}
    \left[
      \Gamma \frac{1-\gamma_4}{2} 
    \right]_{\alpha \beta} \delta_{a b}
  \end{equation}

\item (anti-)heavy-light current with the FWT rotation:
  \begin{equation}
    \mbox{\hspace{-2.5cm}}
    \raisebox{-2em}{\psfig{file=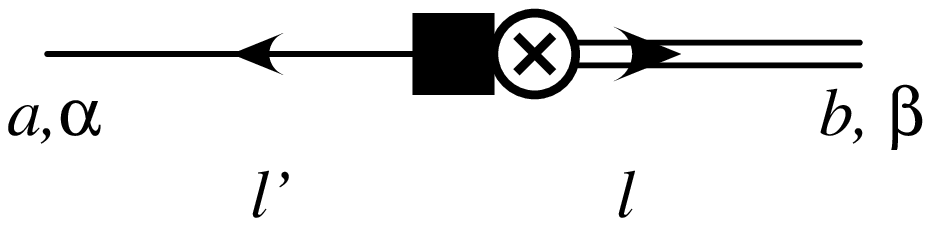,width=4cm}}
    \mbox{\hspace{2cm}}
    \left[
      \Gamma \frac{i}{2M_0} \sum_{i=1}^3 \gamma_i \sin l_i  
      \frac{1-\gamma_4}{2} 
    \right]_{\alpha \beta} \delta_{a b}
  \end{equation}

\item vertex from rotated (anti-)heavy-light current:
  \begin{equation}
    \mbox{\hspace{-1.1cm}}
    \raisebox{-2em}{\psfig{file=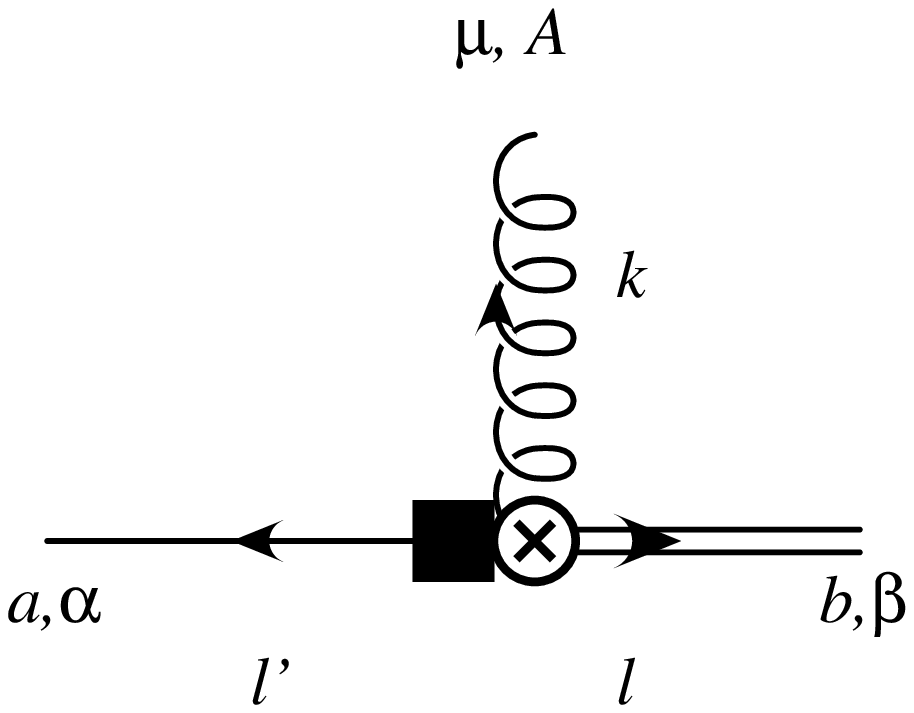,width=4cm}}
    \mbox{\hspace{2cm}}
    \left[
      \Gamma \frac{-i g}{2M_0}  
      \gamma_{\mu}
      \cos\left(l+\frac{k}{2}\right)_{\mu}
      \frac{1-\gamma_4}{2} 
    \right]_{\alpha \beta} (T^A)_{a b}
    \hat{\delta}_{\mu_4}
  \end{equation}

\end{itemize}


\section{One-loop integrals}
\label{sec:One-loop_integrals}
We list the integrals appearing in the one-loop calculations
with the NRQCD action.
\begin{eqnarray}
  \label{eq:I_A}
  I_A & = & 
  \int\frac{d^{4}l}{(2\pi)^4}\
  G(l)S(-l)Q(-l) \nonumber \\
  &&
  \times\left[
    \left(
      \cos\frac{l_4}{2}\sin l_{4}
      +\frac{1}{2}\tilde{l}^{2}\sin\frac{l_4}{2}
    \right) Z(l)
    + \sum_{i=1}^3
    \left(
      \sin^{2}l_{i}
      +\tilde{l}^{2}\sin^{2}\frac{l_{i}}{2}
    \right) \frac{1}{2}X(l) \right.\nonumber\\
  & & \mbox{} \left.
    + \frac{i}{2}\left(1+\frac{c_{sw}}{4}\tilde{l}^2\right)
    \left(
      \sum_{i=1}^3\sin^2l_i
      \sum_{j=1}^3\cos^2\frac{l_j}{2}
      -\sum_{i=1}^3\sin^2l_i \cos^2\frac{l_i}{2} 
    \right) Y(l)
    \right] \nonumber \\
  & &
  - \int\frac{d^4l}{(2\pi)^4}
  \,\frac{\theta(1-l^2)}{(l^2)^2}, \\
  \label{eq:I_B}
  I_B & = & 
  \int\frac{d^4l}{(2\pi)^4}\
  G(l)S(-l)Q(-l) \nonumber \\
  &&
  \times\left[
    -i\left(
      -\sin\frac{l_4}{2}\sin l_{4}
      +\frac{1}{2}\tilde{l}^{2}\cos\frac{l_4}{2}
      -\frac{c_{\rm sw}}{2}\cos\frac{l_4}{2} 
      \sum_{i=1}^3 \sin^2 l_i
    \right) Z(l) \right. \nonumber \\
  & & \mbox{} \left.
    + i \sum_{i=1}^3
    \left(
      \sin^{2}\frac{l_i}{2} - \frac{c_{\rm sw}}{4}\sin^2 l_i 
    \right)
    \sin l_{4}\, X(l)
  \right],
  \\
  \label{eq:I_C}
  I_C & = & 
  \int\frac{d^{4}l}{(2\pi)^4}\
  G(l)S(-l)Q(-l) \nonumber \\
  &&
  \times\left[
    -\frac{i}{6}\left(1+\frac{c_{sw}}{4}\hat{l}^2\right)
    \left(
      \sum_{i=1}^3\sin^2 l_i
      \sum_{j=1}^3\cos^2 \frac{l_j}{2}
      -\sum_{i=1}^3\sin^2 l_i \cos^2 \frac{l_i}{2} 
    \right) Y(l)
  \right],
  \\
  \label{eq:I_D}
  I_D & = & 
  \int\frac{d^4l}{(2\pi)^4}\
  G(l)S(-l)Q(-l)\
  \frac{1}{6M_0} \nonumber \\
  & &
  \times\left[
    \sum_{i=1}^3 \sin^{2}l_i
    \left(
      -\sin\frac{l_4}{2}
      +\frac{c_{\rm sw}}{2}\cos \frac{l_4}{2} \sin l_4
    \right) Z(l) \right.\nonumber \\
  & & \mbox{}
    + \sum_{i=1}^3 \sin^{2}l_i
    \left(
      \sin^{2}\frac{l_4}{2}
      -\frac{c_{\rm sw}}{4} \sin^2 l_4 
    \right) X(l) \nonumber \\
  & & \left. \mbox{} 
    -\frac{i}{2}
    \left(
      \tilde{l}^2 -c_{sw}\sum_{\mu=1}^{4}\sin^2 l_{\mu}
    \right)
    \left(
      \sum_{i=1}^{3} \cos^2 \frac{l_i}{2}
      \sum_{j=1}^{3} \sin^2 l_j
      -\sum_{i=1}^{3}  \sin^2 l_i \cos^2 \frac{l_i}{2}
    \right) Y(l)
  \right],
  \\
  \label{eq:I_E}
  I_E & = &
  \int\frac{d^4l}{(2\pi)^4}\
  G(l)S(-l)Q(-l)\
  \frac{i}{6M_0}
  \left( 1 + \frac{c_{\rm sw}}{4} \tilde{l}^2 \right)
  \nonumber \\
  & &
  \times\left[
    \left( \cos \frac{l_4}{2} Z(l)
      -\frac{1}{2} \sin l_4 X(l)
    \right)
    \sum_{i=1}^{3} \sin^{2}l_i \right. \nonumber\\
  & & \mbox{} \left.
    +\,i \left(
      \sum_{i=1}^{3} \sin^2 l_i 
      \sum_{j=1}^{3} \cos^2 \frac{l_j}{2}
      -\sum_{i=1}^{3} \sin^2 l_i \cos^2 \frac{l_i}{2}
      \right) \sin l_4 \, Y(l)
  \right],
  \\
  \label{eq:I_F}
  I_F & = & -
  \int\frac{d^4l}{(2\pi)^4}\
  G(l)S(-l) \frac{1}{12M_0} \nonumber\\
  & &
  \times \sum_{i=1}^3
  \left(
    \tilde{l}^2\cos^{2}\frac{l_i}{2}
    -\sin^{2}l_i
    +c_{\rm sw} \sin^2 l_i \cos^{2}\frac{l_i}{2}
    -c_{\rm sw} \sum_{\mu=1}^4 \sin^2 l_{\mu} 
    \cos^{2}\frac{l_i}{2}
  \right),
  \\
  \label{eq:I_G}
  I_G & = &
  -\int\frac{d^4l}{(2\pi)^4}\
  G(l)Q(-l)^2 
  \left(
    \sum_{i=1}^{3} \sin^{2}\frac{l_i}{2} X(l)^2 + Z(l)^2
  \right) \nonumber \\
  & &
  + \int\frac{d^4l}{(2\pi)^4}
  \left[\frac{2M_0}{-2 i M_0 l_4 + l^2}\right]^2
  \frac{\theta(1-l^2)}{l^2}
  - \frac{4}{(4\pi)^2}\sinh^{-1}\frac{1}{2M_0}, \\
  \label{eq:I_H}
  I_H & = & 
  - \int\frac{d^4l}{(2\pi)^4}\ 
  G(l) Q(-l)^2\
  \frac{1}{3} \left(
    \sum_{i=1}^3\sin^{2}l_i
    \sum_{j=1}^3\cos^{2}\frac{l_j}{2}
    -\sum_{i=1}^3\sin^{2}l_i\cos^{2}\frac{l_i}{2}
  \right) Y(l)^2,
  \\
  \label{eq:I_I}
  I_I & = & 
  - \int\frac{d^4l}{(2\pi)^4}\ 
  G(l) Q(-l)^2 \frac{1}{4M_0^2} \nonumber \\
  & &
  \times\left[
    \frac{1}{3} Z(l)^2 \sum_{i=1}^3\sin^{2}l_i
    +\frac{1}{3} X(l)^2 
    \sum_{i=1}^3\sin^{2}\frac{l_i}{2} 
    \sum_{j=1}^3\sin^{2}l_j
    \right. \nonumber \\
  & & \mbox{} \left.
    -\frac{1}{3} Y(l)^2 
    \left( 
      \sum_{i=1}^3\sin^{2}l_i \sum_{j=1}^3\sin^{2}l_j
      \sum_{k=1}^3\cos^{2}\frac{l_k}{2}
       - \sum_{i=1}^3\sin^{2}l_i\cos^{2}\frac{l_i}{2}
         \sum_{j=1}^3\sin^{2}l_j
    \right)
  \right],
  \\
  \label{eq:I_J}
  I_J & = & 
  - \int\frac{d^4l}{(2\pi)^4}\ 
  G(l)\ \frac{1}{12M_0^2}\sum_{i=1}^3 \cos^{2}\frac{l_i}{2}\\
  \label{eq:I_K}
  I_K & = & 
  \int\frac{d^4l}{(2\pi)^4}\ 
  G(l) Q(-l) \frac{1}{4M_0^2} \nonumber \\
  & &
  \times\left[
    \frac{1}{6} \sum_{i=1}^{3} \sin^{2}l_i X(l)
    -\frac{i}{3}
    \left( 
      \sum_{i=1}^{3}\sin^{2}l_i
      \sum_{j=1}^{3}\cos^{2}\frac{l_j}{2}
      -\sum_{i=1}^{3}\sin^{2}l_i\cos^{2}\frac{l_i}{2} 
    \right) Y(l)
  \right],
  \\
  \label{eq:I_L}
  I_L & = &
  - \int\frac{d^4l}{(2\pi)^4}\ 
  G(l)S(-l)^2 
  \nonumber\\
  & &
  \times\frac{1}{12}\left(
    \sum_{\alpha,\beta=1}^{4}
    \sin^2 l_{\alpha}\cos^2\frac{l_{\beta}}{2}
    -\sum_{\alpha=1}^4
    \sin^2 l_{\alpha}\cos^2\frac{l_{\alpha}}{2}
  \right)
  \left(
    1 +\frac{1}{16}{c_{\rm sw}}^2 (\tilde{l}^2)^2
    +\frac{1}{2}c_{\rm sw}\tilde{l}^2
  \right) \nonumber \\
  & &
  + \int\frac{d^4l}{(2\pi)^4}\
  \frac{\theta(1-l^2)}{4(l^2)^2}, \\
  \label{eq:I_M}
  I_M &=& 
  \int\frac{d^4l}{(2\pi)^4} 
  G(l)S(-l)^2 \nonumber \\
  & & 
  \times\frac{1}{4}
  \left[
    -\sum_{\mu=1}^{4}\sin^2 l_{\mu} 
    \sum_{\nu=1}^{4}\sin^2\frac{l_{\nu}}{2}
    -\frac{1}{4}{c_{\rm sw}}^2
    \sum_{\mu=1}^{4}\sin^2 l_{\mu} 
    \sum_{\nu=1}^{4}\sin^2 l_{\nu} 
    \sum_{\rho=1}^{4}\cos^2\frac{l_{\rho}}{2}
    \right. \nonumber \\
  & & \mbox{}
    +\frac{1}{4}{c_{\rm sw}}^2
    \sum_{\mu=1}^{4}\sin^2 l_{\mu}
    \sum_{\nu=1}^{4}\sin^2 l_{\nu}\cos^2\frac{l_{\nu}}{2}
    -\frac{1}{4}(\tilde{l}^2)^2
    \sum_{\mu=1}^{4}\cos^2\frac{l_{\mu}}{2}
    +\frac{1}{2}\tilde{l}^2
    \sum_{\mu=1}^{4}\sin^2 l_{\mu} \nonumber \\
  & & \mbox{} \left.
    +\frac{1}{2}{c_{\rm sw}}\tilde{l}^2
    \sum_{\mu=1}^{4}\sin^2 l_{\mu}
    \sum_{\nu=1}^{4}\cos^2\frac{l_{\nu}}{2}
    -\frac{1}{2}{c_{\rm sw}}\tilde{l}^2
    \sum_{\mu=1}^{4}\sin^2 l_{\mu}\cos^2\frac{l_{\mu}}{2}
  \right],
  \\
  \label{eq:I_N}
  I_N &=& 
  \int\frac{d^4l}{(2\pi)^4}\
  G(l)S(-l)^2 
  \left[
    -\frac{1}{2}\tilde{l}^2 
    \sum_{\mu=1}^4\sin^2 l_{\mu}
    -\frac{1}{4}(\tilde{l}^2)^2
    \sum_{\mu=1}^4\sin^2\frac{l_{\mu}}{2} 
  \right. \nonumber\\
  & & \mbox{}
    +\frac{1}{3}
    \left(
      \sum_{\mu=1}^4\sin^2 l_{\mu}
      \sum_{\nu=1}^4\cos^2\frac{l_{\nu}}{2}
      -4\sum_{\mu=1}^4\sin^2 l_{\mu}\cos^2\frac{l_{\mu}}{2}
    \right) \nonumber \\
  & & \mbox{} \left.
    +\frac{1}{3}
    \left(
      \frac{1}{2}c_{sw}\tilde{l}^2
      +\frac{1}{16}c_{sw}^2 (\tilde{l}^2)^2
    \right)
    \left(
      \sum_{\mu=1}^4\sin^2 l_{\mu}
      \sum_{\nu=1}^4\cos^2\frac{l_{\nu}}{2}
      -\sum_{\mu=1}^4
      \sin^2 l_{\mu}\cos^2\frac{l_{\mu}}{2}
    \right)
  \right],
\end{eqnarray}
where functions $X(l)$, $Y(l)$ and $Z(l)$ in the integrands
are
\begin{eqnarray}
  \label{eq:Type_A_current_func_X}
  X(l) & = &
  \left[
    e^{i l_4}A^{(0)}(l)^n + 1
  \right]
  \frac{1}{2 M_0 n} \sum_{m=0}^{n-1}A^{(0)}(l)^m, \\
  \label{eq:Type_A_current_func_Y}
  Y(l) & = & 
  A^{(0)}(l)^n \left[e^{i l_4} + 1\right]
  \frac{i c_4}{4M_0}, \\
  \label{eq:Type_A_current_func_Z}
  Z(l) & = &
  A^{(0)}(l)^n \left[-i e^{i \frac{l_4}{2}}\right]. \\
\end{eqnarray}
There are infra-red divergences in the integrals
$I_A$ (\ref{eq:I_A}), $I_G$ (\ref{eq:I_G}) and 
$I_L$ (\ref{eq:I_L}),
for which we subtract an continuum expression from their
integrand in the region $l^2<1$.
We, then, add back their analytic integral except for the
$\ln(a\lambda)$ term, so that $I_X$ becomes finite.
When those integrals appear in the expressions of on-shell
amplitude, the infra-red divergences will be added.

\section{One-loop four-quark amplitudes}
\label{sec:One-loop_four-quark_amplitudes}

The lattice one-loop expression of the perturbative on-shell
amplitudes is
\begin{eqnarray}
  \langle \bar{b}\Gamma q~\bar{b}\Gamma q \rangle
  & = & 
  Z_h^{lat} Z_l^{lat}
  \Biggl[ 
    \langle \bar{b}\Gamma q~\bar{b}\Gamma q \rangle_0 
    \nonumber \\
  & &
  + \frac{\alpha_s}{4\pi}
  \left(
      X_{heavy-light}^{singlet}
    + X_{heavy-light}^{octet}
    + X_{heavy-heavy}^{octet}
    + X_{light-light}^{octet} 
  \right), 
  \Biggr]
\end{eqnarray}
where $Z_l^{lat}$ and $Z_h^{lat}$ are light and heavy quark
wave function renormalizations respectively, and give by
\begin{eqnarray}
  Z_l^{lat} & = & 
  1 +\frac{\alpha_s}{4\pi} C_F 
  \left[\ln(a^2\lambda^2) + C_l \right], \\
  Z_h^{lat} & = & 
  1 +\frac{\alpha_s}{4\pi} C_F
  \left[-2 \ln(a^2\lambda^2) + C_h \right].
\end{eqnarray}

\begin{figure}
  \begin{center}
    \begin{tabular}{ccc}
      \raisebox{0em}{\psfig{file=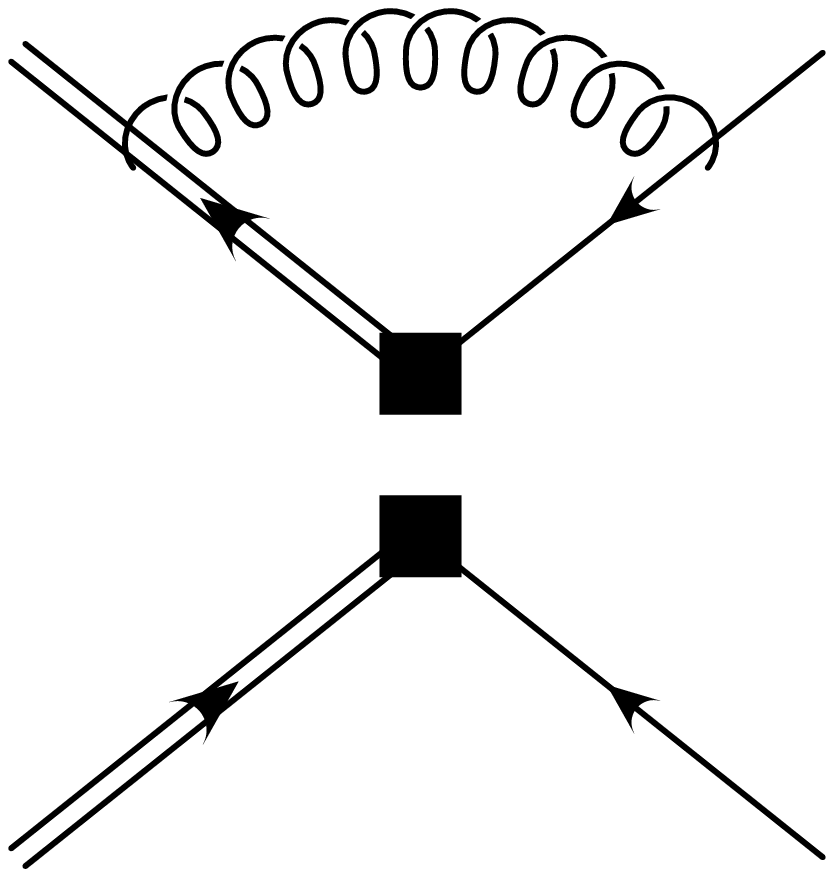,width=4.5cm}} &
      \raisebox{0em}{\psfig{file=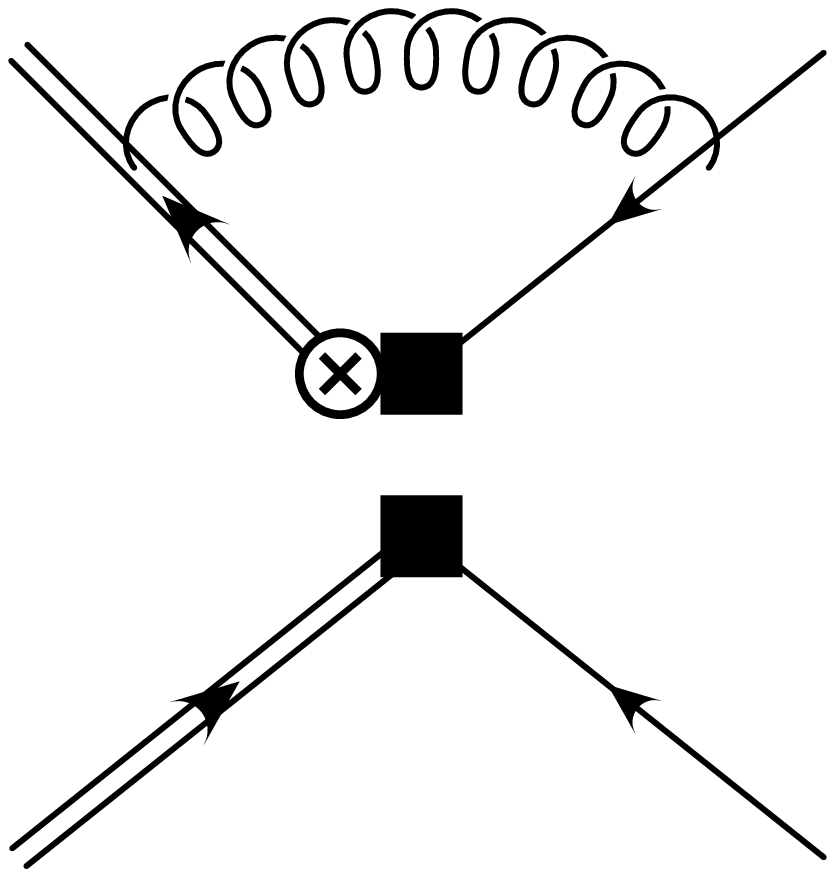,width=4.5cm}} &
      \raisebox{0em}{\psfig{file=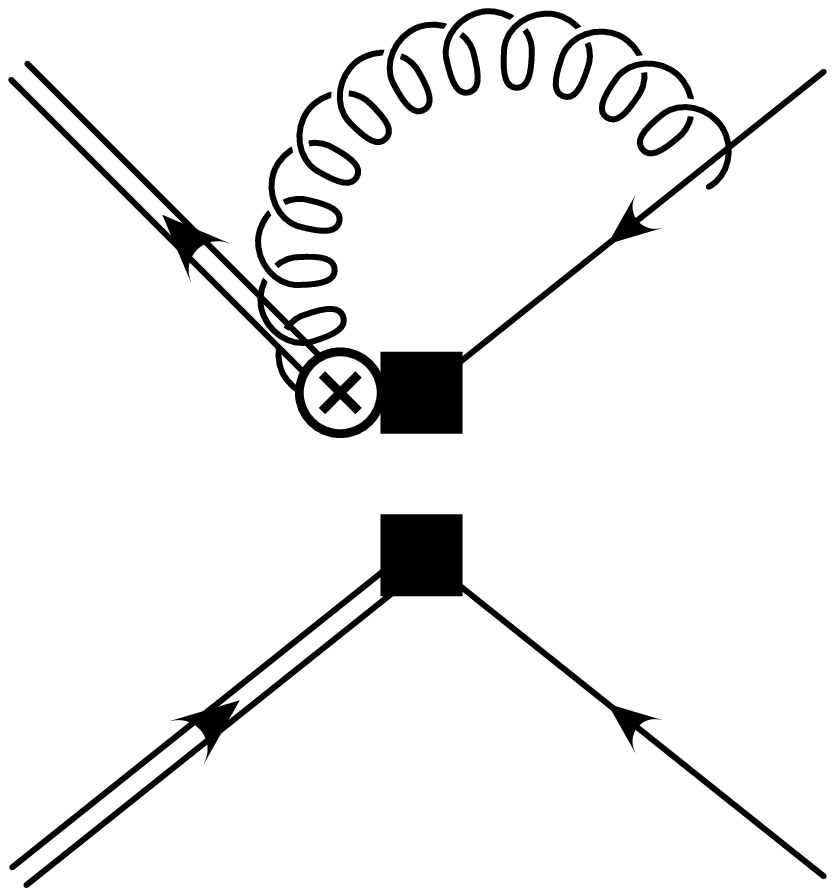,width=4.5cm}}\\
      \raisebox{0em}{\psfig{file=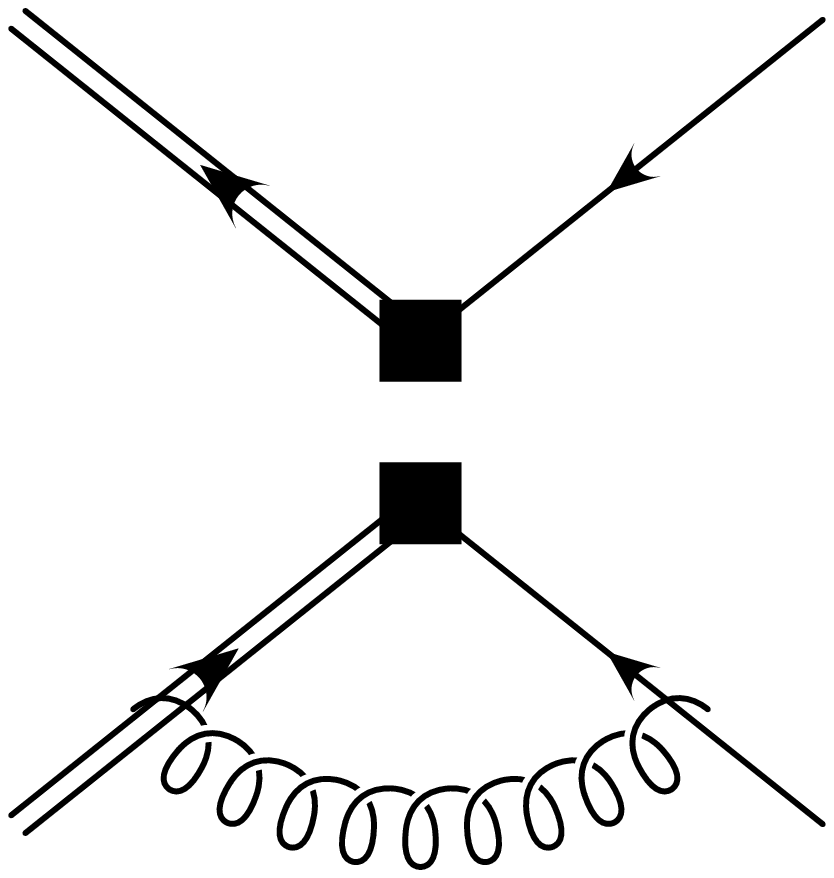,width=4.5cm}} &
      \raisebox{0em}{\psfig{file=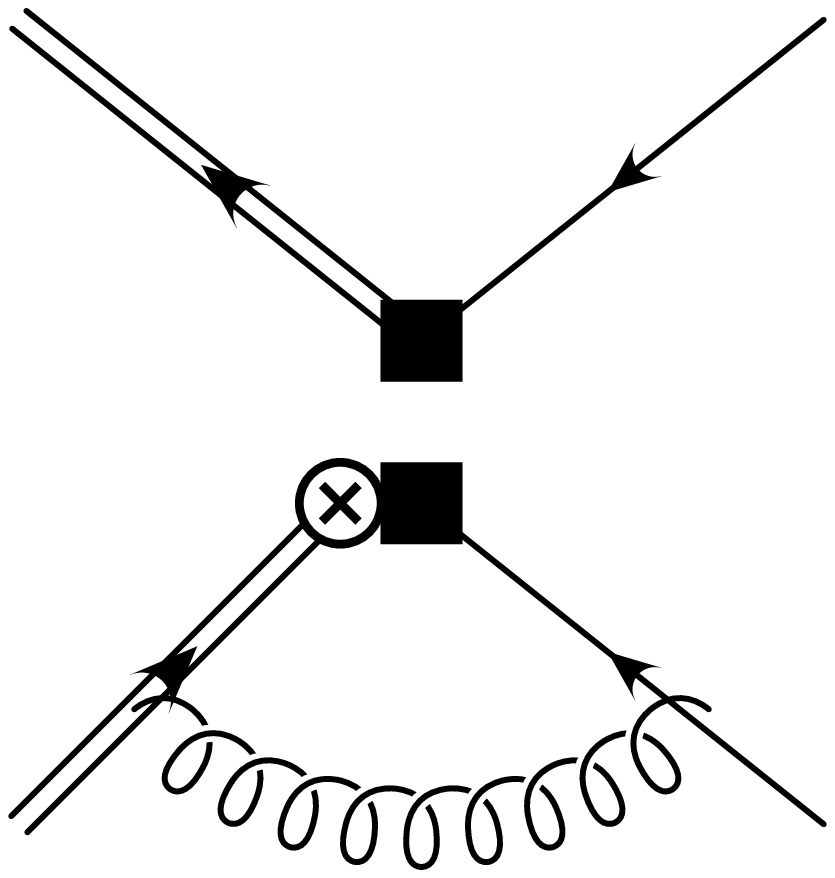,width=4.5cm}} &
      \raisebox{0em}{\psfig{file=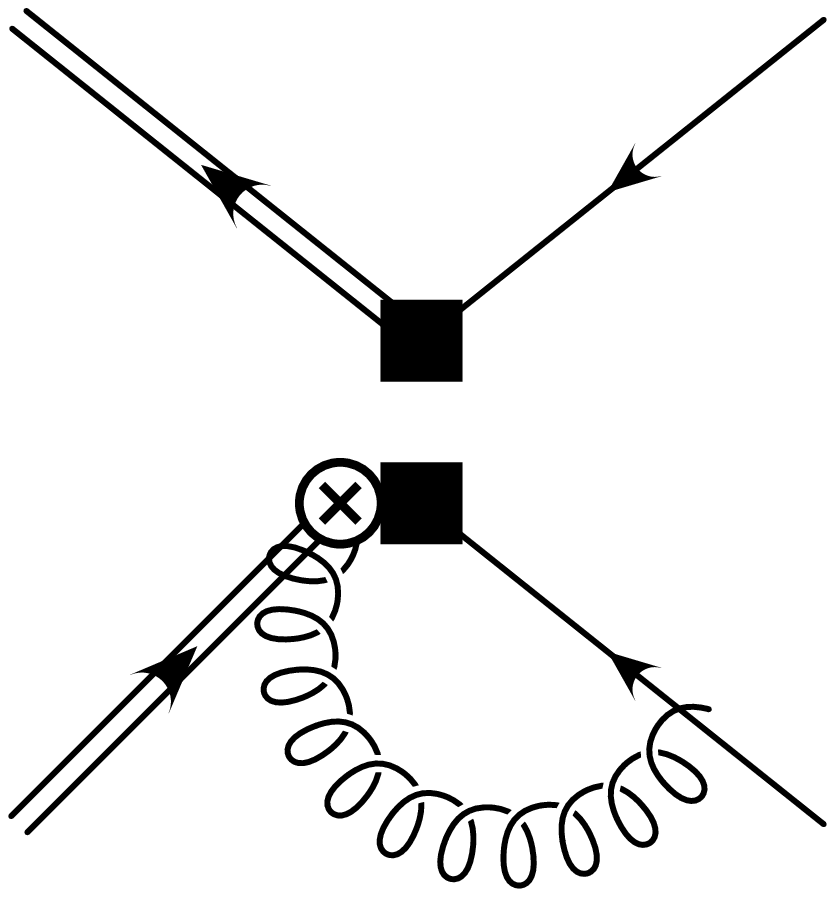,width=4.5cm}}
    \end{tabular}
  \end{center}
  \caption{heavy-light vertex corrections in color singlet channel}
  \label{fig:vertex4_12}
\end{figure}

\begin{figure}
  \begin{center}
    \begin{tabular}{ccc}
      \raisebox{0em}{\psfig{file=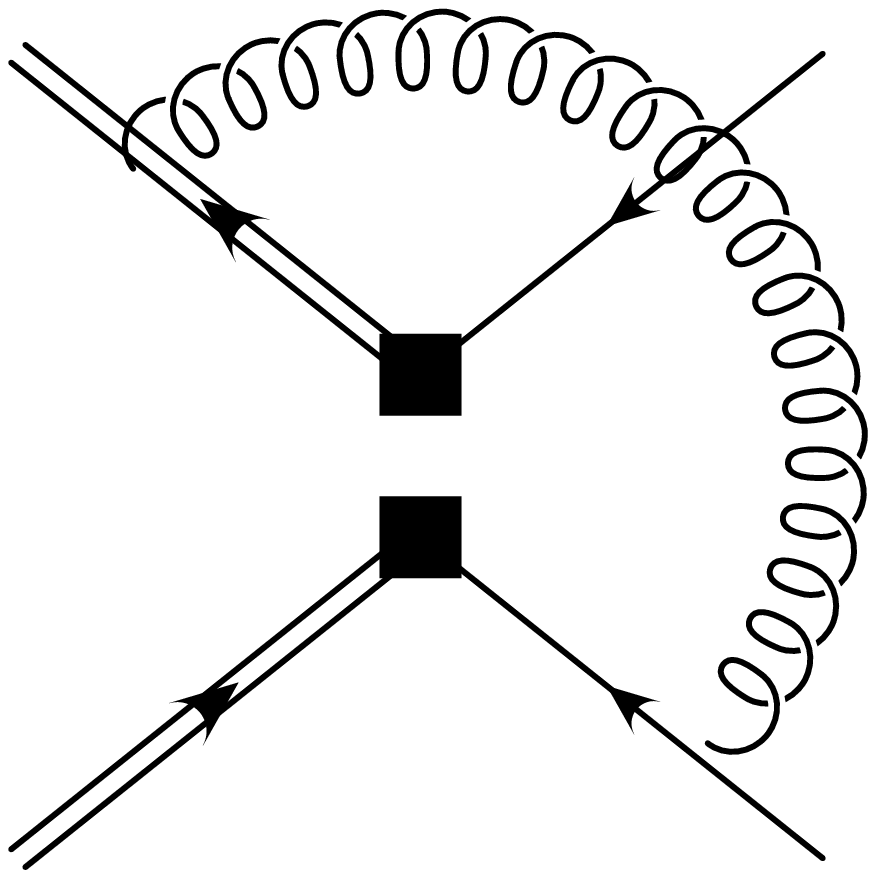,width=4.5cm}} &
      \raisebox{0em}{\psfig{file=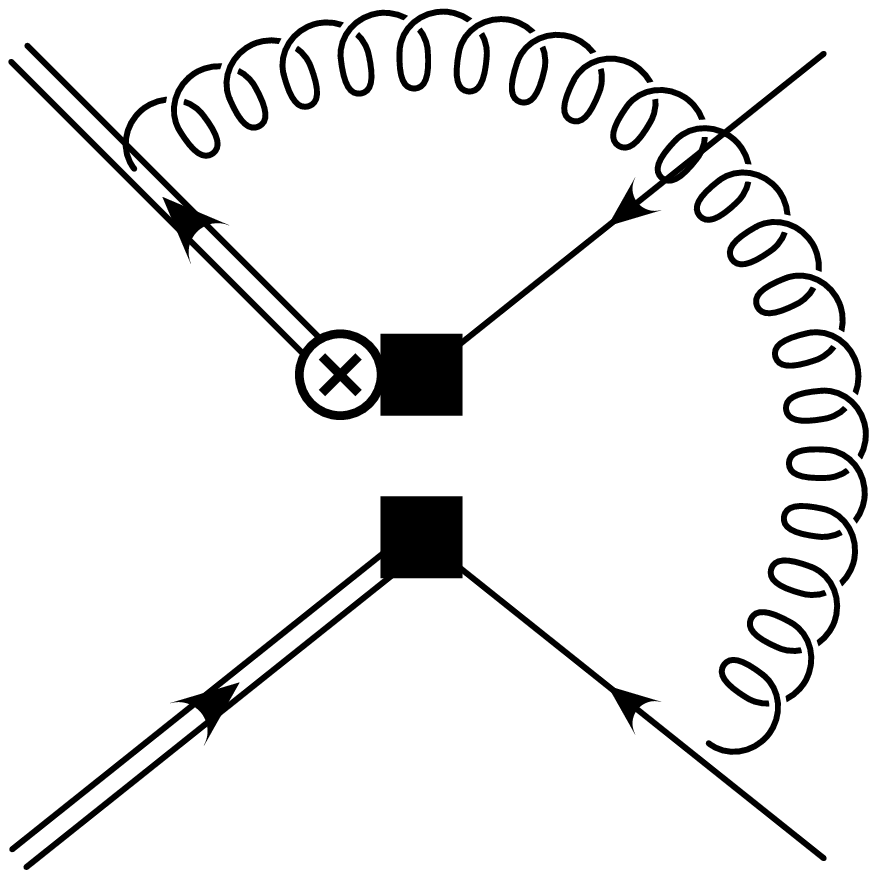,width=4.5cm}} &
      \raisebox{0em}{\psfig{file=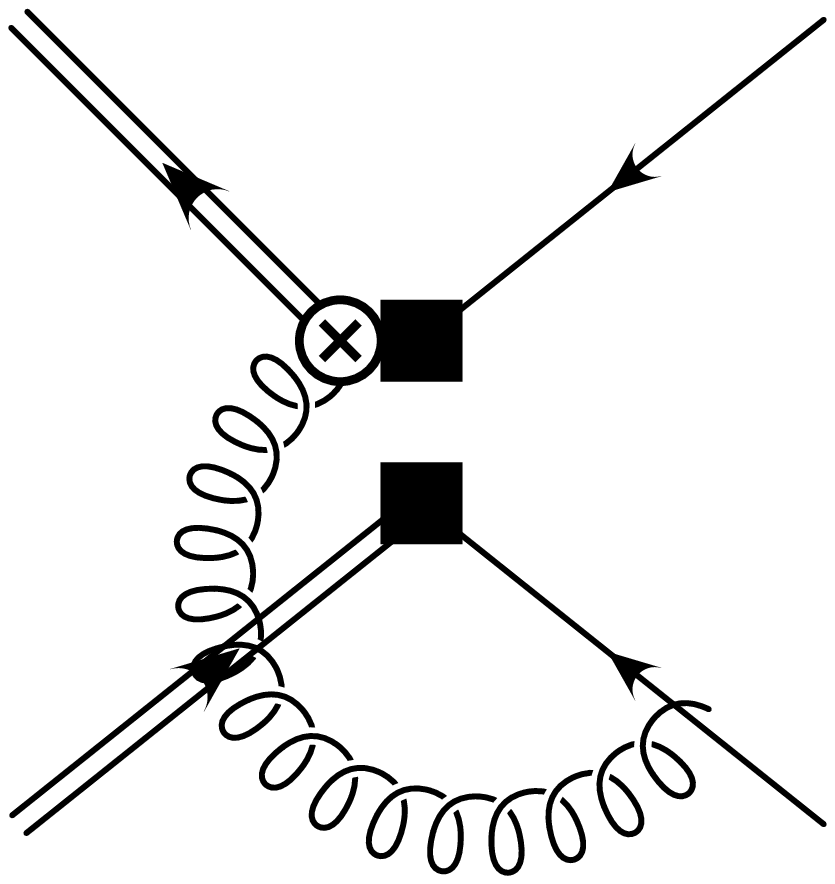,width=4.5cm}}\\
      \raisebox{0em}{\psfig{file=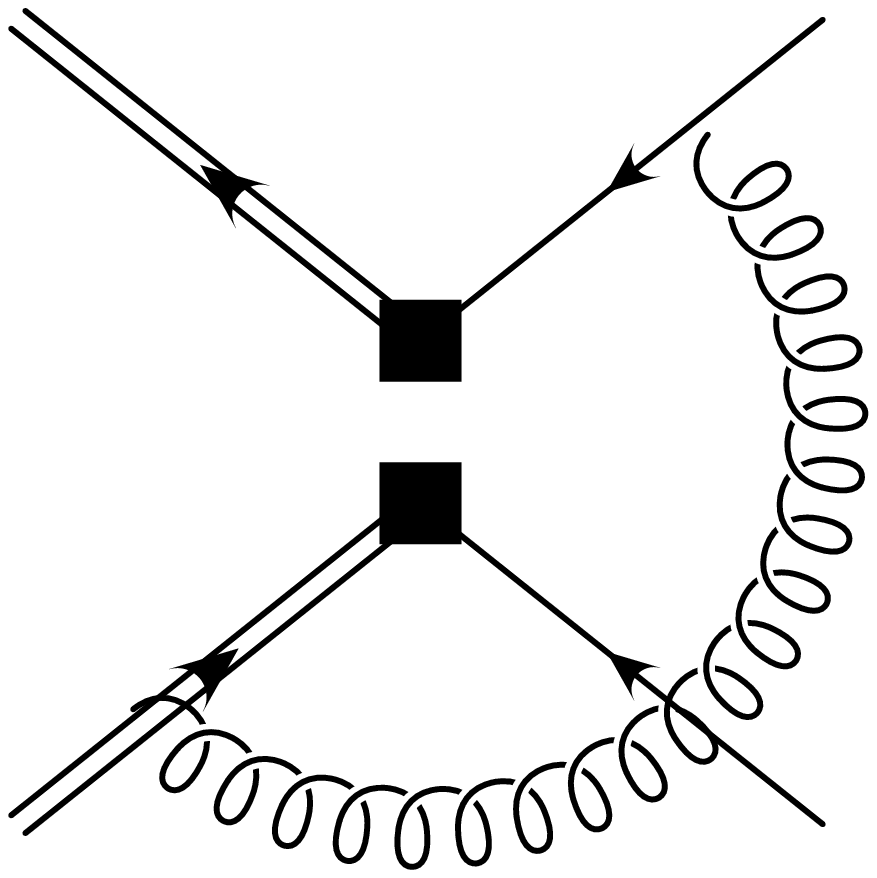,width=4.5cm}} &
      \raisebox{0em}{\psfig{file=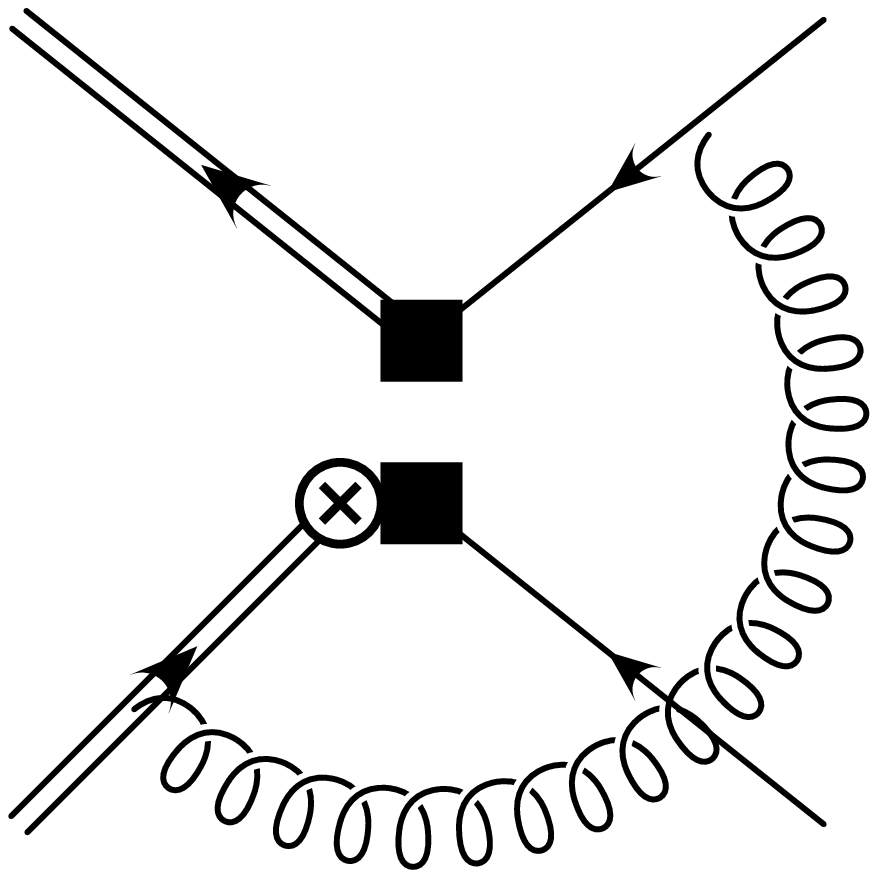,width=4.5cm}} &
      \raisebox{0em}{\psfig{file=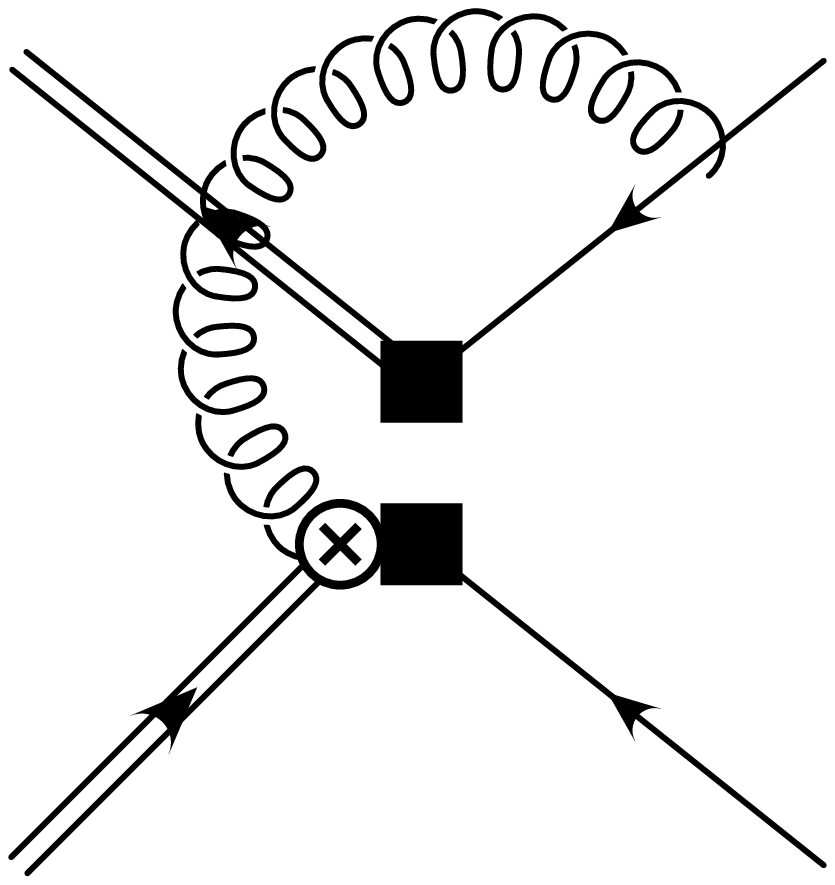,width=4.5cm}}
    \end{tabular}
  \end{center}
  \caption{heavy-light vertex corrections in color octet channel}
  \label{fig:vertex4_34}
\end{figure}

\begin{figure}
  \begin{center}
    \begin{tabular}{ccc}
      \raisebox{0em}{\psfig{file=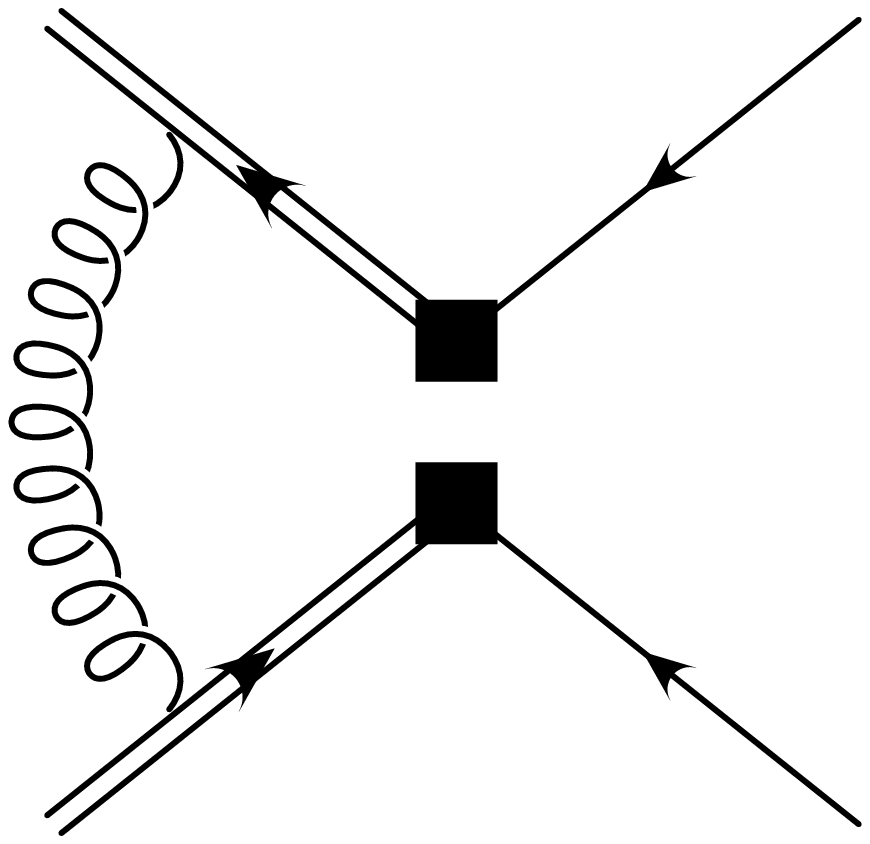,width=4.5cm}} &
      \raisebox{0em}{\psfig{file=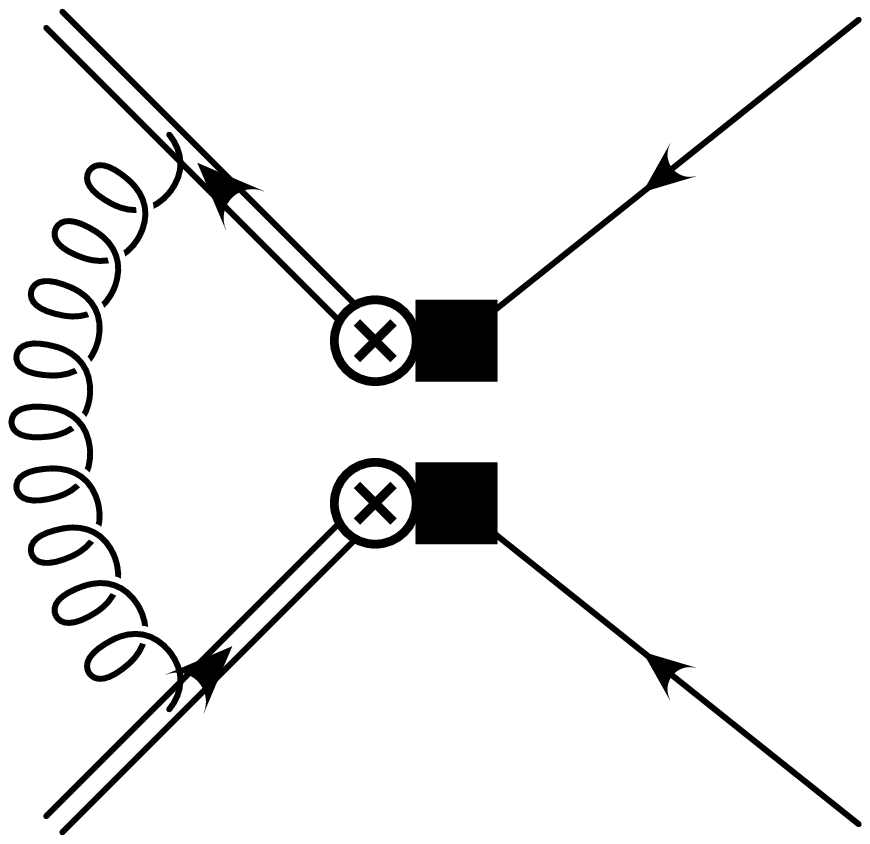,width=4.5cm}} &
      \raisebox{0em}{\psfig{file=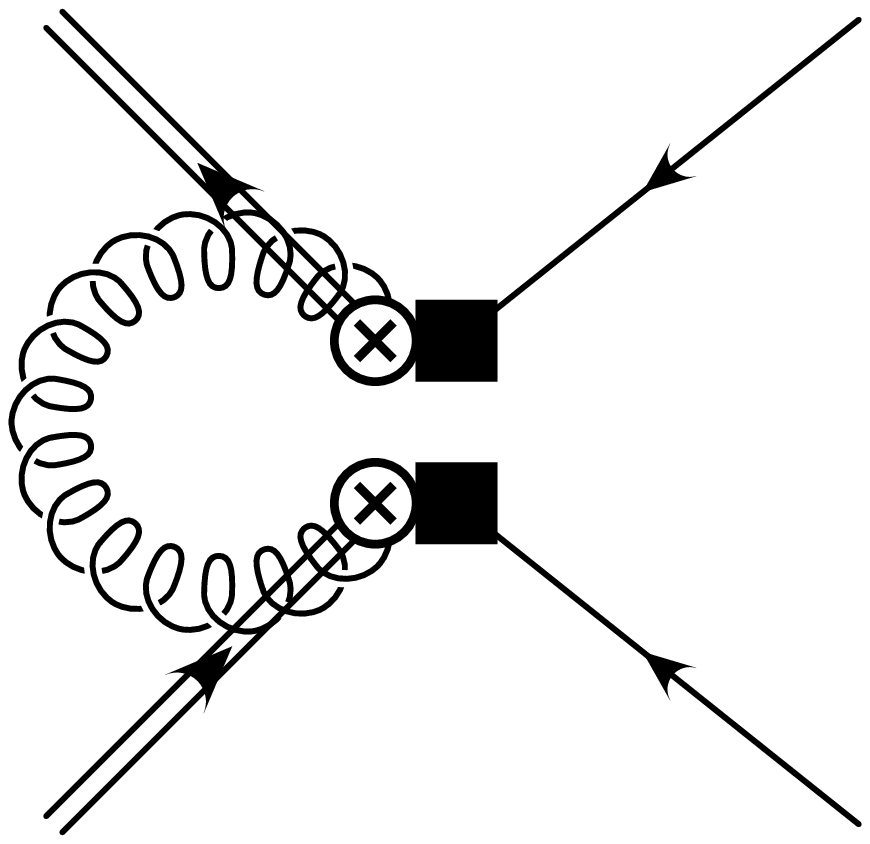,width=4.5cm}}\\
      \raisebox{0em}{\psfig{file=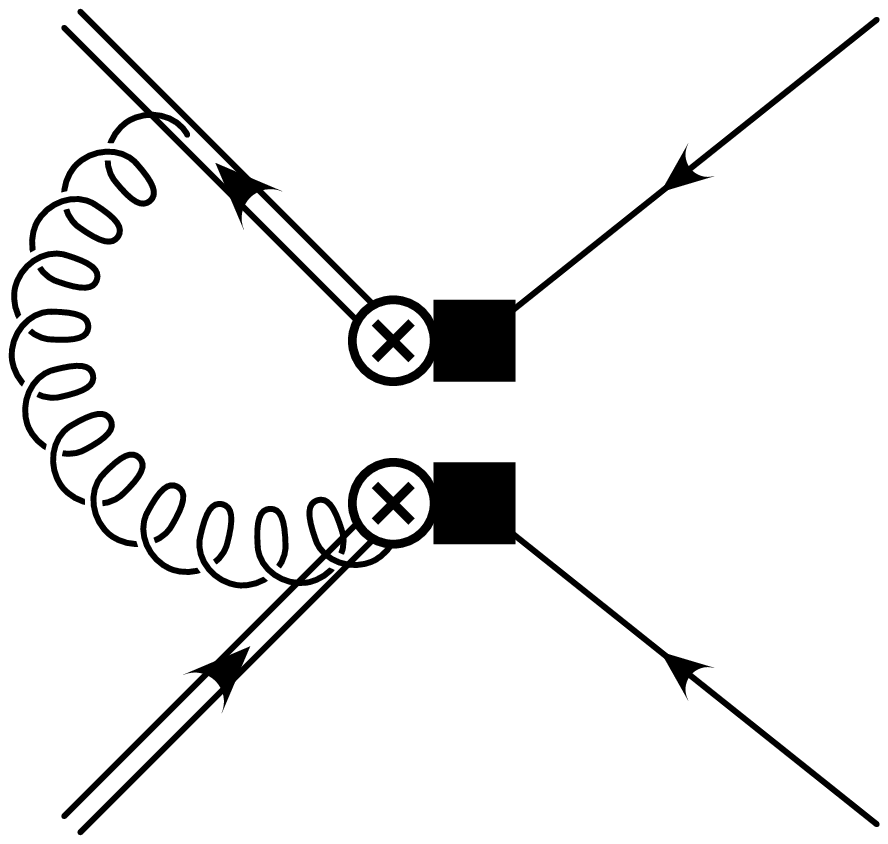,width=4.5cm}} &
      \raisebox{0em}{\psfig{file=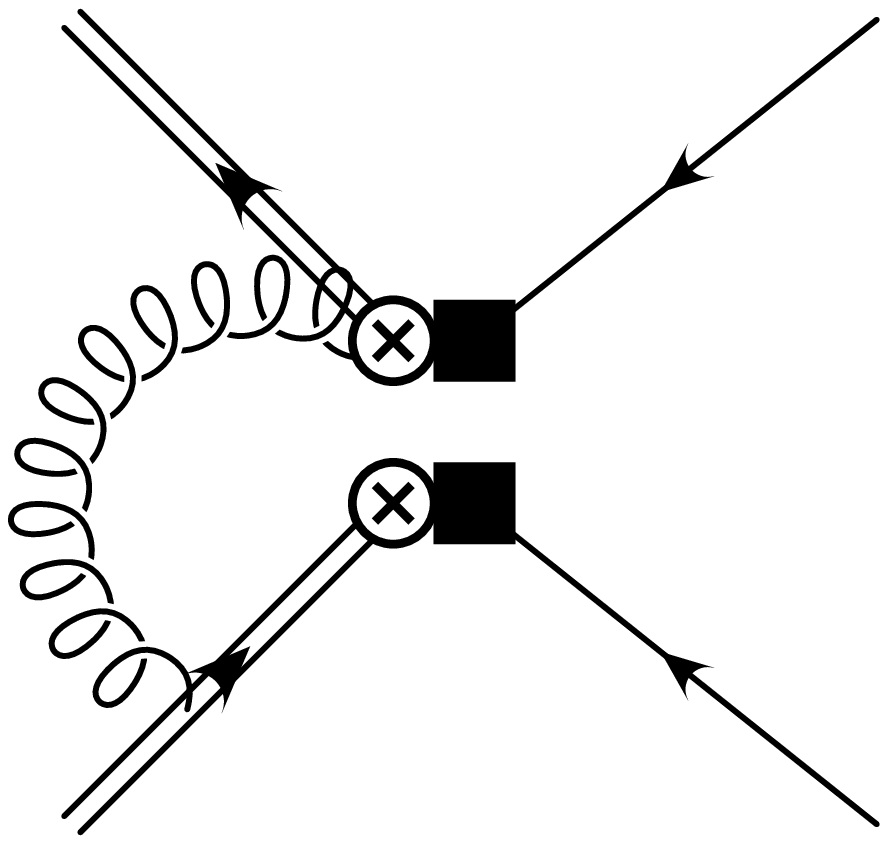,width=4.5cm}}
    \end{tabular}
  \end{center}
  \caption{heavy-heavy vertex corrections in color octet channel}
  \label{fig:vertex4_5}
\end{figure}

\begin{figure}
  \begin{center}
    \begin{tabular}{ccc}
      \raisebox{0em}{\psfig{file=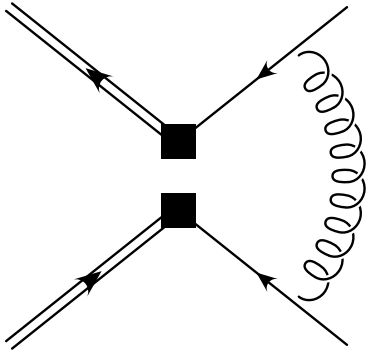,width=4.5cm}} & 
      \makebox[4.5cm]{} & \makebox[4.5cm]{}
    \end{tabular}
  \end{center}
  \caption{light-light vertex correction in color octet
    channel} 
  \label{fig:vertex4_6}
\end{figure}

The vertex corrections $X$'s are classified by the topology
of Feynman diagrams.
Figure \ref{fig:vertex4_12} shows the diagrams in which the
gluon line connects heavy and light quarks and the flow of
color is closed.
The amplitude of these diagrams is denoted as
$X_{heavy-light}^{singlet}$. 
In Figure \ref{fig:vertex4_34} the gluon line connects heavy
and light quarks, but the color flow is not closed, which we
call $X_{heavy-light}^{octet}$.
Figures \ref{fig:vertex4_5} and \ref{fig:vertex4_6}
represent the diagrams in which the gluon line mediates
between two heavy quarks or between two light quarks
respectively. 
The color flow cannot close in these diagrams, and we denote
them as $X_{heavy-heavy}^{octet}$ and
$X_{light-light}^{octet}$, respectively.
The expressions of the one-loop amplitudes are the
following.
\begin{eqnarray}
  \label{eq:X_heavy-light^singlet}
  X_{heavy-light}^{singlet} & = &
  (4\pi)^2 C_F \Biggl[ 
  2 \left( I_A-\frac{1}{16\pi^2} \ln(a^2\lambda^2) \right)
  \langle 
    \overline{b} \Gamma q ~
    \overline{b} \Gamma q 
  \rangle_0
  \nonumber \\
  & &
  +2 I_B 
  \langle 
    \overline{b} \gamma_4 \Gamma \gamma_4 q ~
    \overline{b} \Gamma q
  \rangle_0
  \nonumber \\         
  & &  
  + 2 I_C \sum_{i,j=1}^{3}
  \langle 
    \overline{b} 
    \gamma_i \gamma_j \Gamma \gamma_j \gamma_i q ~
    \overline{b} \Gamma q
  \rangle_0 
  \nonumber \\
  & &
  + 2 (I_D+I_F) \sum_{i=1}^{3} 
  \langle 
    \overline{b} \gamma_i \Gamma \gamma_i q ~
    \overline{b} \Gamma q
  \rangle_0
  \nonumber \\         
  & &  
  + 2 I_E \sum_{i=1}^{3} 
  \langle 
    \overline{b} \gamma_4 \gamma_i \Gamma \gamma_i \gamma_4 q~
    \overline{b} \Gamma q
  \rangle_0 
  \Biggr]
  \\
  \label{eq:X_heavy-light^octet}
  X_{heavy-light}^{octet} & = &
  (4\pi)^2 \Biggl[  
  2\left(I_A-\frac{1}{16\pi^2} \ln(a^2\lambda^2) \right)
  \langle 
    \overline{b} \Gamma T^a q ~
    \overline{b} \Gamma T^a q
  \rangle_0
  \nonumber \\
  & & 
  + 2 I_B 
  \langle 
    \overline{b} \Gamma \gamma_4 T^a q ~
    \overline{b} \gamma_4 \Gamma T^a q
  \rangle_0
  \nonumber \\         
  & &  
  + 2 I_C \sum_{i,j=1}^{3}
  \langle 
    \overline{b} \Gamma \gamma_j \gamma_i T^a q ~
    \overline{b} \gamma_i \gamma_j \Gamma T^a q
  \rangle_0
  \nonumber \\
  & &
  + 2 (I_D+I_F) \sum_{i=1}^{3} 
  \langle 
    \overline{b} \Gamma \gamma_i T^a q ~
    \overline{b} \gamma_i \Gamma T^a q
  \rangle_0
  \nonumber \\         
  & &  
  + 2 I_E \sum_{i=1}^{3} 
  \langle 
    \overline{b} \Gamma \gamma_i \gamma_4 T^a q ~
    \overline{b} \gamma_4 \gamma_i \Gamma T^a q
  \rangle_0 
  \Biggr]
  \\
  \label{eq:X_heavy-heavy^octet}
  X_{heavy-heavy}^{octet} & = &
  (4\pi)^2 \Biggl[  
  \left(I_G - 2 \frac{1}{16\pi^2} \ln(a^2\lambda^2)\right)   
  \langle 
    \overline{b} \Gamma T^a q ~
    \overline{b} \Gamma T^a q
  \rangle_0 
  \nonumber \\
  & &
  + I_H \sum_{i=1}^3
  \langle 
    \overline{b} \Gamma \sigma^i T^a q ~
    \overline{b} \Gamma \sigma^i T^a q
  \rangle_0 
  \nonumber\\
  & &  
  + (I_I+I_J+2I_K) \sum_{i=1}^3
  \langle 
    \overline{b} \Gamma \gamma^i T^a q ~
    \overline{b} \Gamma \gamma^i T^a q
  \rangle_0
  \Biggr]
  \\
  \label{eq:X_light-light^octet}
  X_{light-light}^{octet} & = &
  (4\pi)^2 \Biggl[ 
  \left(I_L + \frac{1}{16\pi^2} \ln(a^2\lambda^2)\right)
  \sum_{\mu,\nu=1}^4
  \langle 
    \overline{b} \Gamma \gamma_{\mu}\gamma_{\nu} T^a q ~
    \overline{b} \Gamma \gamma_{\mu}\gamma_{\nu} T^a q
  \rangle_0 
  \nonumber\\
  & &  
  + I_M \sum_{\mu=1}^4
  \langle 
    \overline{b} \Gamma \gamma_{\mu} T^a q ~
    \overline{b} \Gamma \gamma_{\mu} T^a q
  \rangle_0 
  \nonumber \\
  & &
  + I_N
  \langle 
    \overline{b}\Gamma T^a q ~
    \overline{b}\Gamma T^a q
  \rangle_0
  \Biggr]
\end{eqnarray}

%
%

\begin{table}
  \begin{center}
    \begin{tabular}{ccr} 
      $aM_0$ & n & $C_h$ \\
      \hline
 12.0  & 2 &  2.94(12)\\
 10.0  & 2 &  2.63( 7) \\
 7.0 & 2 &  1.86( 5) \\
 6.5 & 2 &  1.67(12) \\
 5.0 & 2 &  0.89(13) \\
 4.5 & 2 &  0.52(12) \\
 4.0 & 2 &  0.07(11) \\
 3.8 & 2 & $-$0.13(12)\\
 3.5 & 2 & $-$0.49(11) \\
 3.0 & 2 & $-$1.20( 9) \\
 2.6 & 2 & $-$1.93( 8) \\
 2.1 & 3 & $-$2.97( 8) \\
 1.5 & 3 & $-$5.10( 6) \\
 1.3 & 3 & $-$6.10( 6) \\
 1.2 & 3 & -6.67( 6) \\
 0.9 & 4 & $-$8.68( 5) \\
    \end{tabular}
    \caption{Wave function renormalization constants for the
      NRQCD heavy quark}
    \label{tab:C_h}
  \end{center}
\end{table}

\begin{table}
 \begin{center}
  \begin{tabular}{cccccccc} 
 $aM_0$ & $n$ & $I_A$  &   $I_B$   &  $I_C$   &   $I_D$   &  $I_E$    & $I_F$ \\
   \hline
 12.0 & 2 & 0.030932(15) & -0.016562(9) & 0.000794(1) & -0.000029(0) & 0.001020(1) & -0.000519(0) \\
 10.0 & 2 & 0.030277(15) & -0.016041(8) & 0.000936(1) & -0.000031(0) & 0.001209(1) & -0.000623(0) \\
  7.0 & 2 & 0.028754(14) & -0.014812(8) & 0.001282(1) & -0.000032(0) & 0.001678(2) & -0.000890(0) \\
  6.5 & 2 & 0.028388(14) & -0.014512(8) & 0.001367(1) & -0.000031(0) & 0.001791(2) & -0.000958(0) \\
  5.0 & 2 & 0.026993(13) & -0.013325(7) & 0.001702(2) & -0.000024(0) & 0.002256(2) & -0.001246(0) \\
  4.5 & 2 & 0.026371(13) & -0.012802(7) & 0.001853(2) & -0.000019(0) & 0.002469(2) & -0.001384(0) \\
  4.0 & 2 & 0.025697(24) & -0.012179(7) & 0.002033(2) & -0.000010(0) & 0.002723(3) & -0.001557(0) \\
  3.8 & 2 & 0.025327(12) & -0.011894(6) & 0.002116(2) & -0.000006(0) & 0.002843(3) & -0.001639(0) \\
  3.5 & 2 & 0.024791(12) & -0.011422(6) & 0.002252(2) &  0.000003(0) & 0.003039(3) & -0.001780(0) \\
  3.0 & 2 & 0.023692(12) & -0.010492(5) & 0.002524(2) &  0.000025(0) & 0.003423(3) & -0.002076(0) \\
  2.6 & 2 & 0.022733(23) & -0.009579(5) & 0.002793(3) &  0.000052(0) & 0.003826(4) & -0.002396(0) \\
  2.1 & 3 & 0.021122(11) & -0.008532(5) & 0.003253(3) &  0.000111(0) & 0.004524(4) & -0.002966(1) \\
  1.5 & 3 & 0.018483(9)  & -0.006337(3) & 0.003995(4) &  0.000252(0) & 0.005707(6) & -0.004153(1) \\
  1.3 & 3 & 0.017439(9)  & -0.005419(3) & 0.004320(4) &  0.000331(0) & 0.006230(5) & -0.004792(1) \\
  1.2 & 3 & 0.016842(17) & -0.004921(3) & 0.004502(4) &  0.000381(0) & 0.006529(6) & -0.005191(1) \\
  0.9 & 4 & 0.014482(17) & -0.003598(2) & 0.005228(5) &  0.000609(0) & 0.007825(7) & -0.006921(1) \\
\end{tabular}
  \end{center}
  \caption{Integrals $I_A$, $I_B$, $I_C$, $I_D$, $I_E$ and 
    $I_F$.} 
  \label{tab:AF}
\end{table}

\begin{table}
  \begin{center}
\begin{tabular}{ccccccc}
$ aM_0 $ & $n$ &  $I_G$   &  $I_H$   &   $I_I$   &  $I_J$   & $I_K$ \\
   \hline
 12.0 & 2 &  0.01858(8) & 0.000075(0) &  0.000029(0) & -0.000160(0) & 0.000008(0) \\
 10.0 & 2 &  0.01674(7) & 0.000104(0) &  0.000041(0) & -0.000231(0) & 0.000013(0) \\
  7.0 & 2 &  0.01223(5) & 0.000193(1) &  0.000073(0) & -0.000471(0) & 0.000037(0) \\
  6.5 & 2 &  0.01107(5) & 0.000220(1) &  0.000081(0) & -0.000547(0) & 0.000045(0) \\
  5.0 & 2 &  0.00653(3) & 0.000338(1) &  0.000118(0) & -0.000924(0) & 0.000095(0) \\
  4.5 & 2 &  0.00439(2) & 0.000400(2) &  0.000134(1) & -0.001140(0) & 0.000127(0) \\
  4.0 & 2 &  0.00183(2) & 0.000474(0) &  0.000153(0) & -0.001443(0) & 0.000176(0) \\
  3.8 & 2 &  0.00063(2) & 0.000515(2) &  0.000160(1) & -0.001599(0) & 0.000203(0) \\
  3.5 & 2 & -0.00138(1) & 0.000580(2) &  0.000171(1) & -0.001885(0) & 0.000254(0) \\
  3.0 & 2 & -0.00545(2) & 0.000718(3) &  0.000185(1) & -0.002566(1) & 0.000385(0) \\
  2.6 & 2 & -0.00964(2) & 0.000862(1) &  0.000182(0) & -0.003416(1) & 0.000563(0) \\
  2.1 & 3 & -0.01528(5) & 0.001154(5) &  0.000146(1) & -0.005237(1) & 0.000977(1) \\
  1.5 & 3 & -0.02721(9) & 0.001680(5) & -0.000234(1) & -0.010264(3) & 0.002270(1) \\
  1.3 & 3 & -0.03274(9) & 0.001932(8) & -0.000613(2) & -0.013665(4) & 0.003201(2) \\
  1.2 & 3 & -0.03592(4) & 0.002081(2) & -0.000926(2) & -0.016038(4) & 0.003865(2) \\
  0.9 & 4 & -0.04662(12) & 0.002760(9) & -0.002752(9) & -0.028512(8) & 0.007507(4) \\
 \end{tabular}
\end{center}
  \caption{Integrals $I_G$, $I_H$, $I_I$, $I_J$ and $I_K$
    for $c_{sw}$=1.} 
  \label{tab:GK}
\end{table}

\begin{table}
  \begin{center}
    \begin{tabular}{ccrrrrr} 
$aM_0$ & n & $\rho_{VLL,VLL}^{lat}$ & $\rho_{VLL,VLR}^{lat}$ 
& $\rho_{VLL,SLR}^{lat}$ & $\rho_{VLL,SLL}^{lat}$ & $\rho_{VLL,VRR}^{lat}$ \\ 
\hline
 12.0 &  2 & 19.78(16) & 10.751(2) & 20.353(4) & -1.4375(4)  & 0.51217(9) \\ 
 10.0 &  2 & 19.10(9) & 10.480(2) & 19.588(4) & -1.6789(4)  & 0.51217(9) \\ 
  7.0 &  2 & 17.47(7) &  9.840(2) & 17.740(4) & -2.2483(5)  & 0.51217(9) \\ 
  6.5 &  2 & 17.07(16) &  9.687(2) & 17.292(4) & -2.3830(5)  & 0.51217(9) \\ 
  5.0 &  2 & 15.47(17) &  9.091(2) & 15.511(3) & -2.9032(6)  & 0.51217(9) \\ 
  4.5 &  2 & 14.74(16) &  8.828(2) & 14.705(3) & -3.1306(7)  & 0.51217(9) \\ 
  4.0 &  2 & 13.85(15) &  8.519(2) & 13.742(3) & -3.3948(7)  & 0.51217(9) \\ 
  3.8 &  2 & 13.46(16) &  8.381(2) & 13.301(3) & -3.5138(8)  & 0.51217(9) \\ 
  3.5 &  2 & 12.78(15) &  8.153(2) & 12.568(3) & -3.7104(8)  & 0.51217(9) \\ 
  3.0 &  2 & 11.44(12) &  7.710(2) & 11.103(3) & -4.0868(9)  & 0.51217(9) \\ 
  2.6 &  2 & 10.10(11) &  7.286(2) &  9.641(2) & -4.4438(9)  & 0.51217(9) \\ 
  2.1 &  3 &  8.26(11) &  6.894(1) &  7.778(2) & -4.9772(10) & 0.51217(9) \\ 
  1.5 &  3 &  4.76(8) &  6.059(2) &  3.902(1) & -5.7882(12) & 0.51217(9) \\ 
  1.3 &  3 &  3.27(8) &  5.773(2) &  2.154(2) & -6.0962(13) & 0.51217(9) \\ 
  1.2 &  3 &  2.47(8) &  5.642(2) &  1.155(1) & -6.2536(13) & 0.51217(9) \\ 
  0.9 &  4 &  0.06(7) &  5.596(1) & -2.098(1) & -6.6608(15) & 0.51217(9) 
    \end{tabular}
  \end{center}
  \caption{Matching coefficients $O^{lat}_{VLL,Y}$ for
    $c_{sw}$=1.
    The infrared divergence $-4\ln(a^2\lambda^2)$ is
    subtracted from $\rho_{VLL,VLL}^{lat}$.
    The tadpole improvement is applied such that the
    normalization of light quark field becomes
    $\sqrt{8\kappa_{crit}}$. 
    }
  \label{tab:rho_VLL}
\end{table}

\begin{table}
  \begin{center}
    \begin{tabular}{ccrrrrr} 
$aM$ & n & $\rho_{SLL,SLL}^{lat}$ & $\rho_{SLL,VLL}^{lat}$ 
& $\rho_{SLL,SLR}^{lat}$ & $\rho_{SLL,VLR}^{lat}$ & $\rho_{SLL,VRR}^{lat}$ \\ 
\hline
12.0 & 2 & 15.26(16) & -1.8396(16)& -8.3367(14) & -1.2359(3) & -0.12804(3) \\ 
10.0 & 2 & 15.50(9) & -1.7380(13)& -8.1851(13) & -1.1811(2) & -0.12804(3) \\ 
 7.0 & 2 & 16.09(7) & -1.4903(9) & -7.8368(15) & -1.0479(3) & -0.12804(3) \\ 
 6.5 & 2 & 16.24(16) & -1.4299(8) & -7.7565(15) & -1.0154(3) & -0.12804(3) \\ 
 5.0 & 2 & 16.81(17) & -1.1879(6) & -7.4509(14) & -0.8857(2) & -0.12804(3) \\ 
 4.5 & 2 & 17.05(16) & -1.0782(5) & -7.3204(13) & -0.8267(2) & -0.12804(3) \\ 
 4.0 & 2 & 17.35(15) & -0.9466(5) & -7.1714(13) & -0.7557(2) & -0.12804(3) \\ 
 3.8 & 2 & 17.50(16) & -0.8862(4) & -7.1061(13) & -0.7230(2) & -0.12804(3) \\ 
 3.5 & 2 & 17.71(15) & -0.7852(4) & -7.0013(12) & -0.6684(2) & -0.12804(3) \\ 
 3.0 & 2 & 18.15(12) & -0.5837(3) & -6.8068(11) & -0.5587(2) & -0.12804(3) \\ 
 2.6 & 2 & 18.59(11) & -0.3823(3) & -6.6347(11) & -0.4481(1) & -0.12804(3) \\ 
 2.1 & 3 & 19.56(11) & -0.1036(3) & -6.5966(10) & -0.2982(1) & -0.12804(3) \\ 
 1.5 & 3 & 20.93(8) &  0.4334(2) & -6.4940(10) &  0.0127(1) & -0.12804(3) \\ 
 1.3 & 3 & 21.57(8) &  0.6692(3) & -6.5593(10) &  0.1589(1) & -0.12804(3) \\ 
 1.2 & 3 & 21.96(8) &  0.8006(3) & -6.6362(10) &  0.2443(1) & -0.12804(3) \\ 
 0.9 & 4 & 23.95(7) &  1.2199(4) & -7.3536(9)  &  0.5465(1) & -0.12804(3) 
    \end{tabular}
  \end{center}
  \caption{Matching coefficients $O^{lat}_{SLL,Y}$ for
    $c_{sw}$=1.
    The infrared divergences $-\frac{4}{3}\ln(a^2\lambda^2)$
    and $\frac{2}{3}\ln(a^2\lambda^2)$ are subtracted from
    $\rho_{SLL,SLL}^{lat}$ and $\rho_{SLL,VLL}^{lat}$
    respectively. 
    The tadpole improvement is applied such that the
    normalization of light quark field becomes
    $\sqrt{8\kappa_{crit}}$. 
    }
      \label{tab:rho_SLL}
\end{table}

\begin{table}
\begin{center}
\begin{tabular}{cclllll}
$aM_0$ & n & $\zeta_{LL}$ & $\zeta_{LN}$ & $\zeta_{LM}$ &$\zeta_{LS}$ & $\zeta_{LR}$\\
\hline
 12.0 & 2 & -31.46(16) & -5.232(3) & -0.1441(0) & -6.572(2) & -0.51217(9) \\ 
 10.0 & 2 & -30.78(9) & -5.068(3) & -0.1720(1) & -6.326(2) & -0.51217(9) \\ 
  7.0 & 2 & -29.14(7) & -4.677(2) & -0.2426(1) & -5.755(2) & -0.51217(9) \\ 
  6.5 & 2 & -28.75(16) & -4.583(2) & -0.2604(1) & -5.623(2) & -0.51217(9) \\ 
  5.0 & 2 & -27.14(17) & -4.211(2) & -0.3343(1) & -5.105(2) & -0.51217(9) \\ 
  4.5 & 2 & -26.41(16) & -4.045(2) & -0.3693(1) & -4.878(1) & -0.51217(9) \\ 
  4.0 & 2 & -25.53(15) & -3.847(2) & -0.4126(1) & -4.613(2) & -0.51217(9) \\ 
  3.8 & 2 & -25.13(16) & -3.757(2) & -0.4330(1) & -4.494(2) & -0.51217(9) \\ 
  3.5 & 2 & -24.46(15) & -3.609(2) & -0.4676(1) & -4.299(2) & -0.51217(9) \\ 
  3.0 & 2 & -23.12(12) & -3.315(2) & -0.5400(2) & -3.923(2) & -0.51217(9) \\ 
  2.6 & 2 & -21.78(11) & -3.026(2) & -0.6169(2) & -3.564(2) & -0.51217(9) \\ 
  2.1 & 3 & -19.93(11) & -2.695(2) & -0.7515(2) & -3.029(2) & -0.51217(9) \\ 
  1.5 & 3 & -16.45(8) & -2.002(1) & -1.0267(3) & -2.221(2) & -0.51217(9) \\ 
  1.3 & 3 & -14.95(8) & -1.712(1) & -1.1739(3) & -1.913(2) & -0.51217(9) \\ 
  1.2 & 3 & -14.15(8) & -1.555(1) & -1.2660(3) & -1.756(2) & -0.51217(9) \\ 
  0.9 & 4 & -11.74(7) & -1.136(1) & -1.6612(4) & -1.351(3) & -0.51217(9)
\end{tabular}
\end{center}
\caption{Matching coefficients $\zeta$ for $O_{VLL}$}
\label{tab:zeta_L}
\end{table}

\begin{table}
\begin{center}
\begin{tabular}{ccll}
$aM_0$ & n &       $\zeta_A$  & $\zeta_P$  \\
\hline
12.0 &  2 & -14.59(8) &  -11.26(8)\\
10.0 &  2 & -14.23(5) &  -11.26(5)\\
 7.0 &  2 & -13.33(3) &  -11.29(3)\\
 6.5 &  2 & -13.11(8) &  -11.30(8)\\
 5.0 &  2 & -12.21(9) &  -11.34(9)\\
 4.5 &  2 & -11.79(8) &  -11.36(8)\\
 4.0 &  2 & -11.28(7) &  -11.40(7)\\
 3.8 &  2 & -11.05(8) &  -11.40(8)\\
 3.5 &  2 & -10.66(7) &  -11.42(7)\\
 3.0 &  2 &  -9.86(6) &  -11.45(6)\\
 2.6 &  2 &  -9.04(5) &  -11.51(5)\\
 2.1 &  3 &  -7.89(5) &  -11.89(5)\\
 1.5 &  3 &  -5.47(4) &  -12.27(4)\\
 1.3 &  3 &  -4.31(4) &  -12.49(4)\\
 1.2 &  3 &  -3.63(4) &  -12.67(4)\\
 0.9 &  4 &  -1.14(3) &  -13.66(3)
\end{tabular}
\end{center}
\caption{Matching coefficients $\zeta_{A,P}$ for $A_4,P$}
\label{tab:zeta_A,P}
\end{table}

\begin{table}
\begin{center}
\begin{tabular}{cclllll}
$aM_0$ & n & $\zeta_{SS}$ &$\zeta_{SL}$ & $\zeta_{SP}$ & $\zeta_{ST}$ & $\zeta_{SR}$ \\
\hline
 12.0 & 2 &  -5.26(16) & 3.3396(16)& 0.65405(10)& -0.014414(1) & 0.12804(3) \\ 
 10.0 & 2 &  -5.50(9) & 3.2380(13)& 0.63362(10)& -0.017204(1) & 0.12804(3) \\ 
  7.0 & 2 &  -6.09(7) & 2.9903(9) & 0.58467(9) & -0.024260(2) & 0.12804(3) \\ 
  6.5 & 2 &  -6.24(16) & 2.9299(8) & 0.57291(9) & -0.026043(2) & 0.12804(3) \\ 
  5.0 & 2 &  -6.81(17) & 2.6879(6) & 0.52643(8) & -0.033428(2) & 0.12804(3) \\ 
  4.5 & 2 &  -7.05(16) & 2.5782(5) & 0.50564(8) & -0.036930(2) & 0.12804(3) \\ 
  4.0 & 2 &  -7.35(15) & 2.4466(5) & 0.48095(7) & -0.041261(3) & 0.12804(3) \\ 
  3.8 & 2 &  -7.50(16) & 2.3862(4) & 0.46972(7) & -0.043297(3) & 0.12804(3) \\ 
  3.5 & 2 &  -7.71(15) & 2.2852(4) & 0.45114(7) & -0.046764(3) & 0.12804(3) \\ 
  3.0 & 2 &  -8.16(12) & 2.0837(3) & 0.41443(6) & -0.054001(3) & 0.12804(3) \\ 
  2.6 & 2 &  -8.59(11) & 1.8823(3) & 0.37830(6) & -0.061688(4) & 0.12804(3) \\ 
  2.1 & 3 &  -9.56(11) & 1.6036(3) & 0.33695(5) & -0.075151(5) & 0.12804(3) \\ 
  1.5 & 3 & -10.93(8) & 1.0666(2) & 0.25032(4) & -0.102676(7) & 0.12804(3) \\ 
  1.3 & 3 & -11.57(8) & 0.8308(3) & 0.21406(4) & -0.117394(8) & 0.12804(3) \\ 
  1.2 & 3 & -11.96(8) & 0.6994(3) & 0.19437(3) & -0.126601(9) & 0.12804(3) \\ 
  0.9 & 4 & -13.95(7) & 0.2801(4) & 0.14209(3) & -0.166128(11)& 0.12804(3)
\end{tabular}
\end{center}
\caption{Matching coefficients $\zeta$ for $O_{SLL}$}
\label{tab:zeta_S}
\end{table}

%
%
\begin{figure}
  \begin{center}
    \begin{tabular}{ccc}
      \raisebox{0em}{\psfig{file=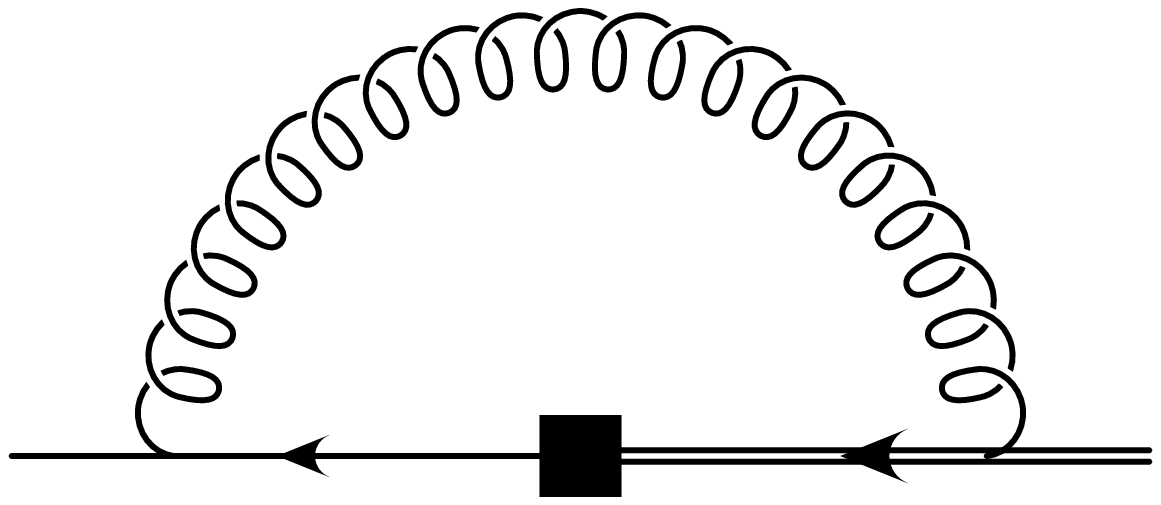,width=4.5cm}} &
      \raisebox{0em}{\psfig{file=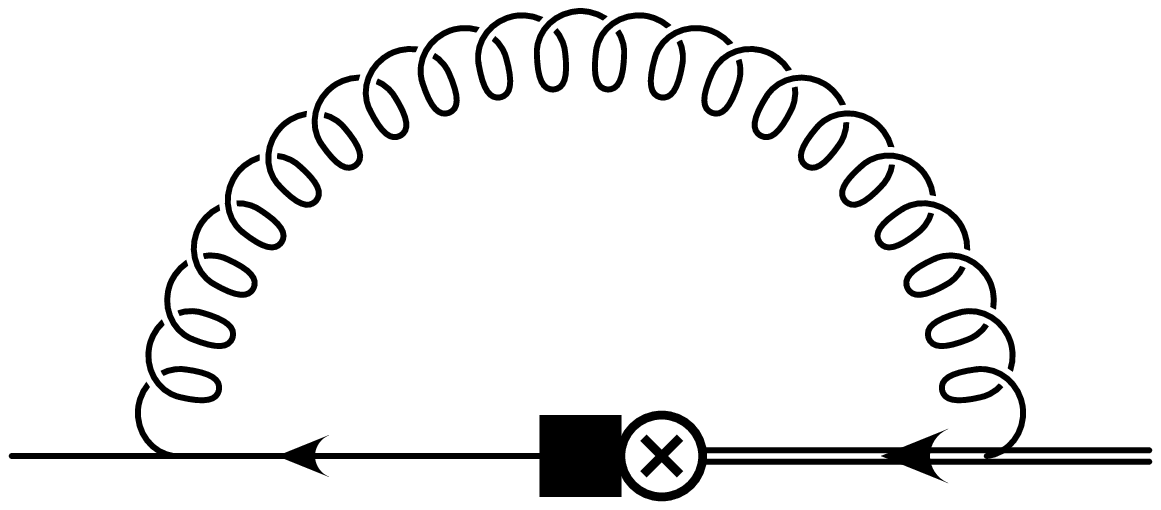,width=4.5cm}} &
      \raisebox{0em}{\psfig{file=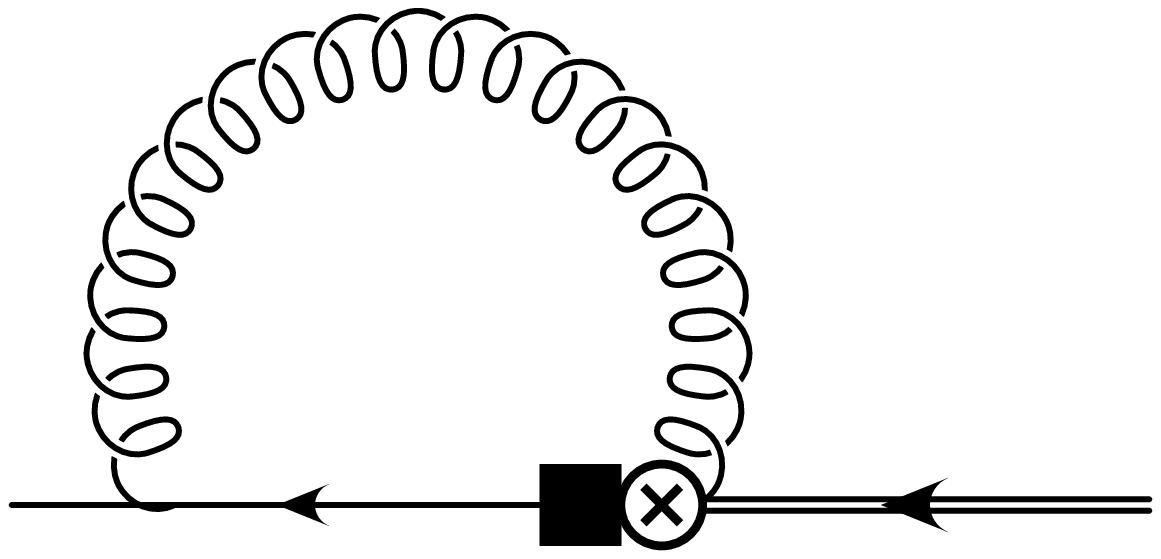,width=4.5cm}}
    \end{tabular}
  \end{center}
  \caption{vertex correction for the bilinear operator}
  \label{fig:vertex2}
\end{figure}

\clearpage

\begin{figure}
  \begin{center}
    \leavevmode\psfig{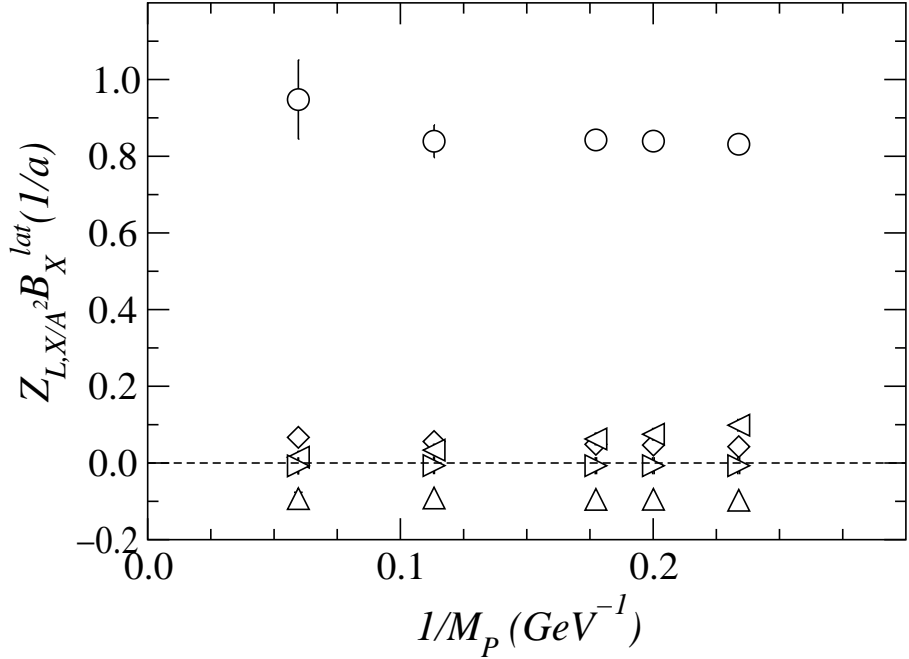}
    \caption{Contribution of individual operators
      $Z_{L,X/A^2} B_X^{lat}(1/a)$ to $B_L(\mu)$.
      Symbols are $X$ = $L$ (circle), $S$ (diamond), 
      $R$ (triangle right), $N$ (triangle up),
      $M$ (triangle left).
      }
    \label{fig:Z_L,X*B_X}
  \end{center}
\end{figure}

\begin{figure}
  \begin{center}
    \leavevmode\psfig{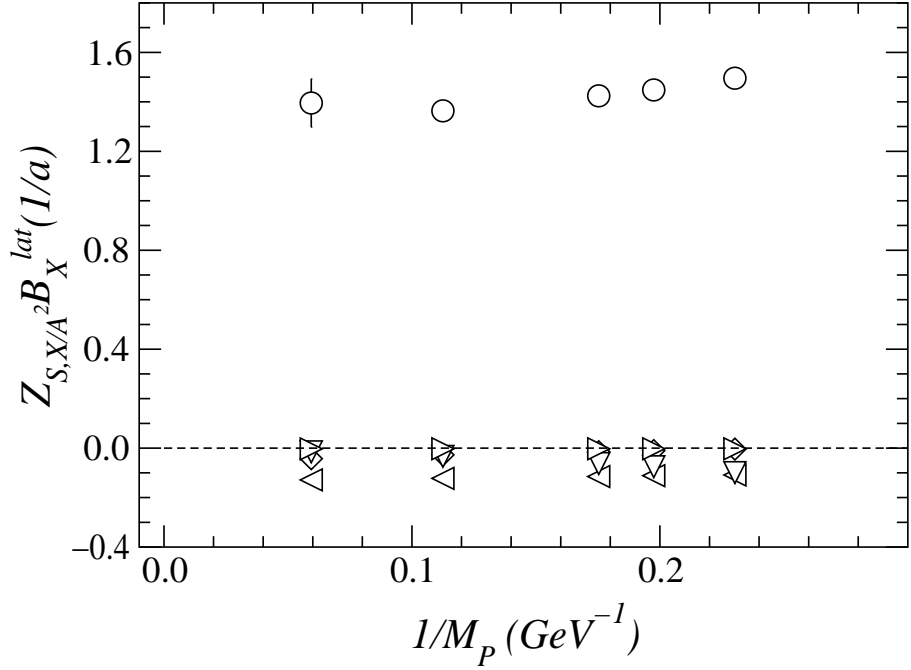}
    \caption{Contribution of individual operators
      $Z_{S,X/A^2} B_X^{lat}(1/a)$ to 
      $B_S(\mu)/{\cal R}(\mu)^2$.
      Symbols are $X$ = $S$ (circle), $L$ (diamond), 
      $R$ (triangle right), $P$ (triangle left),
      $T$ (triangle down).
      }
    \label{fig:Z_S,X*B_X}
  \end{center}
\end{figure}

\end{document}